\documentclass{aa}
\usepackage[varg]{txfonts}
\usepackage{epstopdf}
\usepackage{graphicx}
\usepackage{subfig}
\usepackage{natbib}
\usepackage{longtable}
\usepackage{array}
\usepackage{sidecap}
\usepackage[T1]{fontenc} 
\usepackage{textcomp}
\sidecaptionvpos{figure}{c}
\begin{document}
\title{A solar-like magnetic cycle on the mature K-dwarf 61 Cyg A (HD 201091)}
\titlerunning{61 Cyg A}
\author{S.~Boro Saikia\inst{\ref{inst1}}, S.~V.~Jeffers\inst{\ref{inst1}}, J.~Morin\inst{\ref{inst2}}, P.~Petit\inst{\ref{inst3}, \ref{inst4}}, C.~P.~Folsom\inst{\ref{inst5}}, S.~C.~Marsden\inst{\ref{inst6}}, J.-F.~Donati\inst{\ref
{inst3}, \ref{inst4}}, R.~Cameron\inst{\ref{inst7}}, J.~C.~Hall\inst{\ref{inst8}}, V.~Perdelwitz\inst{\ref{inst9}}, A.~Reiners\inst{\ref{inst1}} and A.~A.~Vidotto\inst{\ref{inst10},\ref{inst11}}
}
\authorrunning{Boro Saikia et al.}
\institute{Institut f\"ur Astrophysik, Universit\"at G\"ottingen, Friedrich Hund Platz 1, 37077 G\"ottingen, Germany\label{inst1}
\and LUPM-UMR 5299,CNRS \& Universit\'e Montpellier, Place Eug\'ene Bataillon, 34095 Montpellier Cedex 05, France \label{inst2}
\and CNRS, Institut de Recherche en Astrophysique et Plan\'etologie, 14 Avenue Edouard Belin, F-31400 Toulouse, France \label{inst3}
\and Universit\'e de Toulouse, UPS-OMP, Institut de Recherche en Astrophysique et Plan\'etologie, Toulouse, France \label{inst4}
\and IPAG, UJF-Grenoble 1/CNRS-INSU, UMR 5274, 38041 Grenoble, France \label{inst5}
\and Computational Engineering and Science Research Centre, University of Southern Queensland, Toowoomba 4350, Australia \label{inst6}
\and Max-Planck-Institut f\"ur Sonnensystemforschung,Justus-von-Liebig-Weg 3, 37077 G\"ottingen, Germany\label{inst7}
\and Lowell Observatory, 1400 West Mars Hill Road, Flagstaff, AZ 86001, USA \label{inst8}
\and Hamburger Sternwarte, Gojenbergsweg 112, 21029 Hamburg, Germany \label{inst9}
\and Observatoire de Gen\`eve, Universit\'e de Gen\`eve, Chemin des Maillettes 51, Sauverny, CH-1290, Switzerland \label{inst10}
\and School of Physics, Trinity College Dublin, The University of Dublin, Dublin-2, Ireland \label{inst11}
}
\abstract{The long-term monitoring of magnetic cycles in cool stars is a key diagnostic in understanding how dynamo generation and amplification of magnetic fields occur in stars similar in structure to the Sun.}
{To investigate the temporal evolution of a possible magnetic cycle, determined from its large-scale field, of 61 Cyg A over its activity cycle using spectropolarimetric observations and compare it to the solar case.} {We use the 
tomographic technique Zeeman Doppler imaging (ZDI) to reconstruct the large-scale magnetic geometry of 61 Cyg A over multiple observational epochs spread over a time span of nine years. We investigate the time evolution of the different 
components of the large-scale field and compare it with the evolution of its chromospheric activity by measuring the flux in three different chromospheric indicators: Ca II H$\&$K, H$\alpha$ and Ca II infra red triplet lines. We also
compare our results with the star's coronal activity using \textit{XMM-Newton} observations.}{The large-scale magnetic geometry of 61 Cyg A exhibits polarity reversals in both poloidal and toroidal field components, in phase with the 
chromospheric activity cycle. We also detect weak solar-like differential rotation with a shear level similar to the Sun. During our observational time span of nine years, 61 Cyg A exhibits solar-like variations in its large-scale field 
geometry as it evolves from minimum activity to maximum activity and vice versa. During its activity minimum in epoch 2007.59, ZDI reconstructs a simple dipolar geometry which becomes more complex close to activity maximum in epoch 
2010.55. The radial field flips polarity and reverts back to a simple geometry in epoch 2013.61. The field is strongly dipolar and the evolution of the dipole component of the field is reminiscent of the solar behaviour. The polarity 
reversal of the large-scale field indicates a magnetic cycle that is in phase with the chromospheric and coronal cycle.}{}
\maketitle
\section{Introduction}
High resolution observations of the Sun, including over 30 years of synoptic magnetic field maps, have revealed a coherent picture of how the large-scale magnetic field evolves over the course of the 22-year solar magnetic cycle.
At minimum the field is concentrated at the polar regions and is mostly axisymmetric with respect to the rotation axis. At activity maximum the field is no longer concentrated at the polar regions and is restricted to lower latitudes 
\citep{hathaway10}. When the next minimum occurs the polar field reappears but with opposite sign to what was observed in the previous minimum at the start of the cycle. This cyclic process repeats every 11 years and it takes 22 years 
for the polar magnetic fields to revert back to the same polarity, forming a 22 year magnetic cycle. During the course of the solar magnetic cycle  the geometry of the large-scale field of the Sun also changes dramatically. It can be 
seen that the Sun varies from an almost dipolar geometry during cycle minimum to a more complex geometry during cycle maximum \citep{sanderson03, derosa12}. A recent study by \citet{ derosa12} has revealed that during cycle minimum the 
dipolar component of the large-scale solar field dominates over the other components. During polarity reversals the quadrupolar and octopolar component of the large-scale field dominates over the dipolar component.
\paragraph{}
When it comes to other solar-type stars, observations do not have the spatial and temporal resolution of the solar observations. Magnetic cycles in other cool stars have been investigated by using two well known proxies of 
magnetic activity such as chromospheric activity \citep{baliunas95, hall08} and coronal activity \citep{pevtsov03, gudel04}. The first long-term monitoring of chromospheric activity in solar-type dwarfs was carried out at the Mount 
Wilson observatory \citep{wilson78,duncan91,baliunas95}, where emission in the line cores of Ca II H$\&$K lines was measured. These observations revealed that different solar-type stars tend to exhibit different levels of 
activity variation: irregular activity variations in fast rotating young stars, cyclic activity in comparatively older slowly rotating solar-type stars and Maunder minimum like flat activity in some stars \citep{baliunas95}. A further 
analysis on the Maunder-minimum candidates from the Mt Wilson survey by \citet{schroeder13} reveals them to be slightly more evolved than the Sun. Coronal activity of cool stars, on the other hand is not intensively monitored compared to
their chromospheric activity. Since the detection of weak statistical evidence of coronal cycles in cool stars by \citet{hempelmann96}, coronal activity cycles have been detected for a few selected G and K dwarfs. For example 
\citet{favata08} reported the presence of a coronal activity cycle for the binary system HD81809. Cyclic X-ray activity was also detected for our target star 61 Cyg A \citep{robrade12} and the K1 dwarf $\alpha$ Cen B 
\citep{robrade12,ayres15}. The shortest coronal activity cycle of 1.6 years was reported for the young active Sun $\iota$ Horologii \citep{sanz-forcada13}, where the coronal activity cycle is in phase with its short chromospheric 
activity cycle.
\paragraph{}
Although these indirect proxies are reliable indicators of magnetic activity in cool stars, they do not provide any direct information on the strength or orientation of the large-scale magnetic field. Direct measurements of the surface 
magnetic field rely on the Zeeman effect. By measuring the broadening of spectral lines due to Zeeman effect on unpolarised spectra \citep{robinson80,saar96,reiners06}, the integrated unsigned magnetic flux averaged over the entire disk 
of the star can be measured. One advantage of this technique is that the small-scale magnetic features also contribute to the total field measurements. This technique does not provide information on the magnetic field geometry.
On the other hand, the tomographic technique Zeeman Doppler imaging (ZDI) \citep{semel89, donati97} uses spectropolarimetric observations to reconstruct the stellar surface magnetic field geometry. Due to cancellation effects this 
technique is insensitive towards small-scale multipolar features, as the overall polarisation signatures of these multipolar field cancel out and only the large-scale field is reconstructed. As the ZDI technique provides 
information about the vector magnetic field, it can provide invaluable insights into the large-scale field geometry as well as the temporal evolution and polarity reversals of the large-scale field.   
\paragraph{}
The large-scale magnetic field of several solar-type stars reconstructed using ZDI over multiple epochs, has revealed their strongly varying magnetic geometry. For the G7 dwarf $\xi$ Bootis A (0.86 $M_\sun$, $T_\mathrm{eff}$ = 5551 K) 
\citep{morgenthaler12}, the K2 dwarf $\epsilon$ Eridani (0.856 $M_\sun$, $T_\mathrm{eff}$ = 5146 K) \citep{jeffers14}, and the G0 dwarf HN Peg (1.085 $M_\sun$, $T_\mathrm{eff}$ = 5974 K) \citep{borosaikia15} rapidly varying field 
geometry was detected, with surprising appearance and disappearance of the azimuthal field for HN Peg and $\epsilon$ Eridani. A two year magnetic cycle was revealed for the planet hosting F7 dwarf $\tau$ Bootis (1.42 $M_\sun$, $T_\mathrm
{eff}$ = 6360 K) \citep{donati08, fares09}. $\tau$ Boo is a hot Jupiter host where the planet is orbiting at 0.049 AU. No correlation was detected between the large-scale polarity switch and its chromospheric activity 
\citep{fares09}. Further monitoring of $\tau$ Boo over three epochs by \citet{fares13} has confirmed the previously determined 2 year magnetic cycle. Recent new analysis of $\tau$ Boo polarised spectra \citet{mengel16} shows
a 3:1 ratio between the magnetic cycle and chromospheric cycle. A magnetic cycle was also detected for the G0 dwarf HD 78366 (1.34 $M_\sun$,$T_\mathrm{eff}$ =  6014 K) \citep{morgenthaler11}, where the radial component of the magnetic 
field exhibits polarity reversals indicating a possible 3-year cycle. Polarity reversals were observed in the large-scale field of the G2 dwarf HD 190771 (0.96 $M_\sun$, $T_\mathrm{eff}$ = 5834 K) \citep{petit09,morgenthaler11}. 
However, the variability of the large-scale field of HD 190771 is more complex as the polarity reversal in the azimuthal field reported by \citet{petit09} is not detected in subsequent epochs. Instead polarity reversal is reported in 
the radial component of the magnetic field in the following epochs as shown by \citet{morgenthaler11}. Polarity reversals of the large-scale field have been detected only for a select few cool stars, out of which none exhibit a magnetic 
cycle period equivalent to the star's chromospheric cycle period. 
\paragraph{}
This paper investigates the variability of the large-scale magnetic field geometry of the solar-type K5 dwarf 61 Cyg A, using a time series of spectropolarimetric observations over 9 years. We also investigate the correlation between 
its mean magnetic field, chromospheric activity, and coronal activity. The stellar parameters of 61 Cyg A are discussed in Section 2, followed by instrumental setup and data reduction in Section 3. Section 4 covers both direct and 
indirect field detection. The large-scale magnetic geometry is discussed in Section 5, followed by the long-term evolution of the magnetic and activity cycle in Section 6. The results are discussed in Section 7 and finally the Summary in Section 8.
\begin{table}
     \caption{Summary of the physical properties of 61 Cyg A. } 
\label{table:1}  
\begin{tabular}{l l c }
\hline\hline                        
Parameters &  HD 201091 &References\\    
\hline                                   
    Effective temperature, {$T_\text{eff}$}(K)& 4545$\pm$40& this work \\
    Spectral type & K5V & 1 \\
    log \textit{g} & 4.75$\pm$0.10 & this work\\
    Radius, $R_\sun$& 0.665$\pm$0.005 & 1\\
    Mass, $M_\sun$& 0.66  & 1\\
    Rotation period, days&35.7$\pm$1.9&this work\\
    ..& 34.2$\pm$3.7 &this work\\
    inclination&70$^{\circ}$&this work\\
    B-V &1.069& 2\\
    Age, Gyr & 6.0 &3\\
    ..& 3.6&4\\
    ..&2.0&5\\
    ..&1.331 &6\\
     \hline                                             
\end{tabular}
\tablebib{
(1)~\citet{kervella08}; (2)~\citet{perryman97}; (3)~\citet{kervella08}; (4)~\citet{mamajek08}; (5)~\citet{barnes07}; (6)~\citet{marsden14}.
The rotation period of 35.7$\pm$1.9 is derived from chromospheric activity measurements and the period of 34.2$\pm$3.7 is derived from ZDI (See text for more details).}
\end{table}
\section{ Physical properties of 61 Cyg A}
A solar-type dwarf of spectral type K5, 61 Cyg A forms the well known visual binary 61 Cyg together with the K7 dwarf 61 Cyg B. The binary 61 Cyg is a northern system with a semi-major axis of approximately 24$''$ 
\citep{malkov12}. It is the first stellar object to have its parallax measured \citep{bessel38}. 61 Cyg A has a parallax of 287.13$\pm$1.51 mas \citep{perryman97} and is at a distance of approximately 3.5 pc. It is a slow rotator 
with a rotation period of 35.7$\pm$1.9 days as shown in Table \ref{table:1}, where the rotation period is obtained from the long-term chromospheric activity measurements. We also derive a rotation period of 34.2 $\pm$ 3.7 days using ZDI 
(See Section 5.3 for more details). 61 Cyg A has a radius of 0.665$\pm$0.005 $R_\sun$ \citep{kervella08}, where the radius was determined using interferometric data and is in agreement with the radius of 0.62 $R_\sun$ \citep{takeda07} 
calculated by using a stellar evolution code. Using the same code and stellar evolutionary tracks \citet{takeda07} also determined a mass of 0.660 $M_\sun$. The angular diameter of 1.775$\pm$0.013 mas \citep{kervella08} and the 
bolometric flux of 0.3844$\times10^{-9}$ \citep{mann13} can be used to determine a $T_\mathrm{eff}$ of 4374 K. Applying to our highest S/N spectra our automatic spectral classification tool inspired from that of \citet{valentifischer05}
and discussed in a previous paper \citep{donati12}, we find that the photosphere temperature, log \textit{g}, and metallicity of 61 Cyg A are respectively equal to 4545$\pm$40 K, 4.75$\pm$0.10, and -0.18$\pm$ 0.05. 
The $T_\mathrm{eff}$ from our analysis agrees with the literature value of 4525$\pm$140 K \citep{takeda07} with improved accuracy. Using the radius, rotation period from Table \ref{table:1} we determine an equatorial 
$v_e$ of 0.94 kms $^{-1}$. This suggests that the $vsin i$ of 61 Cyg A should be $\leq$ 0.94 kms$^{-1}$. A chromospheric age of 1.331 Gyr was determined by \citet{marsden14} from the activity age relationship of \citet{wright04}. 
Using chromospheric activity and rotation relation \citet{mamajek08} determine an age of 3.6 Gyr. On the other hand, \citet{kervella08} predicted an age of approximately 6 Gyr using evolutionary models combined with interferometric 
radius measurements. The age estimates of 61 Cyg A from gyrochronology puts its age to approximately 2 Gyr \citep{barnes07}. The stellar parameters of 61 Cyg A are summarised in Table \ref{table:1}.
\paragraph{}
61 Cyg A is a moderately active star and was included in the long-term Mount Wilson survey \citep{duncan91}. It was discovered to exhibit an activity cycle of 7.3 $\pm$ 0.1 years \citep{baliunas95}. 61 Cyg A was also included in the 
Lowell Observatory long-term survey, where it was classified as a star with solar type variability \citep{hall07,lockwood07}. A coronal activity cycle of approximately 7 years has also been detected for 61 Cyg A \citep{robrade12} using 
X-ray observations, where the coronal activity cycle is in phase with the chromospheric activity cycle.

\section{Instrumental setup and data reduction}
\subsection{Optical data}
The spectropolarimetric data was collected as part of the BCool collaboration \footnote{\url{http://bcool.ast.obs-mip.fr/Bcool}} using the NARVAL spectropolarimeter, at the 2.0 m Telescope Bernard Lyot at Pic du Midi Observatory.
NARVAL is equipped with a cross dispersed echelle spectrograph, which is a twin of the EsPaDOnS spectropolarimeter \citep{donati_esp03} at the 3.6 m Canada-France-Hawaii telescope (CFHT). It covers the full optical wavelength range 
from 370 nm to 1000 nm, with a resolving power of approximately 65,000 \citep{auriere03}. The spectropolarimetric observations of 61 Cyg A were obtained by combining four successive sub-exposures, taken with different half wave rhomb 
angles. The data is reduced on-site by using a fully automated reduction package LIBRE-ESPRIT which generates both intensity (Stokes \textit{I}) and circularly polarised (Stokes \textit{V}) spectra. LIBRE-ESPIRIT is based on the reduction pipeline 
ESPIRIT developed by \citet{donati97}. The reduction package also calculates a diagnostic Null profile, which is a control profile, by combining the four sub-exposures in such a way that any polarisation signal is cancelled out. The 
output from the reduction pipeline is continuum normalised. Table \ref{kstars} shows the journal of observations of 61 Cyg A.
\begin{longtab}
\begin{longtable}{cccccc}
\caption{\label{kstars} Journal of observations for seven epochs(2007.59, 2008.64, 2010.55, 2012.54, 2013.61, 2014.61 and 2015.54). Column 1 represents the year and date of observations, column 2 is the Julian date, 
column 3 is the exposure time, column 4 is the signal-to-noise ratio of each Stokes V LSD profile and column 5 represents the error bars in Stokes V LSD profile.}\\
\hline\hline
Date & Julian date & Exposure time &S/N &$\sigma_\mathrm{LSD}$\\
&(2450000+)& (s)& & (10$^{-5} \mathrm {I_{c}} $) \\
\hline
\endfirsthead
\caption{continued.}\\
\hline\hline

Date & Julian date &Exposure time & S/N&$\sigma_\mathrm{LSD}$\\
&(2450000+)&(s)&&(10$^{-5} \mathrm{I_{c}}$) \\
\hline
\endhead
\hline
\endfoot
2007 July 26  &4308.49809&1200&35765&2.7960\\
2007 July 30  &4312.52989&800&35807&2.7928\\
2007 July 31  &4313.53121&800&27096&3.6905\\
2007 August 02&4315.53608&800&33873&2.9522\\
2007 August 03&4316.53315&800&35102&2.8489\\
2007 August 04&4317.53462&800&30572&3.2710\\
2007 August 08&4321.50738&800&31550&3.1696\\
2007 August 09&4322.52760&800&28744&3.4790\\
2007 August 10&4323.52773&800&24362&4.1048\\
2007 August 17&4330.50089&1600&48802&2.0491\\
2007 August 18&4331.46147&800&35561&2.8121\\
\hline
2008 August 09&4688.54994&1200&46210&2.1641\\
2008 August 12&4691.48653&1200&40079&2.4951\\
2008 August 17&4696.52126&1200&31218&3.2033\\
2008 August 21&4700.47785&1200&36751&2.7210\\
2008 August 22&4701.48433&1200&21527&4.6454\\
2008 August 23&4702.50793&1200&50016&1.999\\
2008 August 24&4703.47966&1200&49602&2.0161\\
2008 August 25&4704.53225&1200&48263&2.0720\\
2008 August 26&4705.52953&1200&49904&2.0039\\
\hline
2010 June 03&5351.64786&900&32082&3.1171\\
2010 June 21&5369.57210&900&31021&3.2236\\
2010 July 01&5379.58460&900&27037&3.6987\\
2010 July 12&5390.54209&900&25343&3.9458\\
2010 July 13&5391.52388&900&26541&3.7678\\
2010 July 14&5392.54747&900&34955&2.8608\\
2010 July 15&5393.63958&900&31139&3.2114\\
2010 July 18&5396.60809&900&38824&2.5757\\
2010 July 23&5401.65920&900&37562&2.6623\\
2010 July 24&5402.49800&900&37471&2.6687\\
2010 July 25&5403.49549&900&37945&2.6354\\
2010 August 02&5411.49161&900&35432&2.8224\\
2010 August 03&5412.56953&900&40189&2.4883\\
2010 August 06&5415.55556&900&32469&3.0799\\
2010 August 07&5416.54320&900&36301&2.7548\\
2010 August 10&5419.56875&900&36220&2.7610\\
\hline
2012 July 08&6117.63676&900&33129&..\\
2012 July 09&6118.63476&900&28811&..\\
2012 July 15&6124.64113&900&35051&..\\
2012 July 16&6125.61827&900&36914&..\\
2012 July 17&6126.61222&900&34327&..\\
2012 July 18&6127.56778&900&34221&..\\
2012 July 19&6128.55829&900&32787&..\\
2012 July 22&6131.55361&900&34150&..\\
2012 July 23&6132.54263&900&35066&..\\
2012 July 24&6133.56659&900&34955&..\\
\hline
2013 July 14&6488.51321&800&15411&6.4889\\
2013 August 02&6507.65889&800&28131&3.5548\\
2013 August 04&6509.51420&800&34901&2.8653\\
2013 August 05&6510.61746&800&24113&4.1472\\
2013 August 08&6513.57114&800&31163&3.2090\\
2013 August 09&6514.54087&800&35398&2.8251\\
2013 August 10&6515.54442&800&35718&2.7997\\
2013 August 11&6516.56402&800&33745&2.9634\\
2013 August 13&6518.58190&800&36075&2.7720\\
2013 August 14&6519.62914&800&35479&2.8186\\
2013 August 18&6523.51294&800&33442&2.9903\\
2013 August 19&6524.49914&800&34439&2.9036\\
2013 August 20&6525.44457&800&33347&2.9988\\
\hline
2014 July 14&6853.55493&800&21642&4.6207\\
2014 July 17&6856.55743&800&28094&3.5596\\
2014 July 22&6861.57410&800&34595&2.8906\\
2014 July 25&6864.55546&800&34374&2.9092\\
2014 July 26&6865.53839&800&35150&2.8449\\
2014 July 27&6866.58350&800&35645&2.8055\\
2014 August 11&6881.38890&800&27850&3.5907\\
2014 August 19&6889.36216&800&18535&5.3954\\
2014 August 23&6893.46397&800&35628&2.8068\\
2014 August 25&6895.46092&800&29424&3.3987\\
2014 August 27&6897.47963&800&36306&2.7544\\
2014 August 29&6899.49512&800&35859&2.7888\\
2014 August 30&6900.52575&800&31814&3.1433\\
2014 August 31&6901.48870&800&36408&2.7467\\
\hline
2015 June 26&7200.57599&900&26403&3.7874\\
2015 June 28&7202.53874&900&36708&2.7242\\
2015 June 29&7203.59939&900&30539&3.2745\\
2015 June 30&7204.56304&900&30669&3.2607\\
2015 July 07&7211.52277&900&31832&3.1415\\
2015 July 08&7212.52460&900&34962&2.8603\\
2015 July 09&7213.50574&900&37205&2.6878\\
2015 July 10&7214.52800&900&33875&2.9521\\
2015 July 11&7215.56071&900&32176&3.1080\\
2015 July 12&716.54418&900&34501&2.8985\\
2015 July 14&7218.55681&900&32888&3.0407\\
2015 July 15&7219.56343&900&23679&4.2233\\
2015 July 20&7224.56050&900&25336&3.9469\\
2015 August 05&7240.48208&900&24468&4.0871\\
2015 August 10&7245.54843&900&22931&4.3610\\
2015 August 11&7246.57034&900&31250&3.2000\\
\hline
 \end{longtable}
\end{longtab}
\paragraph{}
Seven data sets comprising seven epochs, spanning nine years of both Stokes \textit{I} and V spectra, were obtained covering epochs 2007.59, 2008.64, 2010.55, 2012.54, 2013.61, 2014.61, and 2015.54. Each epoch 
contains between 9 and 16 spectra. The spectropolarimetric data from the 2012.54 epoch was discarded due to polarisation anomalies in one of the polarisation rhombs at NARVAL during 2011-2012 \footnote{\url{http://spiptbl.bagn.obs-mip.fr
/Actualites/Anomalies-de-mesures}}. The azimuthal angle of the rhomb was wrong during that epoch leading to incorrect polarisation measurements.  
\paragraph{}
For active cool stars, Zeeman-induced circular polarisation has too low an amplitude to be detected in individual spectral lines. Hence the technique of least square deconvolution (LSD) is applied to the spectra of 61 Cyg A. LSD is 
a multiline technique which assumes a similar line profile for all magnetically sensitive lines in a spectra and generates an averaged line profile by deconvolving the stellar spectra to a line mask \citep{donati97, kochukhov10}.
We used a K5 mask consisting of approximately 12,000 lines to compute the LSD Stokes \textit{I} and Stokes \textit{V} profiles of 61 Cyg A. The LSD technique used in this paper is detailed in \citet{marsden14}. 
\subsection{X-ray data}
61 Cyg A is a strong X-ray source and has been continuously monitored with {\it XMM-Newton} since 2002 with a cadence of two observations a year (see e.g. \cite{hempelmann06} and \cite{robrade12}), so in order to check 
for periodicity in X-ray luminosity we obtained all public data sets from the {\it XMM-Newton} Science Archive. Data reduction was performed with the {\it XMM-Newton} Science Analysis System version 10.0 and XSPEC 12.7.1. In order to 
avoid contamination from 61 Cyg B source photons were extracted with a 15$''$ aperture, while background levels were estimated with a 45$''$ aperture in source-free regions. Time intervals during which obvious flaring occurred were 
omitted. The spectra of all three EPIC detectors (MOS 1\&2 and PN) were then fitted simultaneously in the 0.2-2.0 keV range with a three-temperature APEC model.
\begin{figure*}
  \includegraphics[scale=0.5]{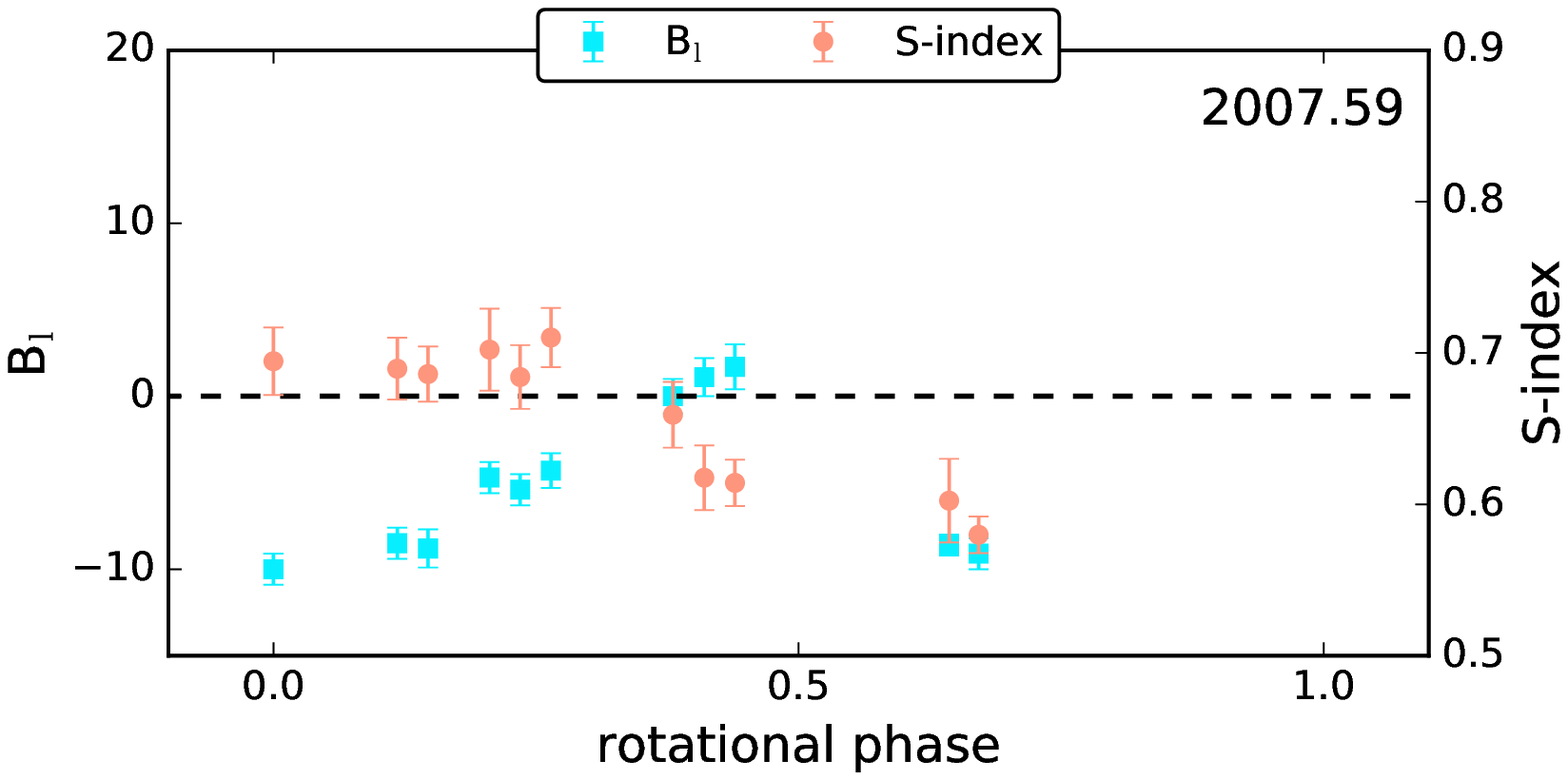}~~~\includegraphics[scale=0.5]{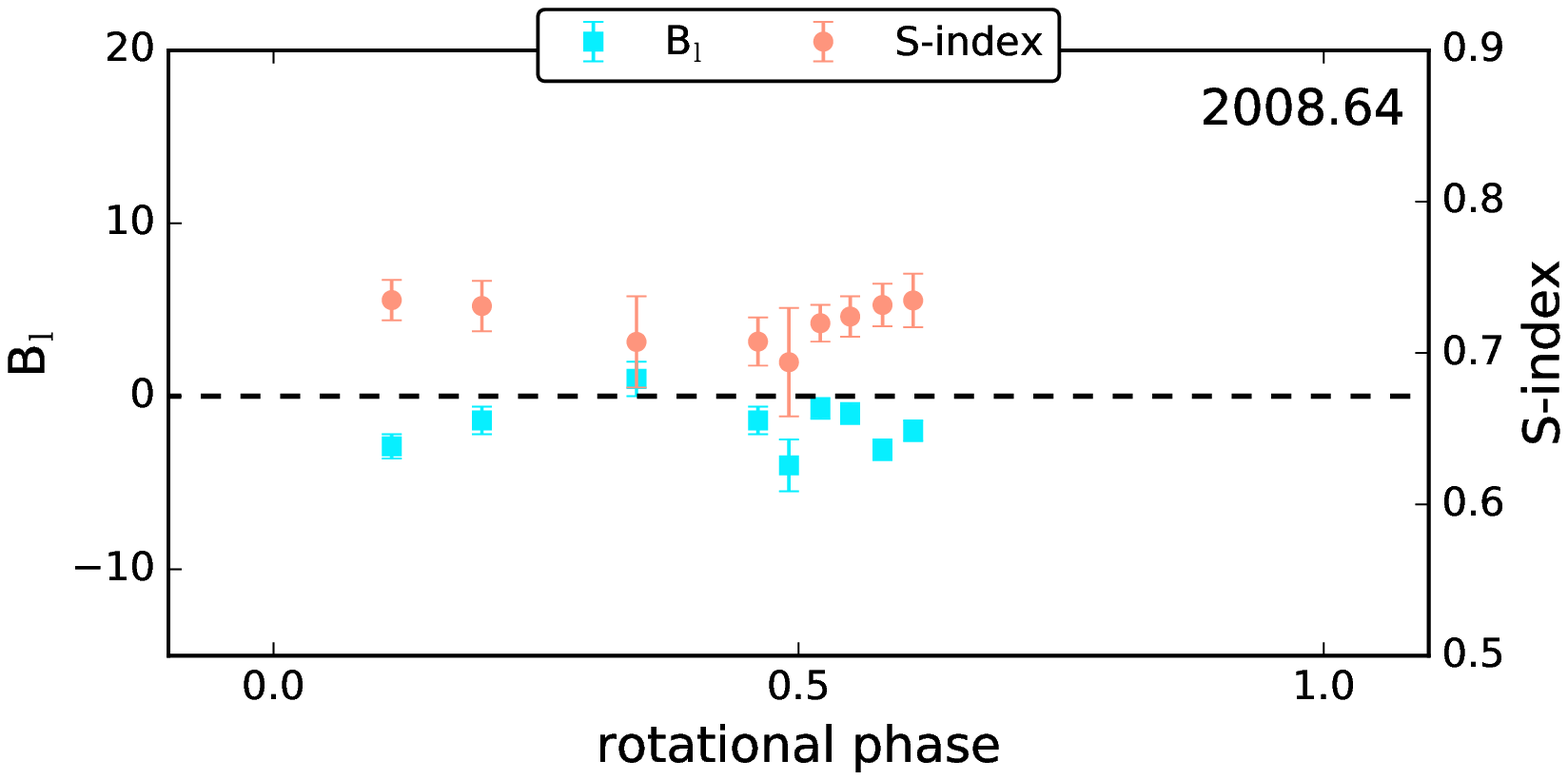}\\[5mm]
  \includegraphics[scale=0.5]{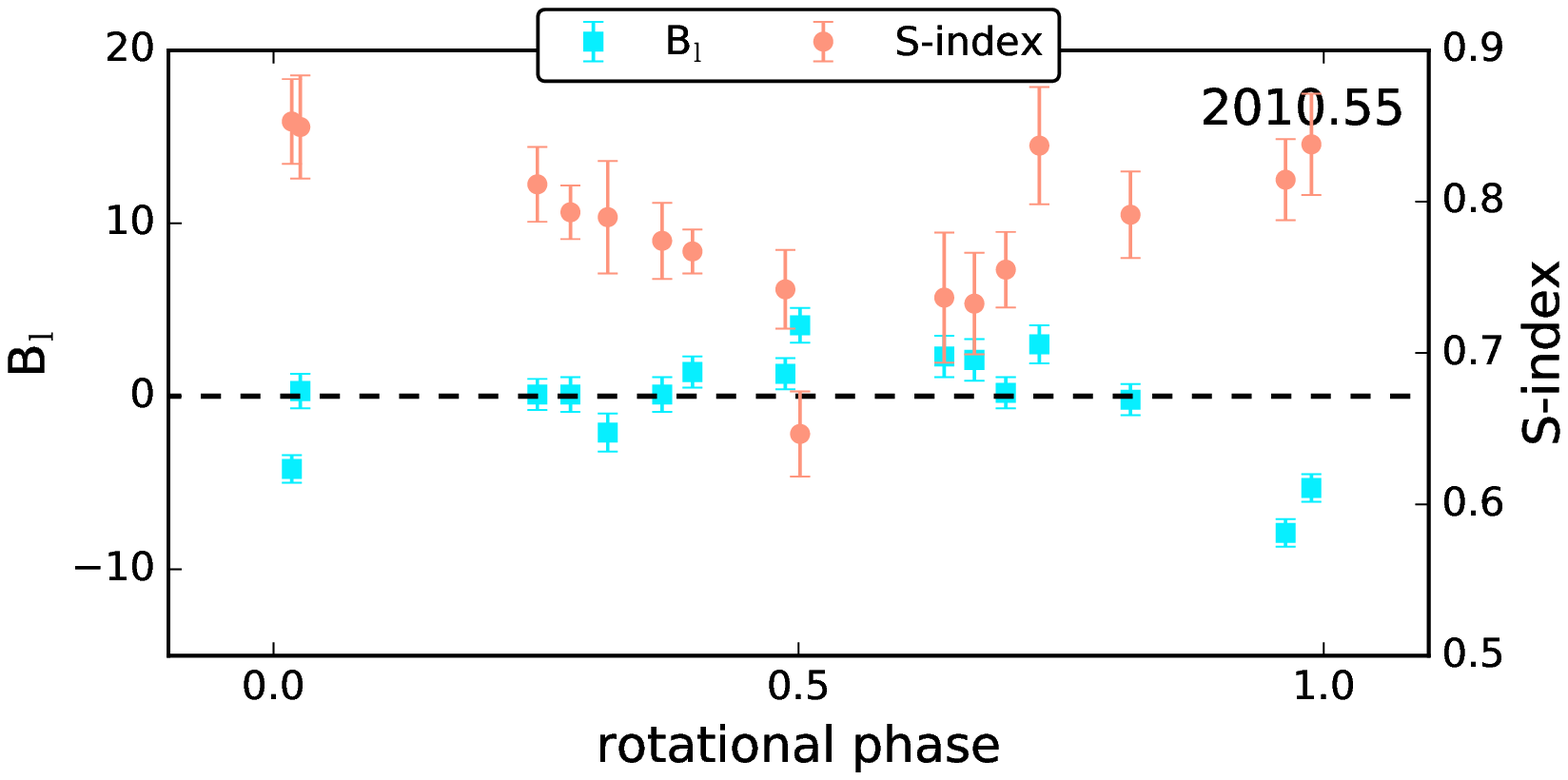}~~~\includegraphics[scale=0.5]{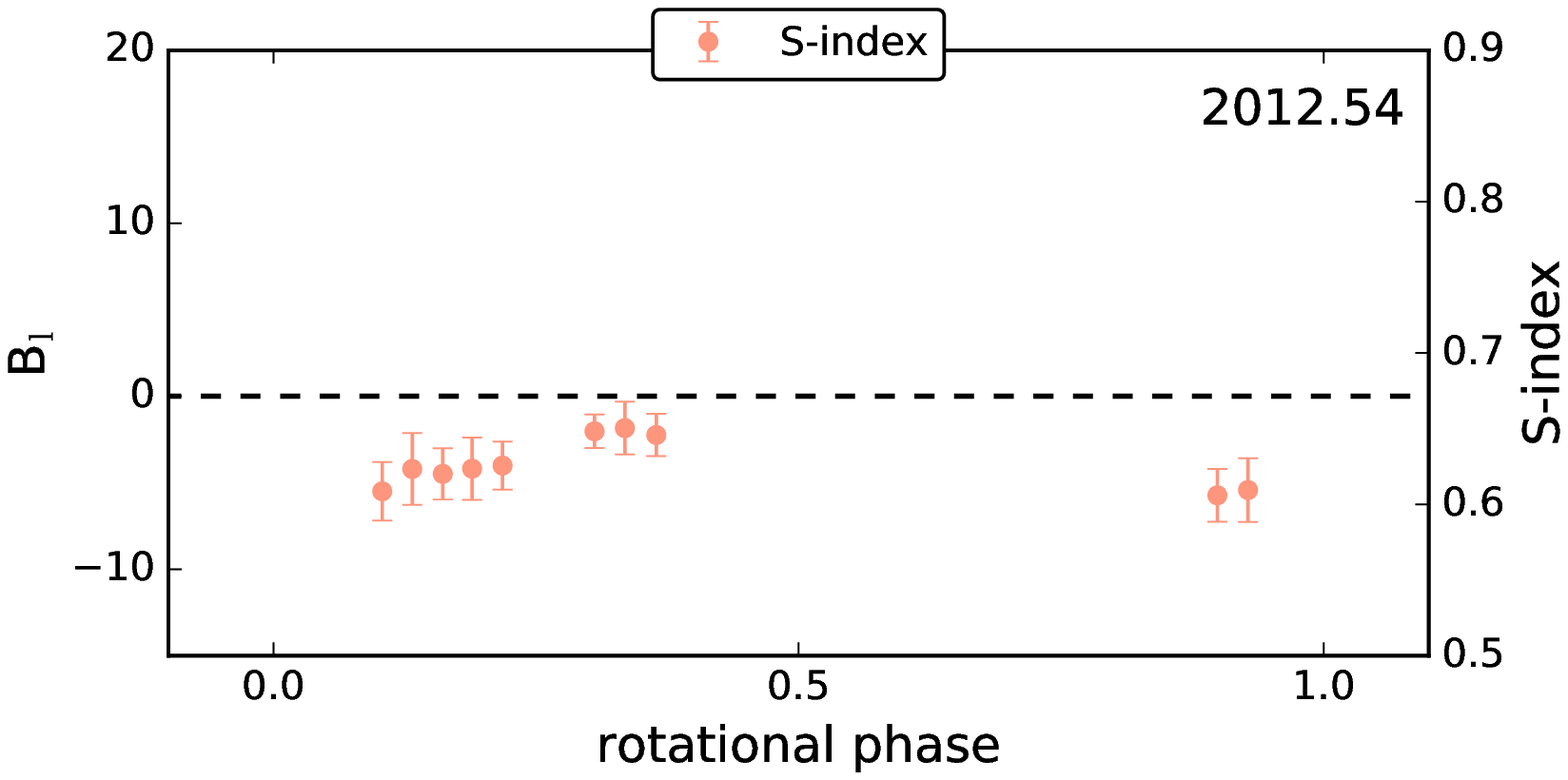}\\[5mm]
  \includegraphics[scale=0.5]{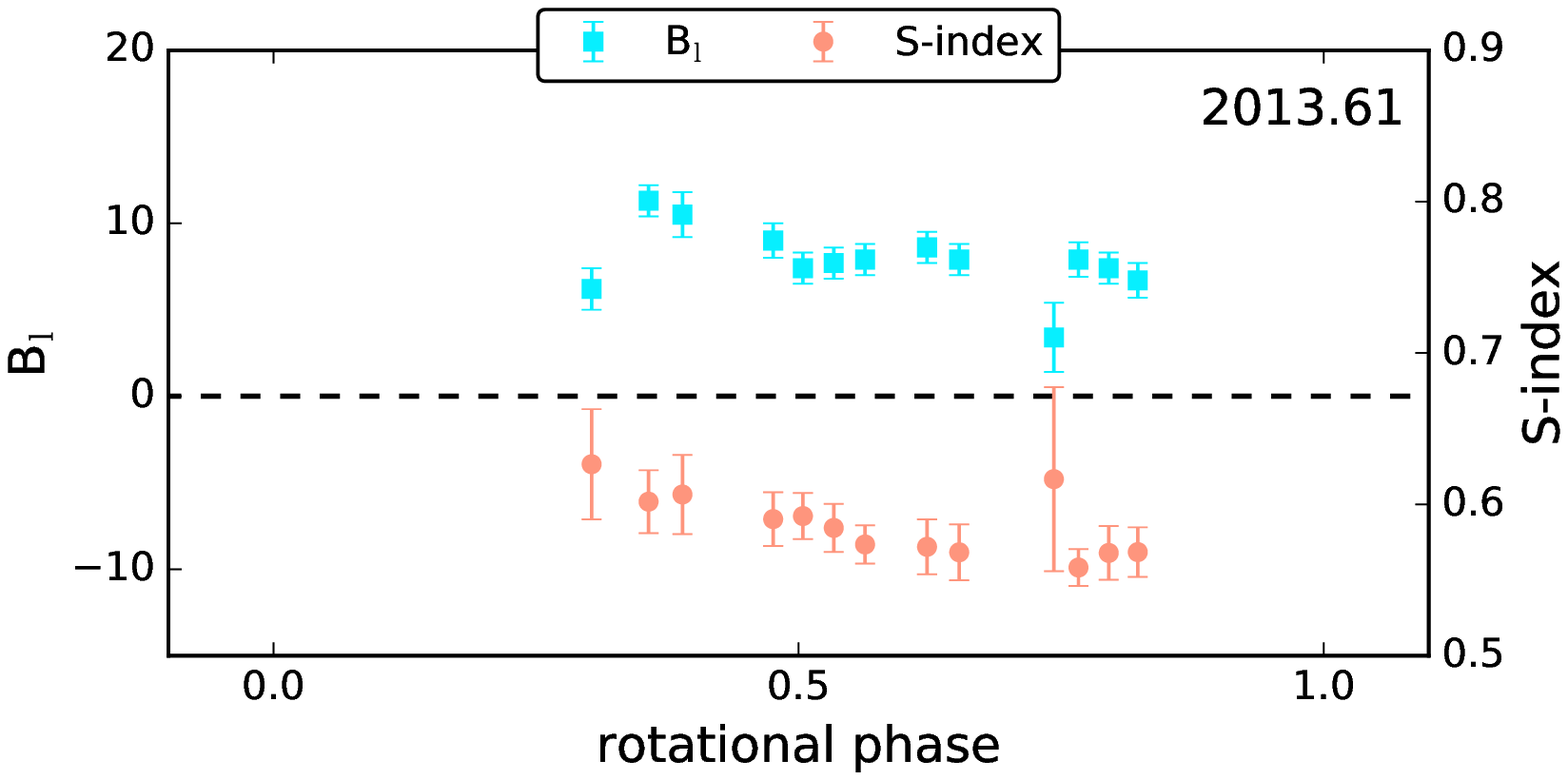}~~~\includegraphics[scale=0.5]{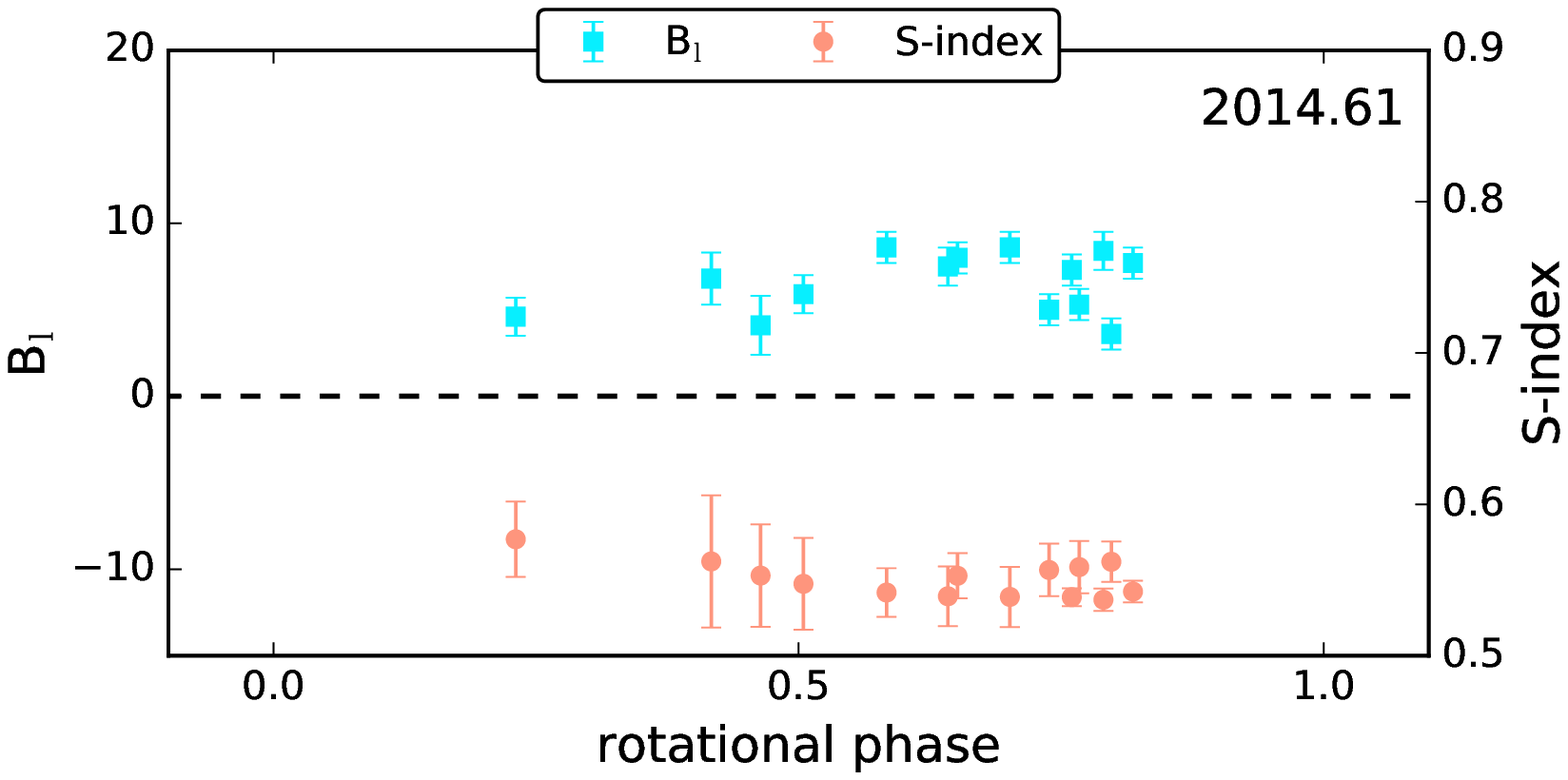}\\[5mm]
\raggedright{
\includegraphics[scale=0.5]{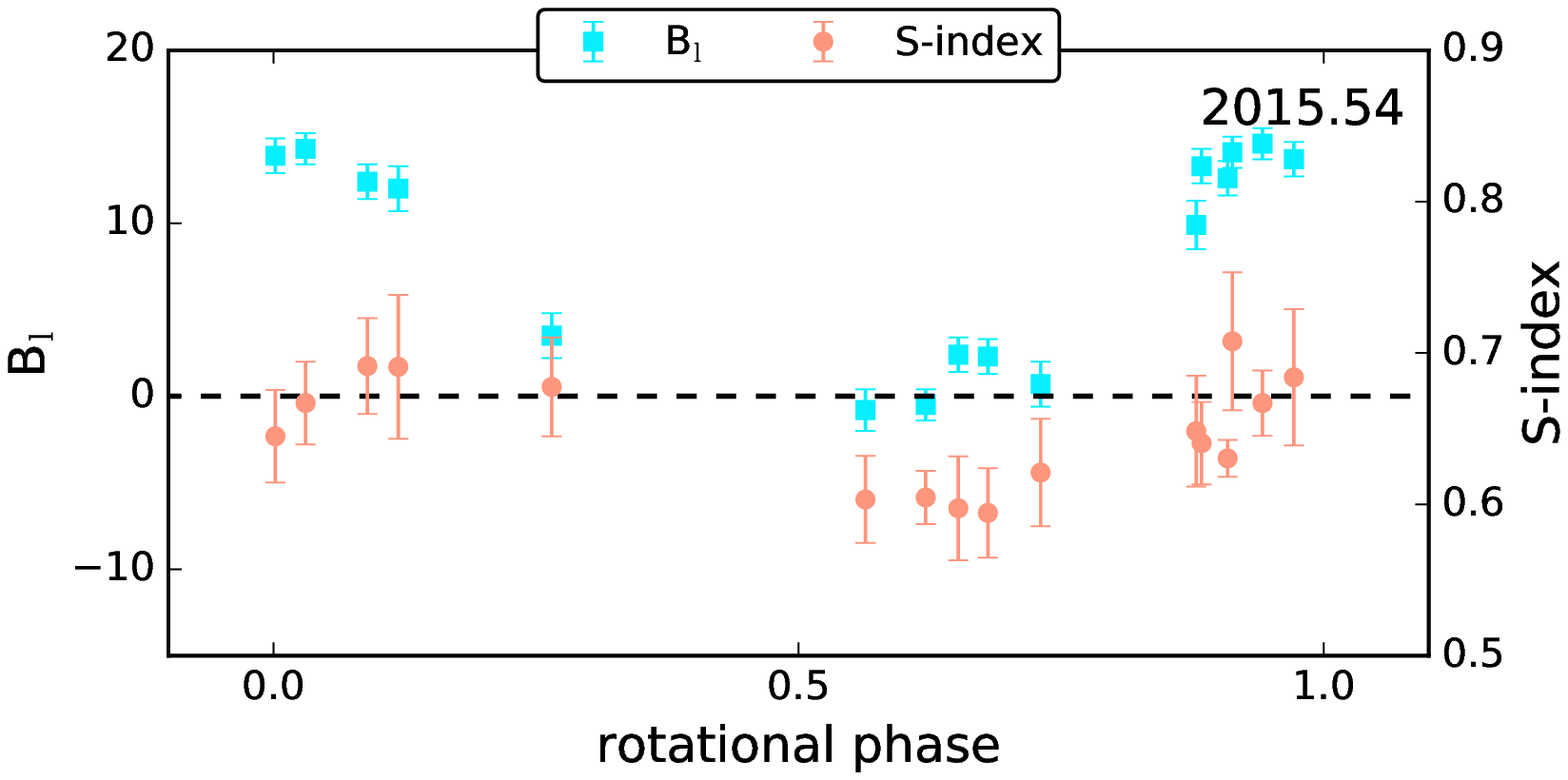}
}
\caption{Variability of the longitudinal field (${B_l}$) (Left y-axis) and S-index (Right y-axis) as a function of rotational phase for each epoch (2007.59, 2008.64, 2010.55, 2012.54, 2013.61, 2014.61 and 2015.54). For epoch 2012.54 
only the S-index is plotted as the Stokes \textit{V} observations were not available (see Section 3). The subplots are all on the same scale for easier comparison between epochs and the dashed line represents $B_l$=0. The scale was chosen based on the minimum and maximum value.}
  \label{index}
\end{figure*}

\begin{figure*}
\centering
\includegraphics[scale = 0.5]{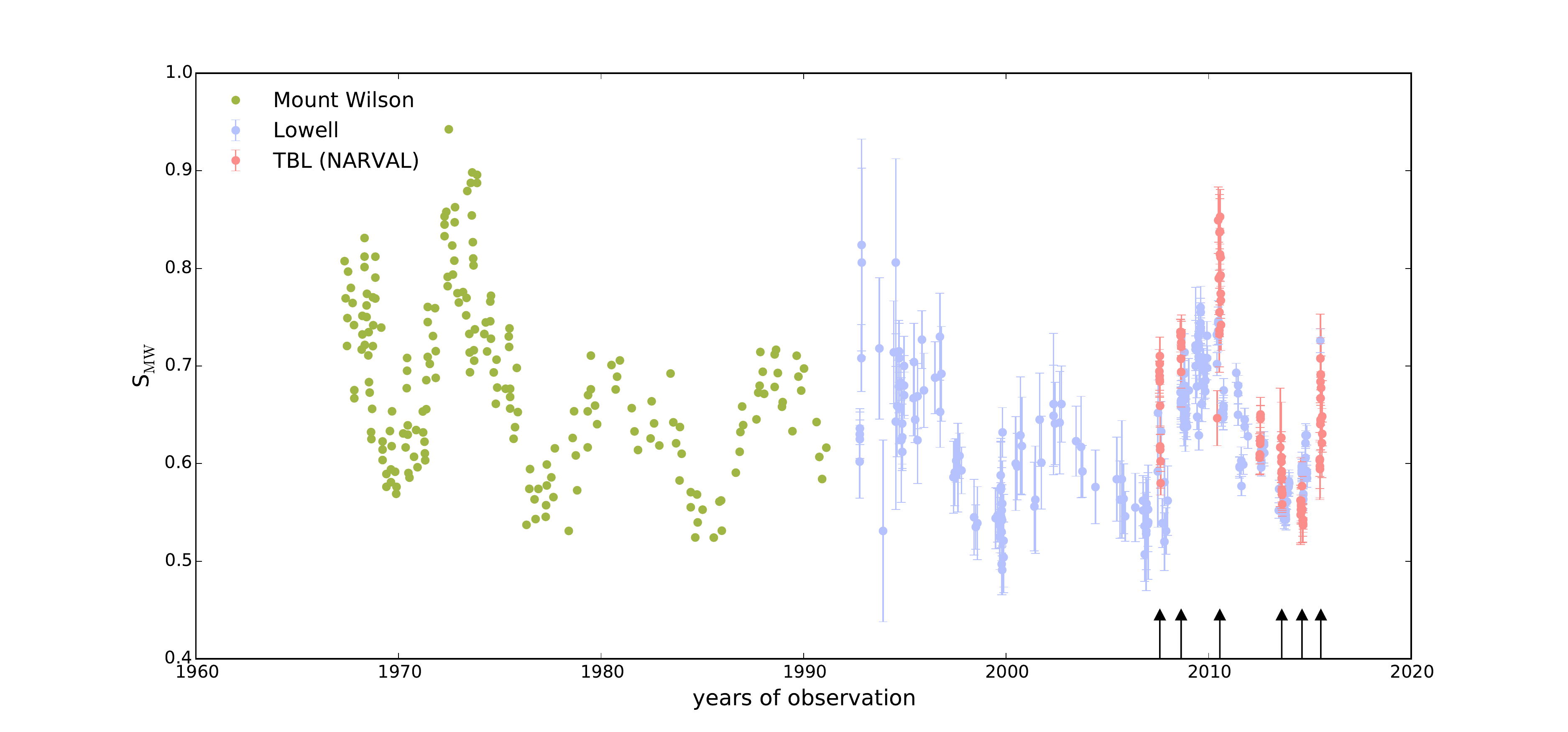}
\caption{Long-term chromospheric activity of 61 Cyg A shown as S-index which is calibrated to the Mount Wilson S-index. The arrow marks represent epochs for which the large-scale field map is reconstructed using ZDI.}
\label{sindex_all}
\end{figure*}

\section{Magnetic field detection: Direct and indirect approach}
\subsection{Mean longitudinal magnetic field}
We measure the line-of-sight component of the magnetic field averaged over the stellar surface for each observation using Stokes \textit{V} and Stokes \textit{I} LSD profile. The mean longitudinal field is measured from the Stokes \textit{V}  and \textit{I} profile 
\citep{donati97} as shown in equation \ref{bl_eq} ,
\begin{equation}
B_{l}\mathrm{(G)} = -2.14 \times 10^{11} \frac{\int v V(v) dv}{\lambda_{0} gc\int(I_c-I(v))dv}
 \label{bl_eq}
\end{equation}

where ${B_l}$ represents the mean longitudinal magnetic field of 61 Cyg A in Gauss, $\lambda_{0}$ = 630 nm is the central wavelength of the LSD profile, \textit{g} = 1.22 is the Land\'e factor of the line list, \textit{c} is
the speed of light in kms$^{-1}$, $v$ is the radial velocity in kms$^{-1}$ and \textit{I$_{c}$} is the normalised continuum intensity. The velocity range of $\pm$ 20 kms$^{-1}$ from the line center was used. The velocity window 
was selected so that the entire signal is included. If a smaller integration window is used then the wings of the signal are ignored which may result in an underestimated value of $B_l$. We also calculated the magnetic field from the 
Null profile ($N_l$). The error estimates were made by propagating the uncertainties calculated by the reduction pipeline as mentioned in \citet{marsden14}. Table \ref{numbers} shows the values of $B_l$, $N_l$ and the associated 
uncertainties for each observation. 
\paragraph{}
\begin{longtab}
\begin{longtable}{cccccccc}
\caption{\label{numbers} The chromospheric activity and longitudinal field measurements of 61 Cyg A for six epochs (2007.59, 2008.64, 2010.55, 2012.54, 2013.61, 2014.61 and 2015.54). 
From left to right it represents: Julian date, rotational phase, S-index, H${\alpha}$-index, Ca$\mathrm{IRT}$-index, longitudinal magnetic field (B$_{l}$) and magnetic field of the Null profile (N$_{l}$).}\\
\hline\hline

Epoch&Julian date&rot.phase&S-index&H$\alpha$-index&Ca$\mathrm{IRT}$-index&B$_{l}$&N$_{l}$\\
&(2454000+)&&&&&(G)&(G)\\
\hline
\endfirsthead
\caption{continued.}\\
\hline\hline

Epoch&Julian date& rot.phase&S-index& H$\alpha$-index& Ca$\mathrm{IRT}$-index& B$_{l}$& N$_{l}$\\
&(2450000+)&&&&&(G)&(G)\\
                                                                                                                                                                                                                                                                                                                                                                                                                                                                                                                                                     
\hline
\endhead
\hline
\endfoot

&4308.49809&0.00000&0.69$\pm$0.02&0.402$\pm$0.001& 0.889$\pm$0.003&-10.0 $\pm$ 0.9&-0.8 $\pm$0.9  \\
&4312.52989&0.11789&0.69$\pm$0.02& 0.402$\pm$0.001&0.895$\pm$0.003&-8.5  $\pm$ 0.9&-0.0 $\pm$0.9 \\
&4313.53121&0.14717&0.69$\pm$0.01& 0.401$\pm$0.001&0.894$\pm$0.004&-8.8  $\pm$ 1.1&-1.1 $\pm$1.1 \\
&4315.53608&0.20579&0.70$\pm$0.02& 0.401$\pm$0.001&0.891$\pm$0.003&-4.7  $\pm$ 0.9&0.1  $\pm$0.9 \\
&4316.53315&0.23494&0.68$\pm$0.02& 0.402$\pm$0.001&0.892$\pm$0.003&-5.4  $\pm$ 0.9&0.1  $\pm$0.9 \\
2007.59&4317.53462&0.26423&0.71$\pm$0.01& 0.400$\pm$0.001&0.889$\pm$0.004&-4.3  $\pm$ 1.0&0.6  $\pm$1.0 \\
&4321.50738&0.38039&0.66$\pm$0.02& 0.399$\pm$0.001&0.888$\pm$0.004&0.0   $\pm$1.0 &-0.2 $\pm$1.0 \\
&4322.52760&0.41022&0.62$\pm$0.02& 0.399$\pm$0.001&0.889$\pm$0.004&1.1   $\pm$1.1 &-0.6 $\pm$1.1 \\
&4323.52773&0.43946&0.61$\pm$0.01& 0.400$\pm$0.001&0.887$\pm$0.005&1.7   $\pm$1.3 &0.6  $\pm$1.3 \\
&4330.50089&0.64336&0.60$\pm$0.02& 0.399$\pm$0.001&0.876$\pm$0.002&-8.6  $\pm$0.6 &-0.4 $\pm$0.6 \\
&4331.46147&0.67144&0.58$\pm$0.01& 0.401$\pm$0.001&0.886$\pm$0.003&-9.1  $\pm$0.9 &0.4  $\pm$0.9 \\
\hline
&4688.54994&0.11263& 0.73$\pm$ 0.01&0.401$\pm$0.001 &0.898 $\pm$0.003&-2.9 $\pm$0.7&0.3 $\pm$0.7 \\
&4691.48653&0.19849& 0.73$\pm$ 0.02&0.401$\pm$ 0.001&0.899$\pm$ 0.003&-1.4$\pm$ 0.8&-0.9$\pm$0.8\\
&4696.52126&0.34571& 0.71$\pm$ 0.03&0.402$\pm$ 0.001&0.900$\pm$ 0.004&1.0 $\pm$ 1.0&0.4 $\pm$1.0  \\
&4700.47785&0.46140& 0.71$\pm$ 0.02&0.400$\pm$ 0.001&0.896$\pm$ 0.004&-1.4$\pm$ 0.8&0.2 $\pm$0.8 \\
2008.64&4701.48433&0.49083& 0.69$\pm$ 0.04&0.402$\pm$ 0.002&0.905$\pm$ 0.006&-4.0$\pm$ 1.5&-0.3$\pm$1.5\\
&4702.50793&0.52076& 0.72$\pm$ 0.01&0.400$\pm$ 0.001&0.889$\pm$ 0.003&-0.7$\pm$ 0.6&0.4 $\pm$0.6 \\
&4703.47966&0.54917& 0.72$\pm$ 0.01&0.401$\pm$ 0.001&0.887$\pm$ 0.003&-1.0$\pm$ 0.6&1.1 $\pm$0.6 \\
&4704.53225&0.57995& 0.73$\pm$ 0.01&0.401$\pm$ 0.001&0.890$\pm$ 0.003&-3.1 $\pm$0.6&-0.1$\pm$0.6\\
&4705.52953&0.60911& 0.73$\pm$ 0.02&0.401$\pm$ 0.001&0.887$\pm$ 0.003&-2.0$\pm$ 0.6&0.2 $\pm$0.6 \\
\hline                                                                            
&5351.64786&0.50146& 0.65$\pm$0.03&0.400$\pm$ 0.001&0.892$\pm$ 0.004&4.1 $\pm$1.0&-0.8$\pm$1.0\\
&5369.57210&0.02556& 0.85$\pm$0.03&0.402$\pm$ 0.001&0.909$\pm$ 0.004&0.3 $\pm$1.0&-0.6$\pm$1.0\\
&5379.58460&0.31832& 0.79$\pm$0.04&0.402$\pm$ 0.001&0.901$\pm$ 0.004&-2.1$\pm$1.1&-1.9$\pm$1.1\\
&5390.54209&0.63871& 0.74$\pm$0.04&0.400$\pm$ 0.001&0.899$\pm$ 0.005&2.3 $\pm$1.2&1.4 $\pm$1.2 \\
&5391.52388&0.66742& 0.73$\pm$0.03&0.399$\pm$ 0.001&0.894$\pm$ 0.005&2.1 $\pm$1.2&1.6 $\pm$1.2 \\
&5392.54747&0.69735& 0.76$\pm$0.02&0.400$\pm$ 0.001&0.893$\pm$ 0.004&0.2 $\pm$0.9&1.0 $\pm$0.9 \\
&5393.63958&0.72928& 0.84$\pm$0.04&0.405$\pm$ 0.001&0.900$\pm$ 0.004&3.0 $\pm$1.1&-0.8$\pm$1.1\\
2010.55&5396.60809&0.81608& 0.79$\pm$0.03&0.401$\pm$ 0.001&0.901$\pm$ 0.003&-0.2$\pm$0.9&-0.7$\pm$0.9\\
&5401.65920&0.96378& 0.81$\pm$0.03&0.401$\pm$ 0.001&0.901$\pm$ 0.004&-7.9$\pm$0.8&0.2 $\pm$0.8\\
&5402.49800&0.98830& 0.84$\pm$0.03&0.401$\pm$ 0.001&0.903$\pm$ 0.004&-5.3$\pm$0.8&-0.3$\pm$0.8\\
&5403.49549&0.01747& 0.85$\pm$0.03&0.403$\pm$ 0.001&0.905$\pm$ 0.004&-4.2$\pm$0.8&0.5 $\pm$0.8\\
&5411.49161&0.25127& 0.81$\pm$0.02&0.402$\pm$ 0.001&0.904$\pm$ 0.004&0.1 $\pm$0.9&-0.3$\pm$0.9\\
&5412.56953&0.28279& 0.79$\pm$0.02&0.402$\pm$ 0.001&0.898$\pm$ 0.003&0.1 $\pm$1.0&-0.6$\pm$1.0\\
&5415.55556&0.37010& 0.77$\pm$0.03&0.402$\pm$ 0.001&0.902$\pm$ 0.004&0.1 $\pm$1.0&-0.6$\pm$1.0\\
&5416.54320&0.39897& 0.77$\pm$0.01&0.402$\pm$ 0.001&0.895$\pm$ 0.004&1.4 $\pm$0.9&-1.5$\pm$0.9\\
&5419.56875&0.48745& 0.74$\pm$0.03&0.399$\pm$ 0.001&0.892$\pm$ 0.004&1.3 $\pm$0.9&0.1 $\pm$0.9 \\
\hline
&6117.63676&0.89879&0.61$\pm$0.02&0.399$\pm$0.001&0.880$\pm$0.004&..&..\\
&6118.63476&0.92797&0.61$\pm$0.02&0.400$\pm$0.001&0.874$\pm$0.005&..&..\\
&6124.64113&0.10360&0.61$\pm$0.02&0.400$\pm$0.001&0.872$\pm$0.004&..&..\\
&6125.61827&0.13217&0.62$\pm$0.02&0.400$\pm$0.001&0.880$\pm$0.003&..&..\\
2012.54&6126.61222&0.16123&0.62$\pm$0.02&0.401$\pm$0.001&0.881$\pm$0.004&..&..\\
&6127.56778&0.18917&0.62$\pm$0.02&0.400$\pm$0.001&0.884$\pm$0.004&..&..\\
&6128.55829&0.21813&0.63$\pm$0.02&0.400$\pm$0.001&0.878$\pm$0.004&..&..\\
&6131.55361&0.30572&0.65$\pm$0.01&0.400$\pm$0.001&0.885$\pm$0.004&..&..\\
&6132.54263&0.33464&0.65$\pm$0.02&0.400$\pm$0.001&0.889$\pm$0.004&..&..\\
&6133.56659&0.36458&0.65$\pm$0.01&0.401$\pm$0.001&0.885$\pm$0.004&..&..\\
\hline
&6488.51321&0.74313&0.62$\pm$0.06&0.399$\pm$0.002&0.889$\pm$0.006&3.4 $\pm$2.0&-2.7 $\pm$2.1 \\
&6507.65889&0.30295&0.63$\pm$0.04&0.398$\pm$0.001&0.884$\pm$0.004&6.2 $\pm$1.2&0.5  $\pm$1.2  \\
&6509.51420&0.35720&0.60$\pm$0.02&0.400$\pm$0.001&0.881$\pm$0.004&11.3$\pm$0.9&0.3  $\pm$0.9 \\
&6510.61746&0.38946&0.61$\pm$0.03&0.398$\pm$0.001&0.890$\pm$0.005&10.5$\pm$1.3&-0.2 $\pm$1.3\\
&6513.57114&0.47582&0.59$\pm$0.02&0.398$\pm$0.001&0.883$\pm$0.004&9.0 $\pm$1.0&-0.1 $\pm$1.0 \\
&6514.54087&0.50417&0.59$\pm$0.02&0.398$\pm$0.001&0.883$\pm$0.004&7.4 $\pm$0.9&-0.0 $\pm$0.9 \\
2013.61&6515.54442&0.53352&0.58$\pm$0.02&0.397$\pm$0.001&0.876$\pm$0.004&7.7 $\pm$0.9&-0.1 $\pm$0.9 \\
&6516.56402&0.56333&0.57$\pm$0.01&0.398$\pm$0.001&0.876$\pm$0.004&7.9 $\pm$0.9&0.7  $\pm$0.9  \\
&6518.58190&0.62233&0.57$\pm$0.02&0.398$\pm$0.001&0.877$\pm$0.004&8.6 $\pm$0.9&0.7  $\pm$ 0.9  \\
&6519.62914&0.65295&0.57$\pm$0.02&0.398$\pm$0.001&0.878$\pm$0.004&7.9 $\pm$0.9&-0.2 $\pm$0.9 \\
&6523.51294&0.76652&0.56$\pm$0.01&0.398$\pm$0.001&0.877$\pm$0.004&7.9 $\pm$1.0&-0.1 $\pm$1.0 \\
&6524.49914&0.79535&0.57$\pm$0.02&0.399$\pm$0.001&0.879$\pm$0.004&7.4 $\pm$0.9&-1.4 $\pm$0.9 \\
&6525.44457&0.82300&0.57$\pm$0.02&0.398$\pm$0.001&0.878$\pm$0.004&6.7 $\pm$1.0&-1.1 $\pm$1.0 \\
\hline
&6853.55493&0.41687&0.56$\pm$0.04&0.399$\pm$0.001&0.892$\pm$0.005&6.8$\pm$1.5&-1.6$\pm$1.5 \\
&6856.55743&0.50466&0.55$\pm$0.03&0.398$\pm$0.001&0.894$\pm$0.004&5.9$\pm$1.1&-1.4$\pm$1.1 \\
&6861.57410&0.65135&0.55$\pm$0.02&0.400$\pm$0.001&0.886$\pm$0.004&8.0$\pm$0.9&0.4 $\pm$0.9  \\
&6864.55546&0.73852&0.56$\pm$0.02&0.400$\pm$0.001&0.885$\pm$0.004&5.0$\pm$0.9&-0.7$\pm$0.9 \\
&6865.53839&0.76726&0.56$\pm$0.02&0.401$\pm$0.001&0.885$\pm$0.003&5.3$\pm$0.9&-1.2$\pm$0.9 \\
&6866.58350&0.79782&0.56$\pm$0.01&0.400$\pm$0.001&0.887$\pm$0.003&3.6$\pm$0.9&1.1 $\pm$0.9  \\
2014.61&6881.38890&0.23073&0.58$\pm$0.02&0.399$\pm$0.001&0.889$\pm$0.004&4.6$\pm$1.1&1.7 $\pm$1.1  \\
&6889.36216&0.46386&0.55$\pm$0.03&0.401$\pm$0.002&0.890$\pm$0.006&4.1$\pm$1.7&0.0 $\pm$1.7  \\
&6893.46397&0.58380&0.54$\pm$0.02&0.400$\pm$0.001&0.882$\pm$0.003&8.6$\pm$0.9&1.2 $\pm$0.9  \\
&6895.46092&0.64219&0.54$\pm$0.02&0.399$\pm$0.001&0.887$\pm$0.004&7.5$\pm$1.1&0.9 $\pm$1.1  \\
&6897.47963&0.70121&0.54$\pm$0.02&0.398$\pm$0.001&0.878$\pm$0.003&8.6$\pm$0.9&-0.3$\pm$0.9 \\
&6899.49512&0.76015&0.54$\pm$0.01&0.400$\pm$0.001&0.883$\pm$0.003&7.3$\pm$0.9&0.7 $\pm$0.9  \\
&6900.52575&0.79028&0.54$\pm$0.01&0.399$\pm$0.001&0.884$\pm$0.004&8.4$\pm$1.1&-0.9$\pm$1.1 \\
&6901.48870&0.81844&0.54$\pm$0.01&0.400$\pm$0.001&0.883$\pm$0.003&7.7$\pm$0.9&-1.4$\pm$0.9 \\ 
\hline
&7200.57599&0.56368&0.60$\pm$0.03&0.400$\pm$0.001&0.893$\pm$0.004&-0.8$\pm$1.2&0.3 $\pm$1.2 \\
&7202.53874&0.62107&0.60$\pm$0.02&0.401$\pm$0.001&0.889$\pm$0.003&-0.5$\pm$0.9&-1.1$\pm$0.9\\
&7203.59939&0.65208&0.59$\pm$0.03&0.400$\pm$0.001&0.886$\pm$0.004&2.4 $\pm$1.0&1.0 $\pm$1.0  \\
&7204.56304&0.68026&0.59$\pm$0.03&0.401$\pm$0.001&0.890$\pm$0.004&2.3 $\pm$1.0&0.4 $\pm$1.0  \\
&7211.52277&0.88376&0.64$\pm$0.03&0.400$\pm$0.001&0.896$\pm$0.004&13.3$\pm$1.0&-1.2$\pm$1.0\\
&7212.52460&0.91306&0.71$\pm$0.05&0.401$\pm$0.001&0.890$\pm$0.005&14.1$\pm$0.9&-0.5$\pm$0.9\\
&7213.50574&0.94174&0.67$\pm$0.02&0.401$\pm$0.001&0.897$\pm$0.003&14.6$\pm$0.9&0.3 $\pm$0.9 \\
2015.54&7214.52800&0.97163&0.68$\pm$0.05&0.401$\pm$0.001&0.903$\pm$0.005&13.7$\pm$1.0&1.0 $\pm$1.0 \\
&7215.56071&0.00183&0.64$\pm$0.03&0.402$\pm$0.001&0.898$\pm$0.004&13.9$\pm$1.0&-0.7$\pm$1.0\\
&7216.54418&0.03059&0.67$\pm$0.03&0.402$\pm$0.001&0.900$\pm$0.004&14.3$\pm$0.9&-0.9$\pm$0.9\\
&7218.55681&0.08943&0.69$\pm$0.03&0.403$\pm$0.001&0.906$\pm$0.004&12.4$\pm$1.0&0.5 $\pm$1.0 \\
&7219.56343&0.11887&0.69$\pm$0.05&0.403$\pm$0.001&0.906$\pm$0.005&12.0$\pm$1.3&0.2 $\pm$1.3 \\
&7224.56050&0.26498&0.68$\pm$0.03&0.401$\pm$0.001&0.904$\pm$0.005&3.5 $\pm$1.3&-0.2$\pm$1.2 \\
&7240.48208&0.73053&0.62$\pm$0.04&0.400$\pm$0.001&0.897$\pm$0.005&0.7 $\pm$1.3&1.1 $\pm$1.3  \\
&7245.54843&0.87866&0.65$\pm$0.04&0.400$\pm$0.001&0.896$\pm$0.005&9.9 $\pm$1.4&0.2 $\pm$1.4  \\
&7246.57034&0.90855&0.63$\pm$0.01&0.399$\pm$0.001&0.892$\pm$0.004&12.6$\pm$1.0&0.3 $\pm$1.0 \\
\hline
\footnote{The longitudinal magnetic field measurements are not included for 2012.54 epoch due to spectropolarimetric errors at NARVAL. See Sect. 3 for details.} 
\end{longtable}
\end{longtab}

\subsection{Chromospheric activity as a proxy of magnetic activity}
61 Cyg A is known to exhibit cyclic chromospheric activity of approximately 7 years \citep{baliunas95}. In order to investigate the chromospheric activity of 61 Cyg A we monitored the fluxes in three chromospheric 
indicators: Ca II H$\&$K, H$\alpha$, and the Ca II infrared triplet. These three chromospheric indicators are well known proxies of stellar activity, where the different indicators have different sensitivities to  the various 
chromospheric features such as plages, networks, flares, and filaments \citep{meunier09}. The Ca II H$\&$K lines are resonance lines which are dependent on the local chromospheric temperature and electron density \citep{jefferies59}. 
Contrary to the Ca II H$\&$K lines, the H$\alpha$ line is a Balmer line which is insensitive to the local chromospheric conditions but more sensitive to the radiation temperature \citep{jefferies59}. Hence, it is important to explore the
different chromospheric tracers to obtain as complete as possible picture of the chromosphere of an active cool star.
\subsubsection{S-index}
The S-index is one of the most widely used proxies of magnetic activity since its introduction in the long-term  Mount Wilson survey \citep{wilson78,duncan91,baliunas95}. We measure the S-index of 61 Cyg A by using two triangular 
band passes with a FWHM of 0.1 nm, centred at the line cores of the Ca II K and H line at 393.3663 nm and 396.8469 nm respectively. The fluxes in the nearby continuum are measured by using two 2 nm wide rectangular band passes at the 
blue and red side of the K and H lines at 390.107 nm and 400.107 nm respectively \citep{duncan91,wright04}. In order to calibrate our measurements to the original Mount Wilson scale, we use the calibration coefficients calculated by 
\citet{marsden14}, which is used in the S-index formula as shown in equation \ref{sindex_eq},

\begin{equation}
 \mathrm {\text{S-index}} = \frac{{\alpha F_H+\beta F_K}}{{\gamma F_R+\delta F_V}}+ {\phi}
 \label{sindex_eq}
\end{equation}
where, ${F_H}$, ${F_K}$, ${F_R}$, and ${F_V}$ are the flux in the H, K, R, and V bands and $\alpha=12.873$, $\beta=2.502$, $\gamma=8.877$, $\delta=4.271$, and $\phi$=1.183e-03 are the 
calibration coefficients respectively. The uncertainties associated with the S-index measurements are obtained by carrying out error propagation and are tabulated in Table \ref{numbers}.
\begin{figure*}
  \includegraphics[scale=0.5]{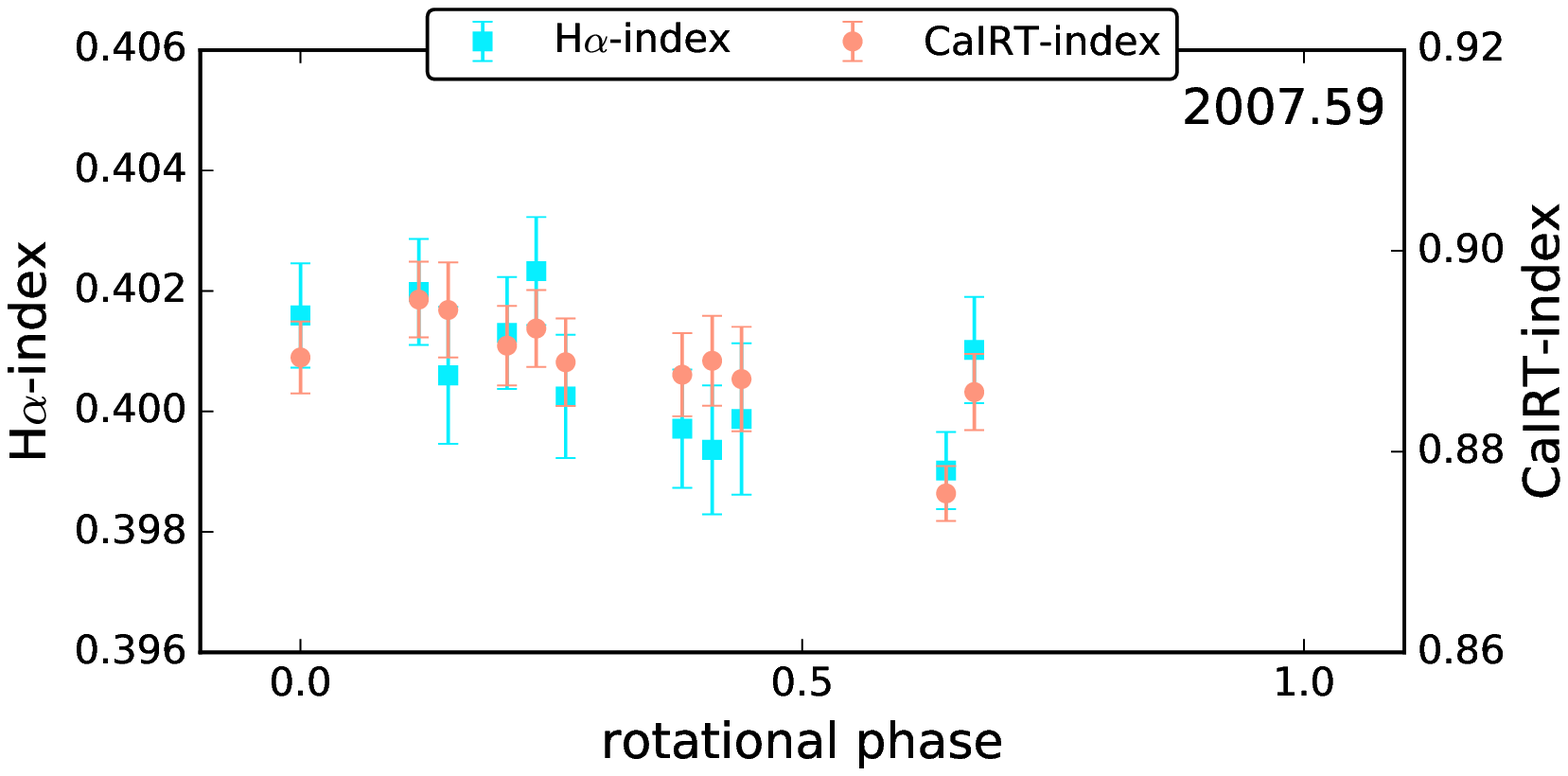}~~~\includegraphics[scale=0.5]{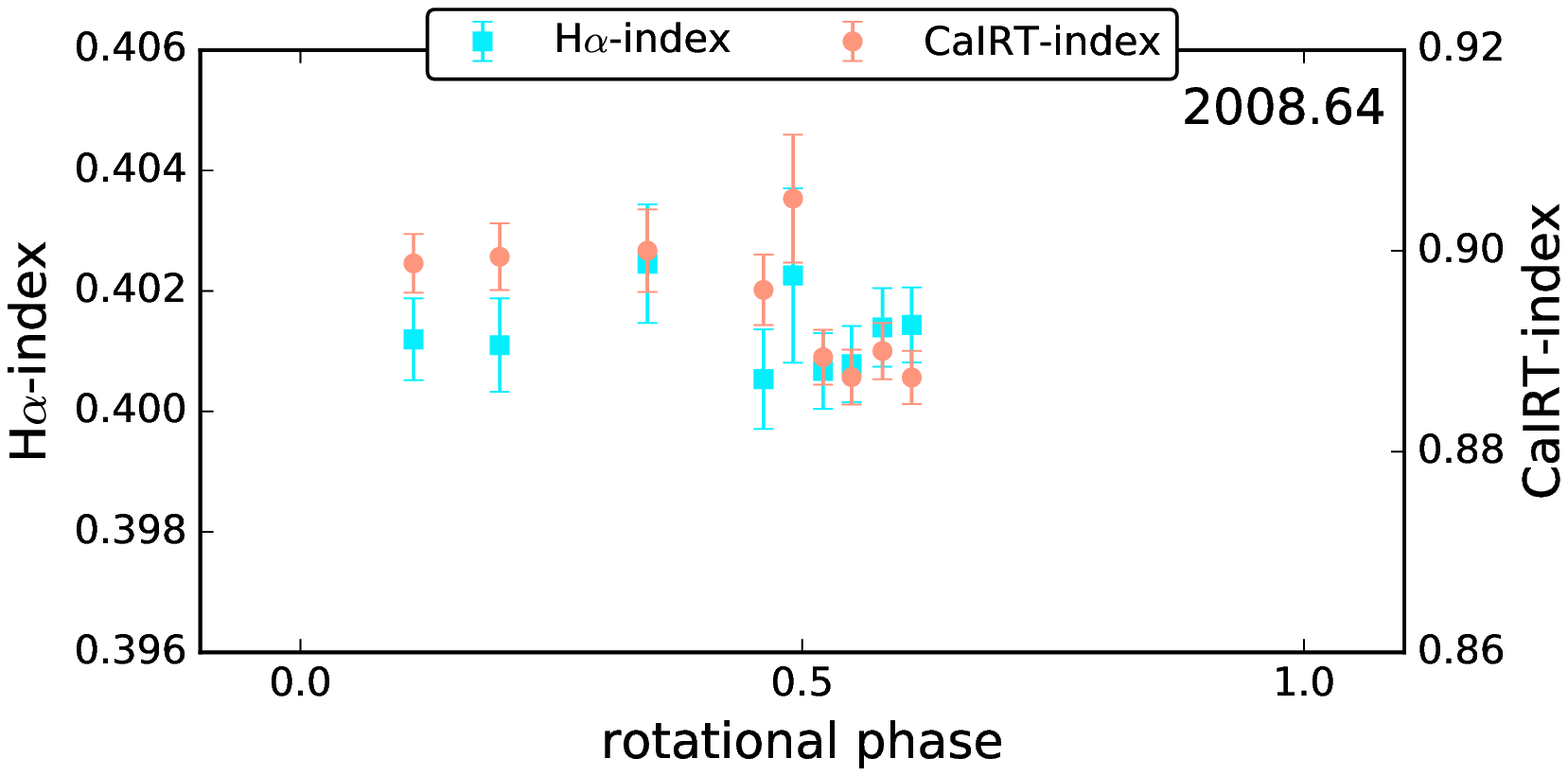}\\[5mm]
  \includegraphics[scale=0.5]{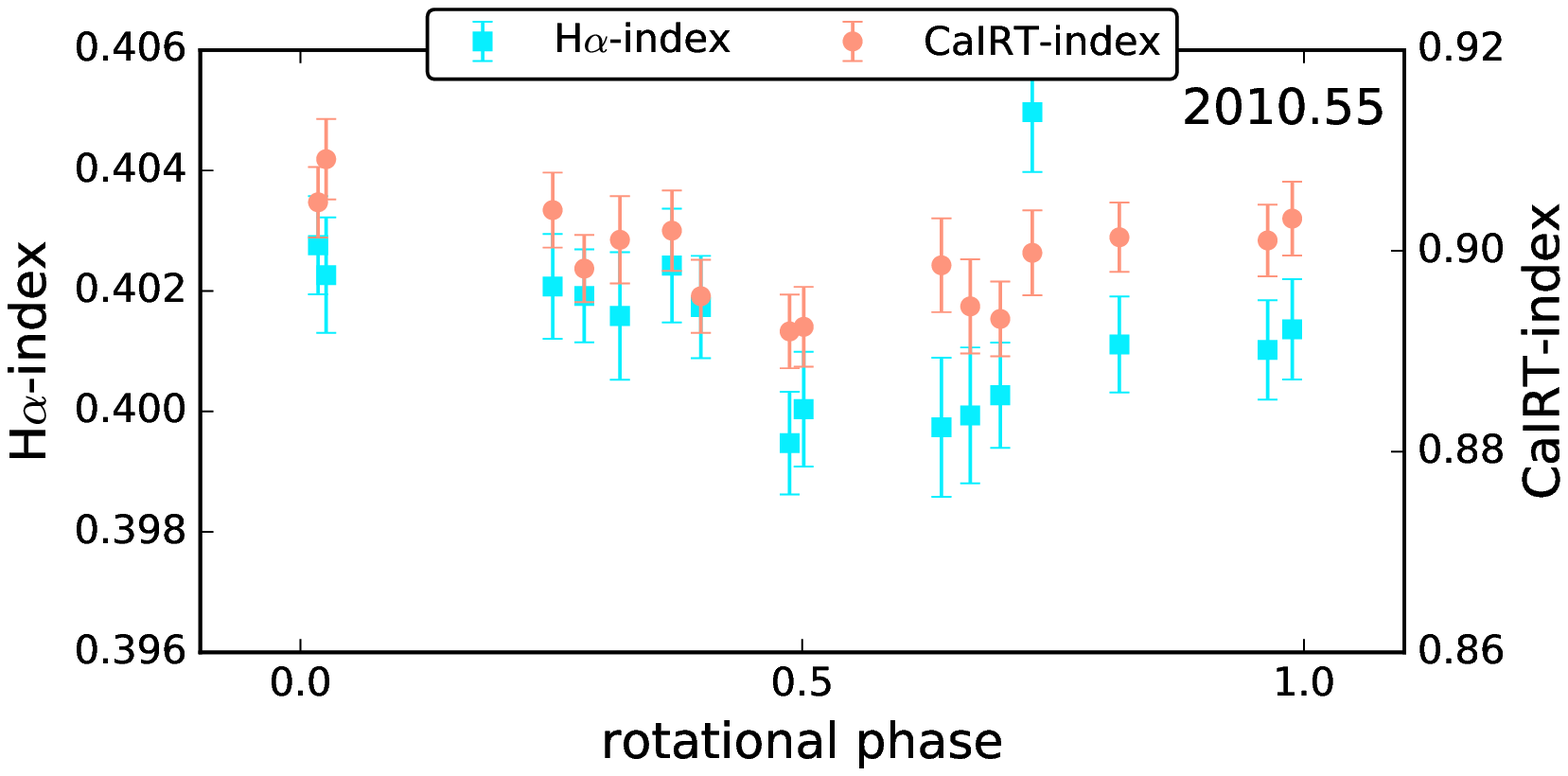}~~~\includegraphics[scale=0.5]{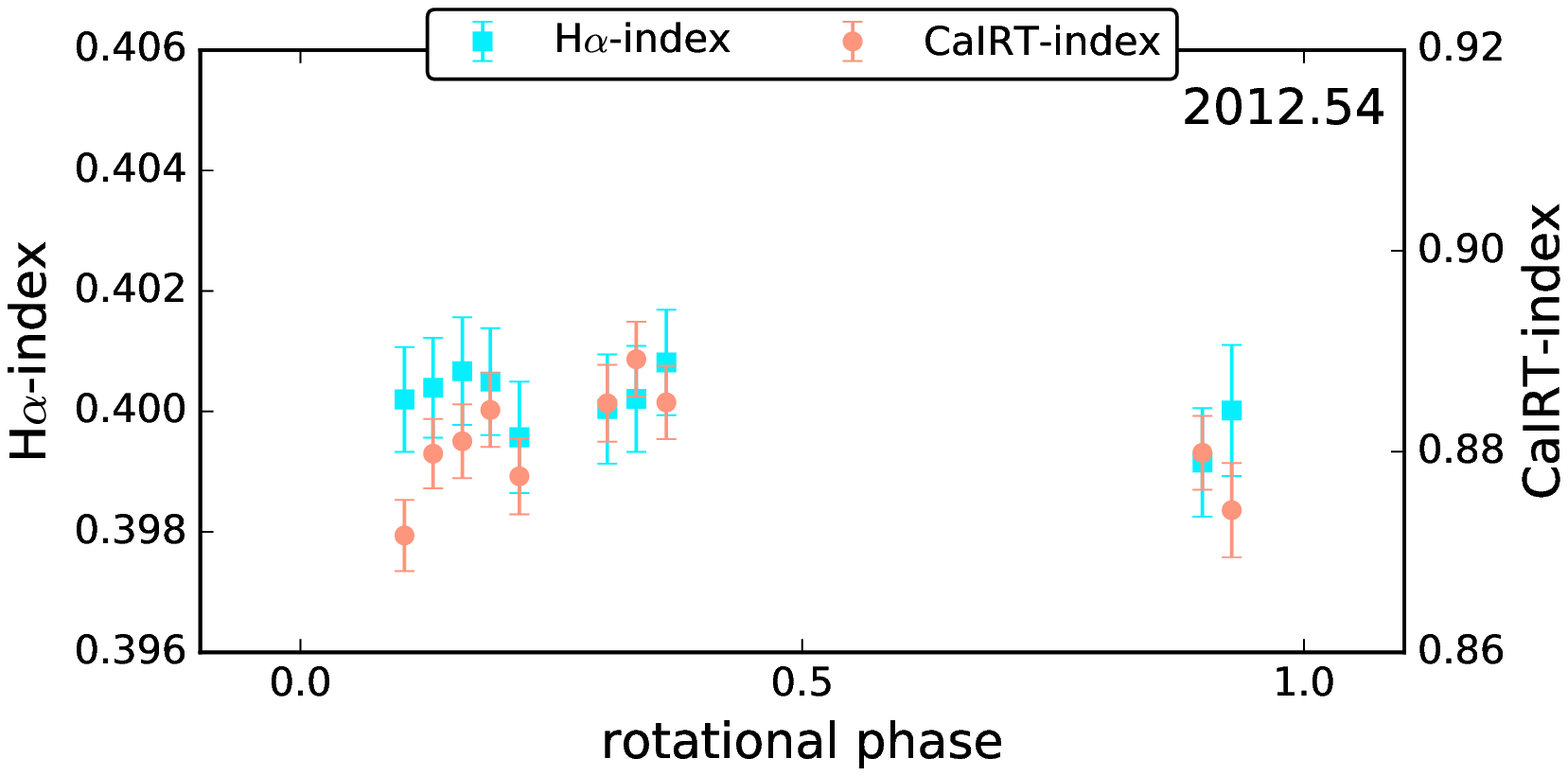}\\[5mm]
  \includegraphics[scale=0.5]{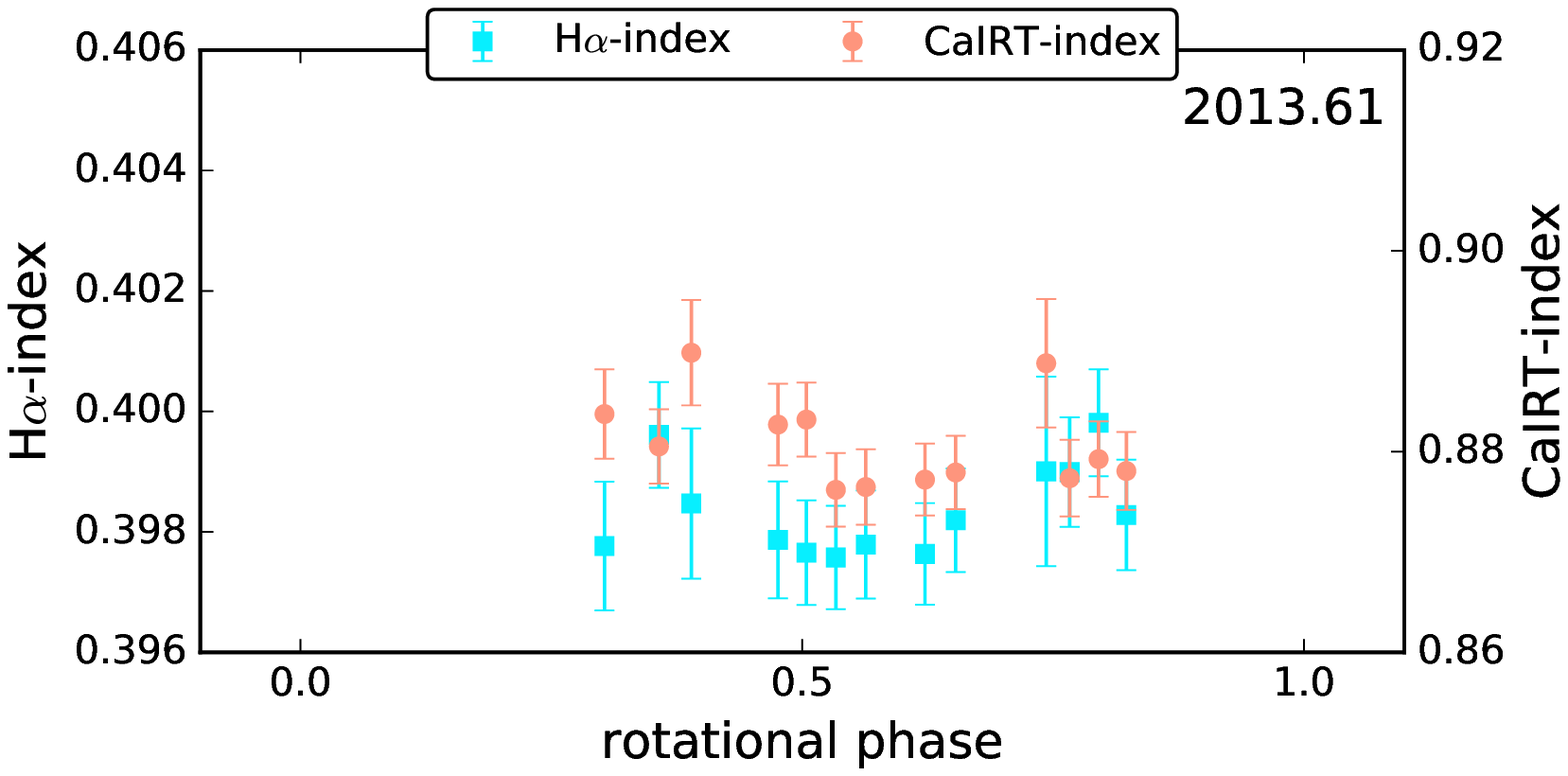}~~~\includegraphics[scale=0.5]{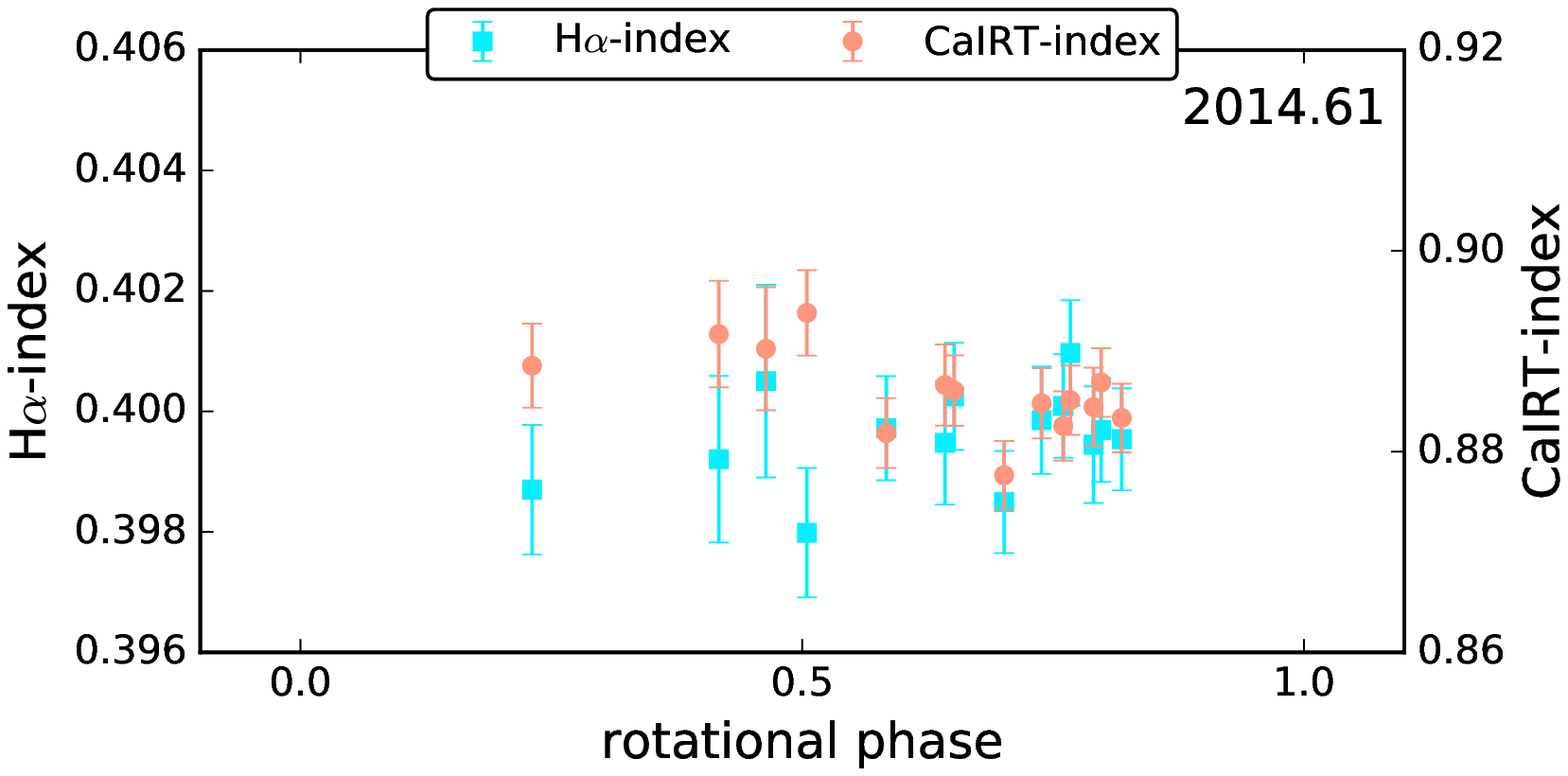}\\[5mm]
\raggedright{
\includegraphics[scale=0.5]{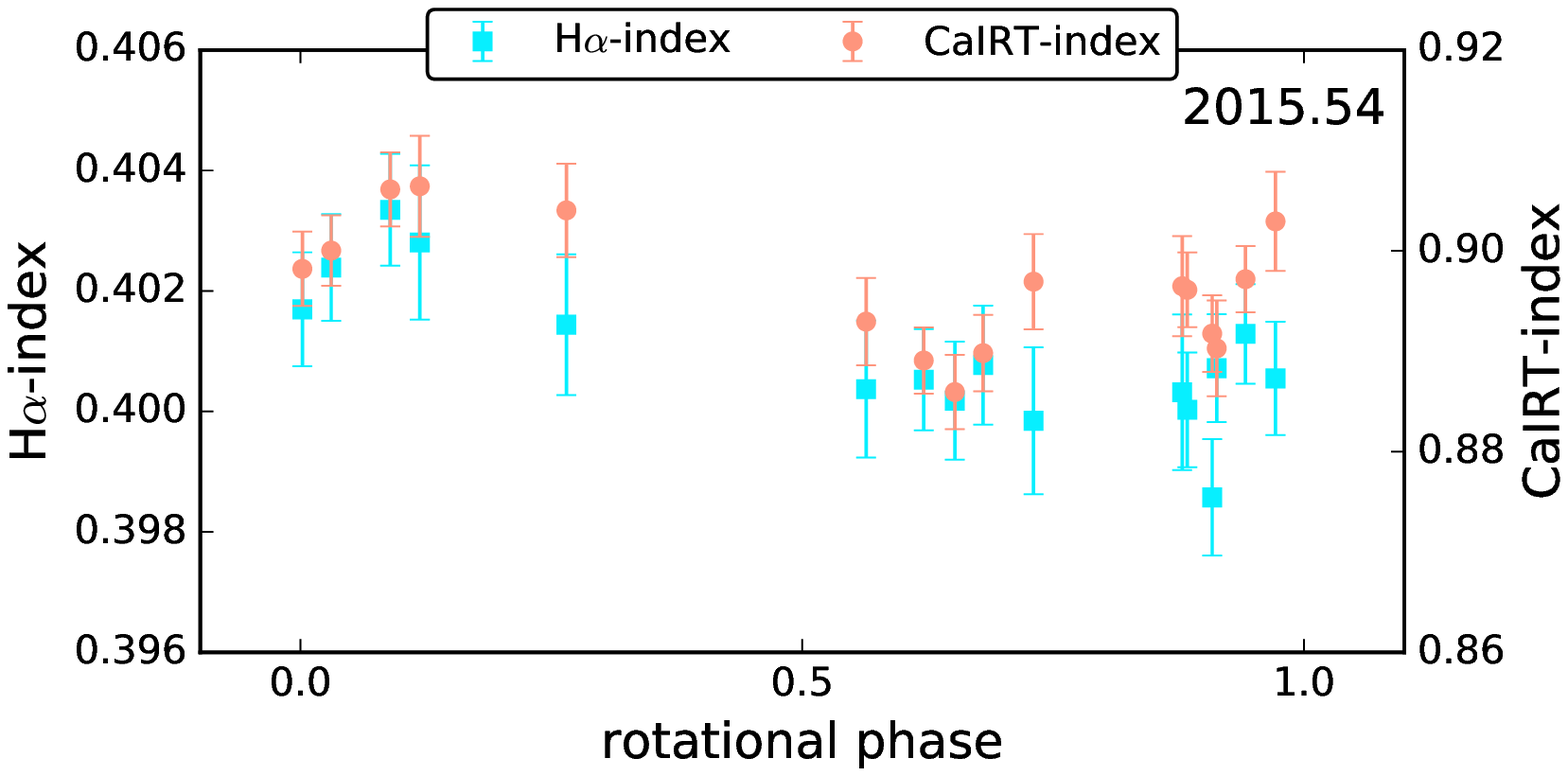}\\
}
 \caption{Same as Fig \ref{index} for H$\alpha$-index (Left y-axis) and CaIRT-index (Right y-axis). The scales are chosen based on the minimum and maximum value.}
  \label{rest_index}
\end{figure*}
\paragraph{}
Fig \ref{index} shows the rotational variability of both ${B_l}$ and S-index as a function of the rotational phase for each epoch, allowing us to study variability on the rotation timescale. Overall 61 Cyg A exhibits a stronger 
rotational variability of its longitudinal magnetic field during epochs 2007.59, 2010.55 and 2015.54 compared to epochs 2008.64, 2013.61 and 2014.61. During our observational timespan the longitudinal field $B_l$ shows a change in sign 
which indicates that the field is non-axisymmetric, except in epochs 2013.61 and 2014.61 showing signs of a more axisymmetric field. The S-index exhibits low rotational dependence indicating a simple magnetic geometry. However one 
should be careful in interpreting the results as the rotational variations might be hidden due to poor phase coverage in certain epochs such as 2012.54. The longitudinal field could not be measured for epoch 2012.54, due to 
unavailability of Stokes \textit{V} observations as already mentioned in the previous Section.
 \begin{figure*}
 \centering
 \includegraphics[scale= 0.45]{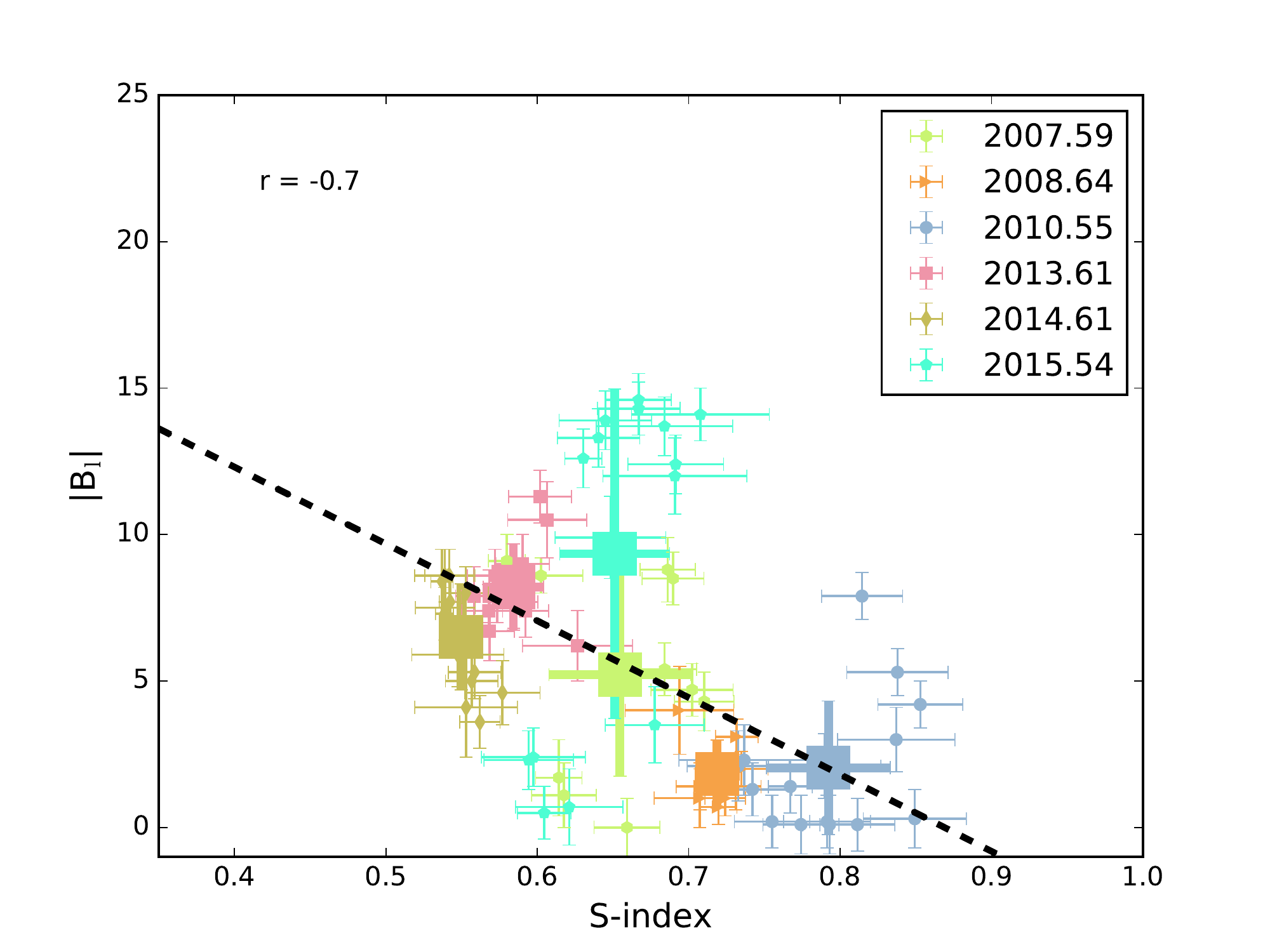}~~\includegraphics[scale= 0.45]{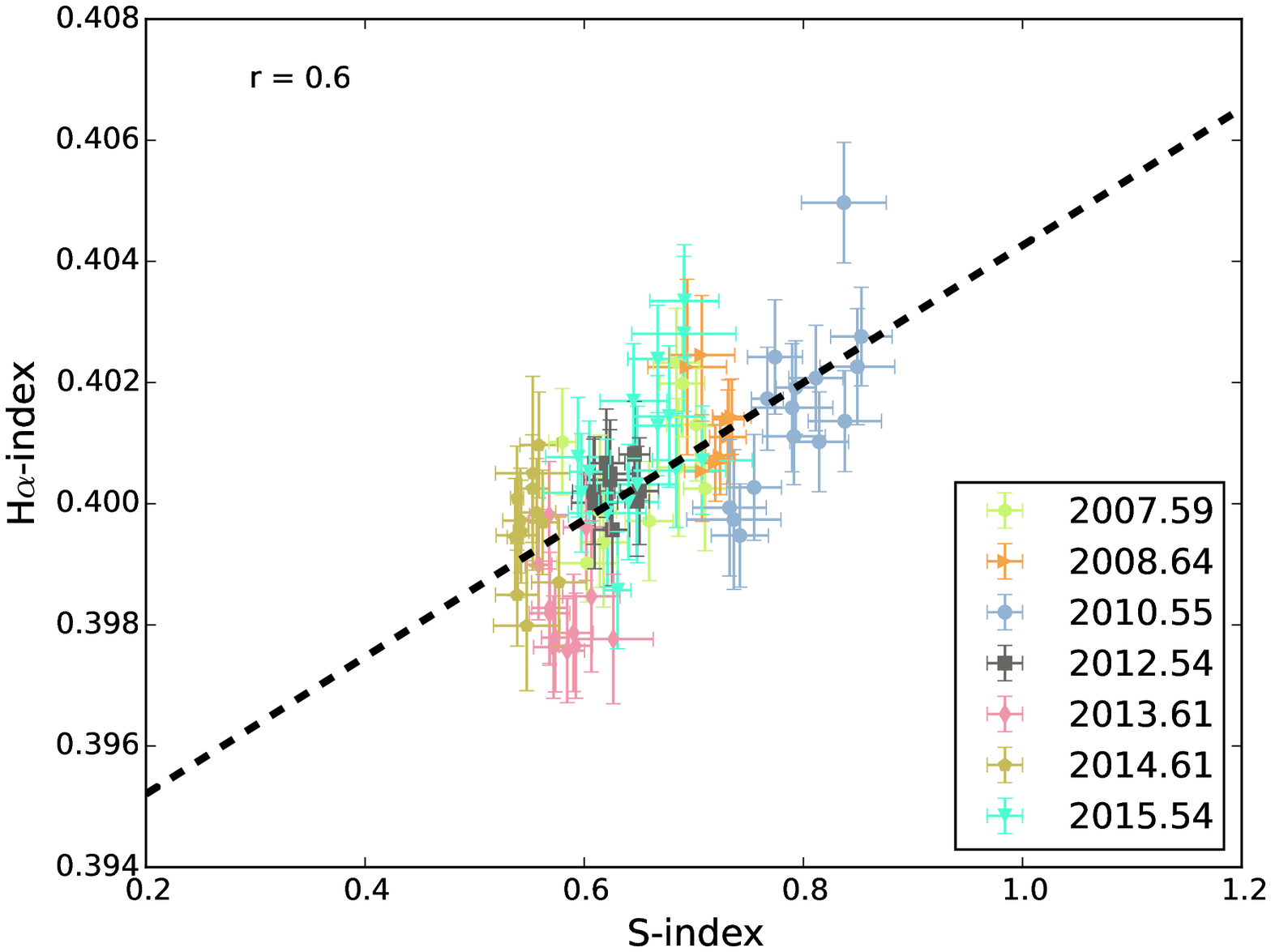}\\
 \includegraphics[scale= 0.45]{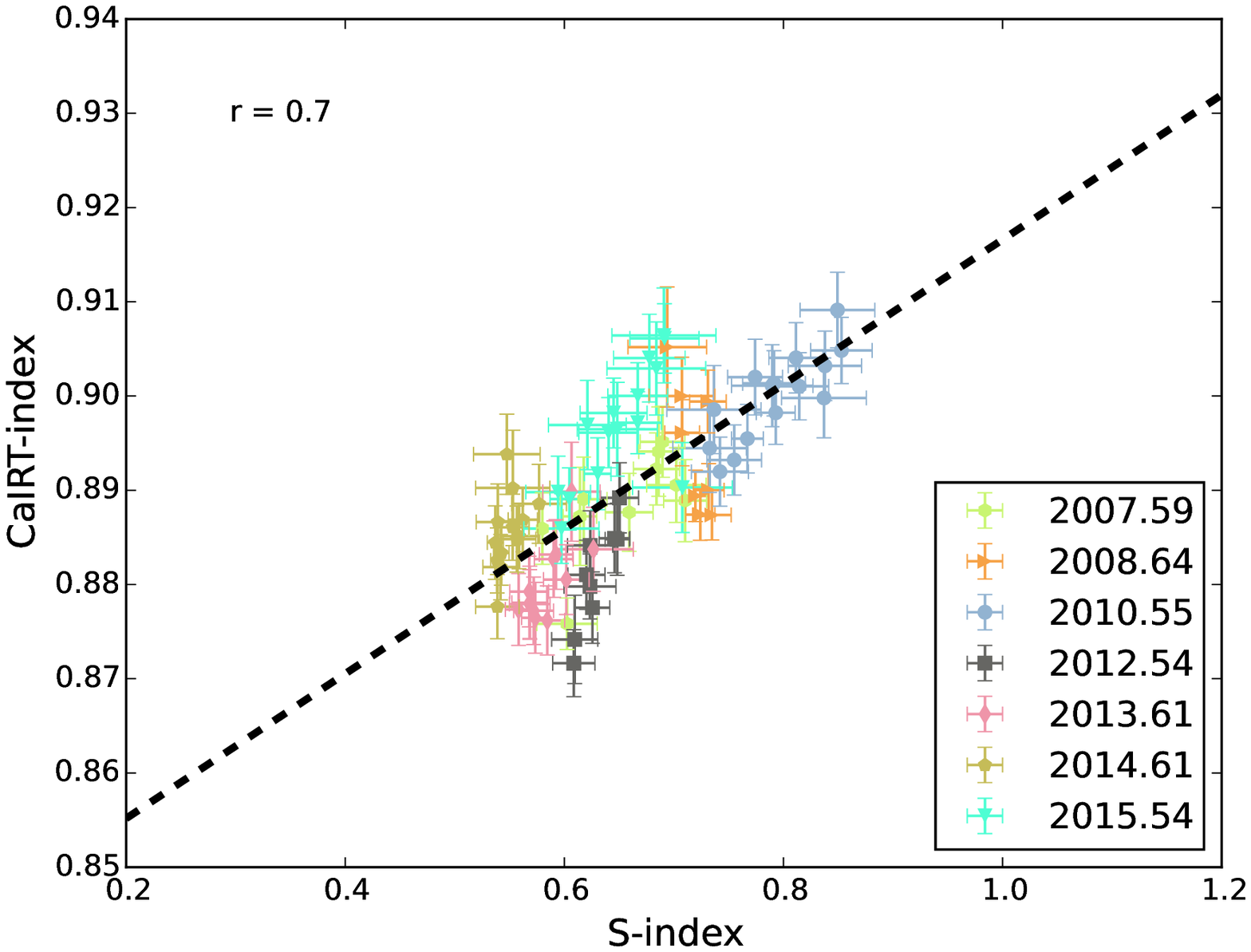}~~ \includegraphics[scale= 0.45]{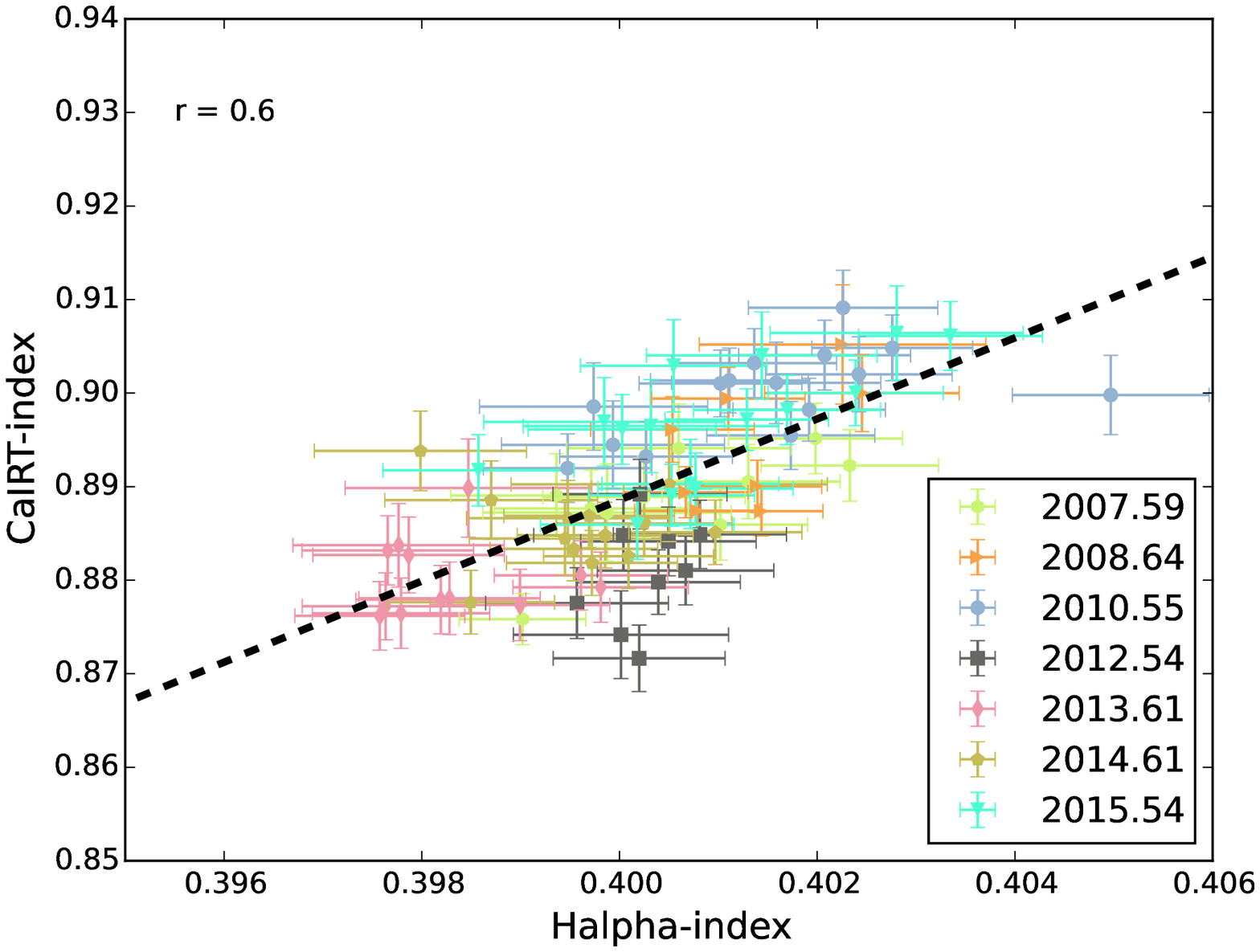}\\
\caption{Correlation between the different activity indicators and the longitudinal field $|B_l|$, where the dashed line represents least square fit to the data and the Pearson's r is shown in the plot. \textit{Top Left}: 
Anti-correlation between S-index and $|B_l|$ over six epochs of observations. The bold crosses represent the mean and standard deviation as dispersion. Only the mean values are included in the fit. \textit{Top Right}: Plot 
showing the correlation between S-index and H$\alpha$-index over seven epochs of observations. \textit{Bottom Left}: Correlation between S-index and CaIRT-index over seven epochs of data. \textit{Bottom Right}: Correlation between 
H$\alpha$-index and CaIRT-index over seven epochs of data. }
\label{corr_plot}
\end{figure*}

\paragraph{}
61 Cyg A was part of the original Mount Wilson survey and was observed for three decades at the Mount Wilson observatory from 1961 to 1991 \citep{baliunas95}. A continuation of the Mount Wilson survey is being currently carried out at 
the Lowell observatory \citep{hall07} and all available S-index data of 61 Cyg A together with our measurements are shown in Fig \ref{sindex_all}. It shows the temporal variation of the S-index of 61 Cyg A calibrated to the Mount Wilson 
scale over a time span of 48 years. At the beginning of our observational data in epoch 2007.59 61 Cyg A exhibits low activity and it increases to its maximum during epoch 2010.55. The chromospheric activity is low in epoch 
2013.61 which indicates it is close to activity minimum. The activity in epoch 2014.61 is also very low but it increases again during epoch 2015.54. An activity cycle period of 7.2$\pm$1.3 yrs was measured by taking a "generalised"
Lomb-Scargle periodogram \citep{lomb76,scargle82,zechmeister09} on all available S-index data. This cycle period agrees with the activity cycle period of 7.3$\pm$0.1 yrs measured using only the Mount Wilson data \citep{baliunas95}. 
\subsubsection{H$\alpha$-index}
We measured the variability of the chromospheric activity of 61 Cyg A using H$\alpha$ as an activity tracer for each epoch of this analysis. We employed the technique used by \citet{gizis02}, where a rectangular band pass of 0.36 nm 
width was used at the H$\alpha$ line at 656.285 nm. Two rectangular band passes $H_\mathrm{blue}$ and $H_\mathrm{red}$ of 0.22 nm width were used to measure the nearby continuum,
\begin{equation}
 \mathrm{\text{H$\alpha$-index}} = \frac{{H\alpha}}{{H_\mathrm{blue}+H_\mathrm{red}}}
 \label{ha_equation}
\end{equation}
where $H\alpha$, $H_\mathrm{blue}$ and $H_\mathrm{red}$ refers to different band passes used.
\subsubsection{CaIRT-index}
The CaIRT-index of 61 Cyg A was measured by taking 0.2 nm wide rectangular band passes centred at the Ca II triplet lines at  849.8023 nm, 854.2091 nm, and 866.241 nm respectively. The nearby continuum is measured by taking two 0.5 nm
band passes $IR_\mathrm{red}$ and $IR_\mathrm{blue}$ at 870.49 nm, and 847.58 nm respectively \citep{marsden14}, as shown in equation \ref{cairt_eq},
\begin{equation}
 \mathrm{\text{CaIRT-index}}={\frac{IR1+IR2+IR3}{IR_\mathrm{red}+IR_\mathrm{blue}}}
 \label{cairt_eq}
\end{equation}
where \textit{IR1, IR2}, and \textit{IR3} represent flux in the Ca II infra-red triplet lines. The continuum flux is represented by $IR_\mathrm{red}$ and $IR_\mathrm{blue}$ respectively. 
\begin{figure}
\centering
\includegraphics[scale=0.45]{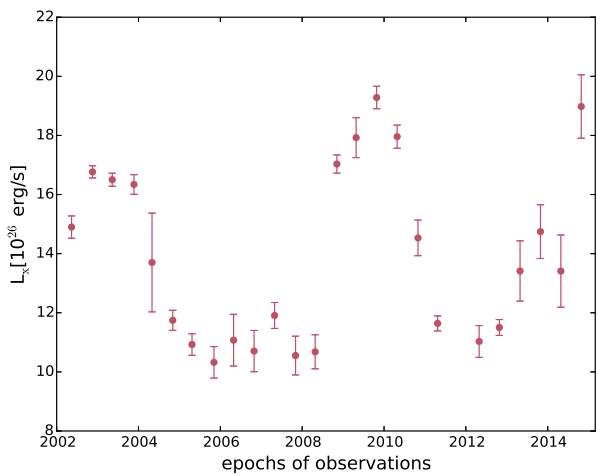}
\caption{Variation of X-ray luminosity of 61 Cyg A determined through \textit{XMM-Newton} observations.}
\label{xrayplot}
\end{figure}

\paragraph{}
The H$\alpha$  and Ca IRT lines are in absorption throughout our observations. The H$\alpha$-index variability as a function of the rotational phase is shown in Fig \ref{rest_index}. The rotational variability of the CaIRT-index 
is also shown in Fig \ref{rest_index}. Surprisingly the H$\alpha$ and Ca II IRT indices do not show the strong variations as shown by the S-index. The  uncertainties are obtained by carrying out error propagation and  shown in Table 
\ref{numbers}.
\subsubsection{Coronal activity}
The X-ray luminosity variations of 61 Cyg A is shown in Fig \ref{xrayplot}, where the data up to 2012 were previously published by \citet{robrade12} and the results from 2013 onward are derived for the first time here. While 
comparison to the results of \citet{robrade12} yields minor discrepancies caused by the different data fitting approaches, the overall picture is consistent. A clear cyclic variation in the X-ray luminosity is recovered with a cycle 
period of 6.6$\pm$0.5 years. This is in good agreement with the chromospheric cycle of 7.2$\pm$1.3 years.
\subsubsection{Correlation between activity indicators and longitudinal field}
As shown in Fig \ref{corr_plot}, a possible weak anti-correlation between absolute value of the longitudinal magnetic field $\mathrm{|B_l|}$ and S-index is detected. The field measurements are the weakest in epoch 2010.55, when 
the activity is at a maximum in our observations and is stronger close to minimum activity. The average of the absolute longitudinal magnetic field ({$\mathrm{|B_l|}$}) per epoch is strongest in epoch 2015.54 with 9.3$\pm$5.9 G and the
weakest in epoch 2010.55 with 0.6$\pm$3.1 G. On the other hand strong correlation is detected between the three chromospheric activity indicators as shown in Fig \ref{corr_plot}. The CaIRT-index exhibits a slightly stronger correlation
with the S-index (r=0.7) compared to the H$\alpha$-index (r=0.6), which is not surprising as those two indicators are sensitive to similar chromospheric features. A strong correlation between the S-index and the CaIRT-index was also 
detected in the G dwarf $\xi$ Bootis A \citet{morgenthaler12}. Correlation between the logR$'_\mathrm{HK}$ and CaIRT-index was also studied for 170 solar-type stars as part of the BCool snapshot survey \citep{marsden14}. The 
logR$'_\mathrm{HK}$ is also an indicator of stellar activity where the photospheric contribution is corrected for the S-index \citep{wright04}.
\section{Large-scale magnetic field geometry}
The large-scale surface magnetic geometry of 61 Cyg A is reconstructed using the ZDI technique. ZDI is an inverse tomographic technique which reconstructs the stellar magnetic geometry from a time series of Stokes \textit{V} LSD profiles using 
maximum entropy as a regularisation technique \citep{skillingandbryan84}. The ZDI version used in this paper reconstructs the magnetic geometry into its poloidal (potential field) and toroidal (non-potential field) components expressed 
as spherical harmonic decomposition \citep[See][]{donati06}.
\paragraph{}
If the total surface magnetic field is a combination of poloidal and toroidal field it can be formalised into three field components $B_r$, $B_\theta$ and B$_\phi$. $B_r$ is the radial component of the field and is entirely poloidal 
described by the spherical harmonics coefficient $\alpha_{l,m}$, where $l$ and $m$ are the order and degree respectively. $B_\theta$ is the meridional component of the total field and $B_\phi$ is the azimuthal component of 
the total field. $B_\theta$ and $B_\phi$ can described by a combination of the spherical harmonics coefficients $\beta_{l,m}$ and $\gamma_{l,m}$ \citep[See Section 5.1]{donati06}. The order of the spherical harmonics coefficients 
determines if the field is simple dipolar or complex and the degree determines if the field is axisymmetric or non-axisymmetric. For example, $l=1,m=0$ denotes the axisymmetric dipolar field. The quadrupolar and octopolar modes are 
denoted by higher order $l=2$ and $l=3$ respectively. Using the spherical harmonics coefficients the surface magnetic map and the associated Stokes \textit{V} spectra of a star can be derived. ZDI tackles the inverse problem and reconstructs the 
large-scale surface magnetic field geometry by comparing model Stokes \textit{V} spectra to the observed Stokes \textit{V} spectra.
 \begin{figure*}
 \centering
 \begin{tabular}{ccc}
 \subfloat[2007.59]{\includegraphics[scale = 0.25]{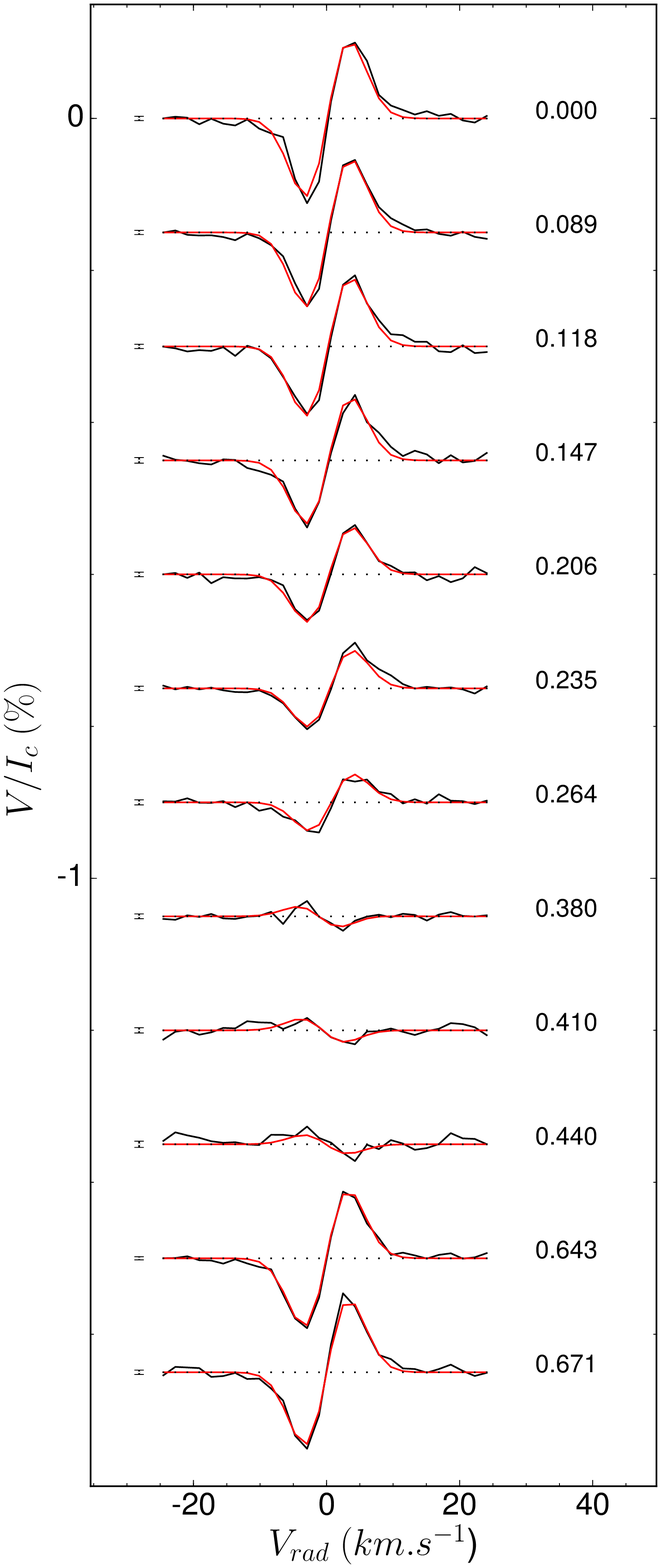}}&
 \subfloat[2008.64]{\includegraphics[scale = 0.25]{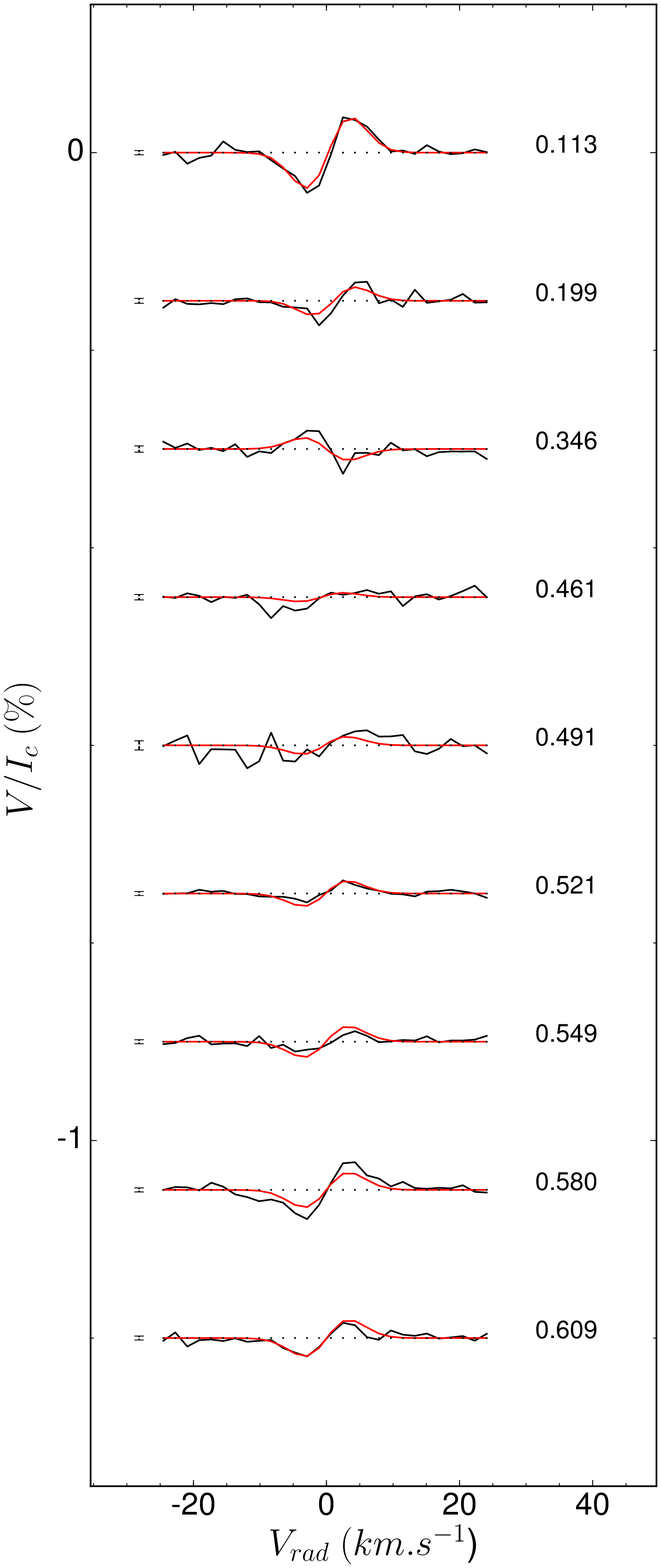}}&
  \subfloat[2010.55]{\includegraphics[scale = 0.25]{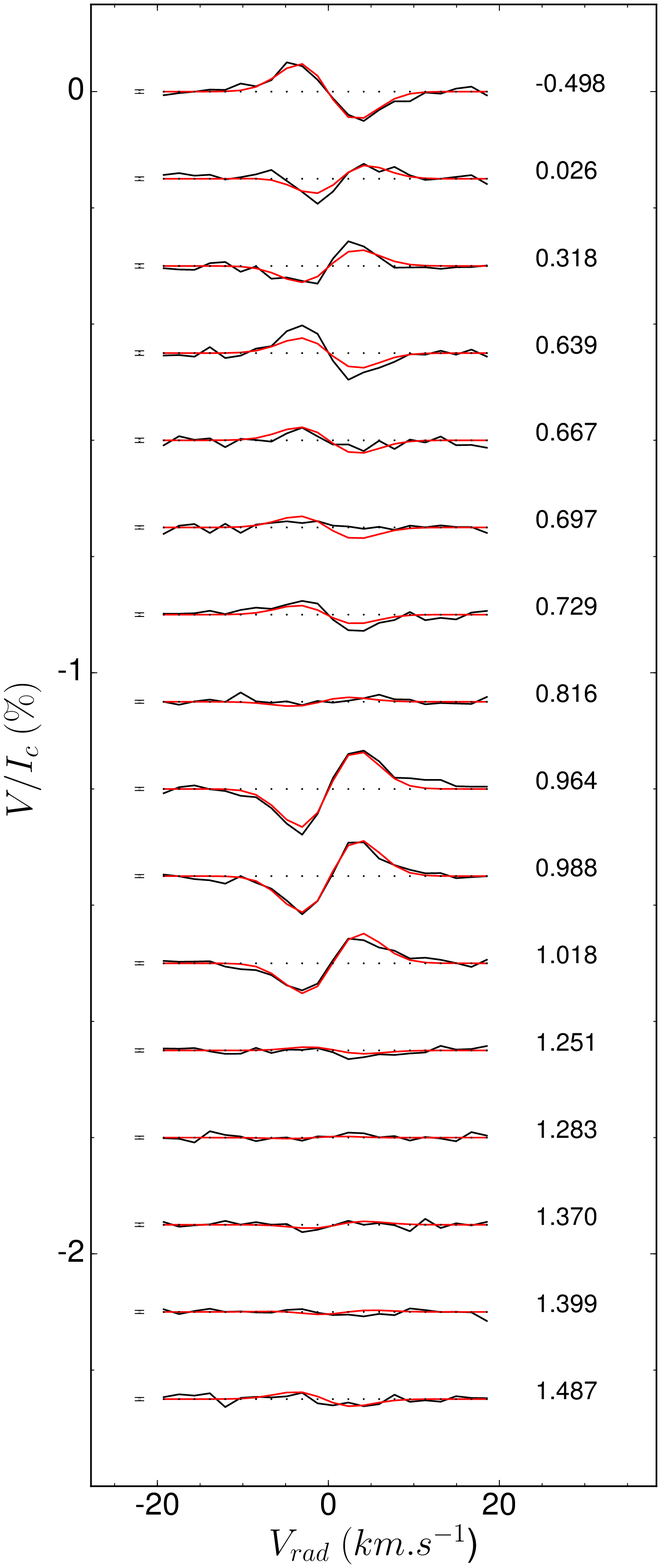}}\\
 \subfloat[2013.61]{\includegraphics[scale = 0.25]{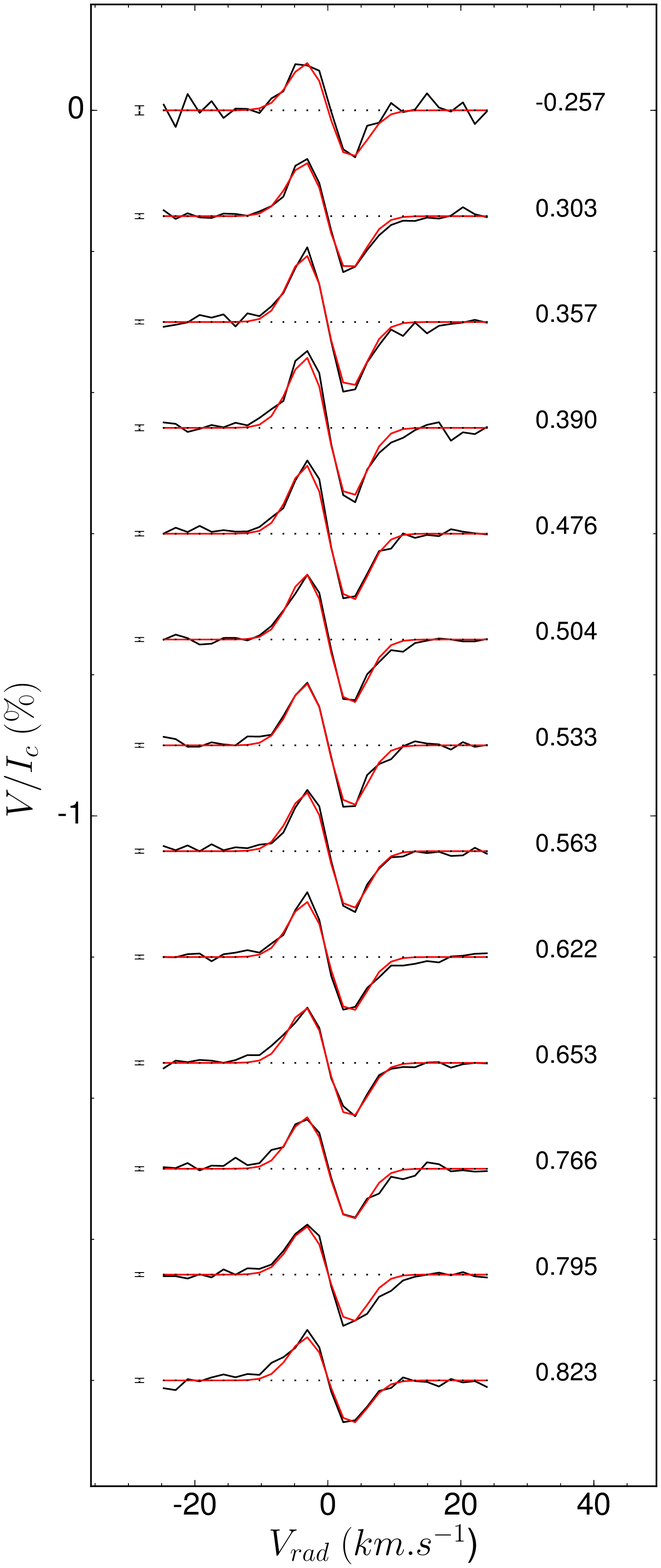}}&
 \subfloat[2014.61]{\includegraphics[scale = 0.25]{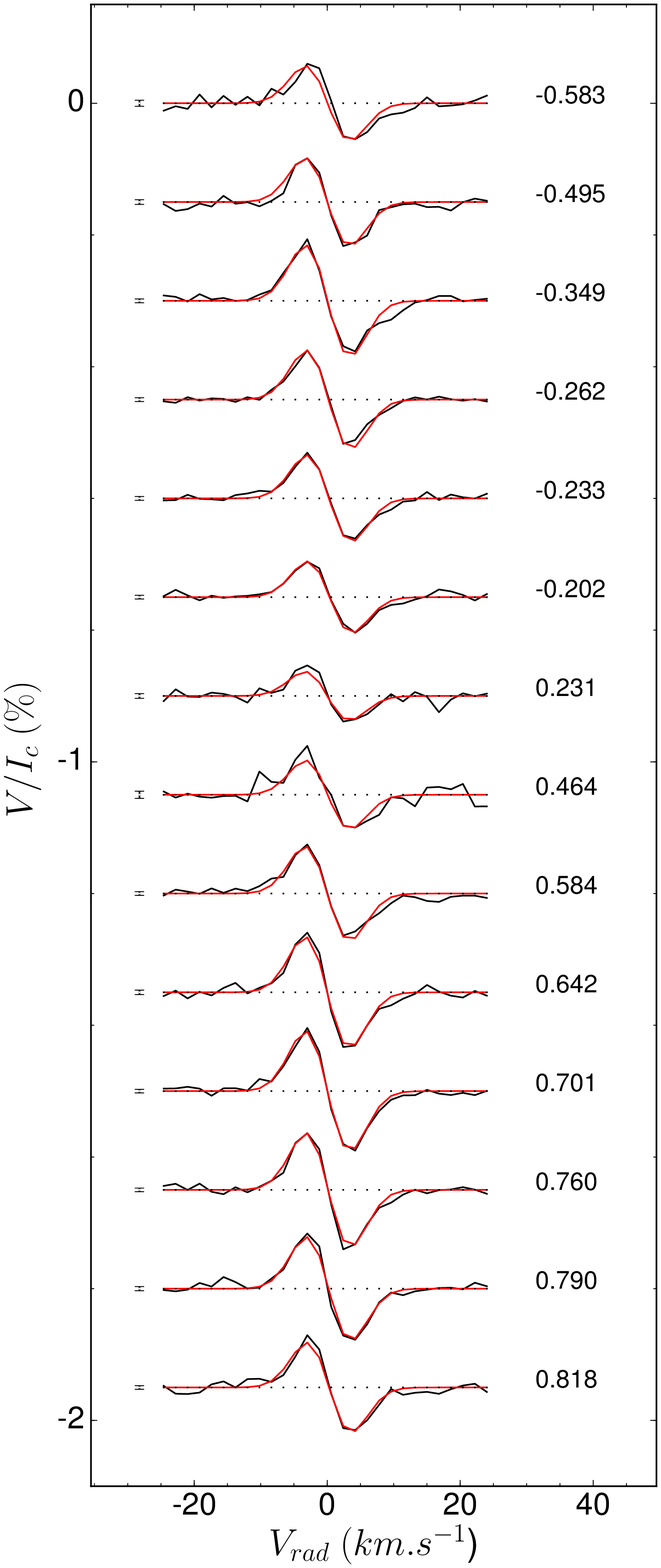}}&
  \subfloat[2015.54]{\includegraphics[scale = 0.25]{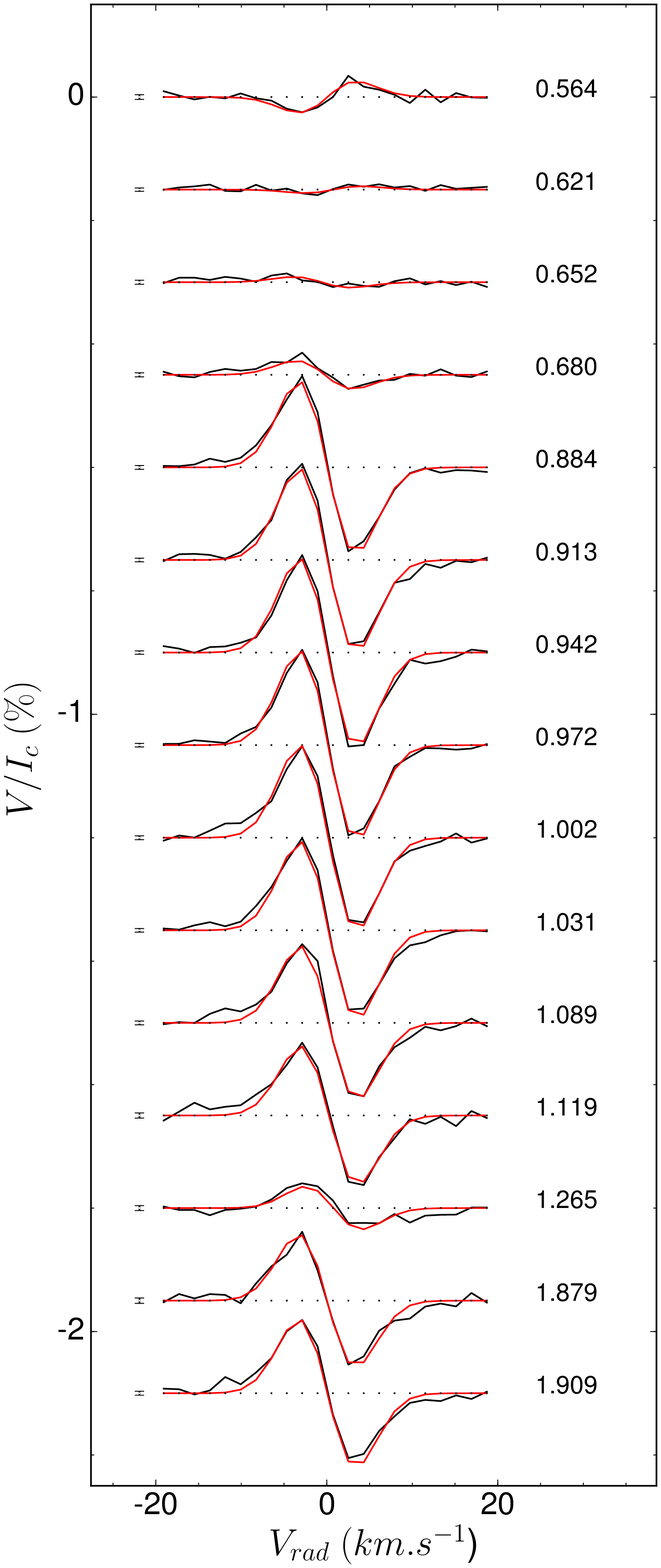}}\\
 \end{tabular}
\caption{Time series of LSD Stokes \textit{V} spectra of 61 Cyg A, where the  observed spectra is shown in black and the corresponding fit is shown in red. For each observed spectra the phase of the rotational cycle is shown on the 
right and on the left side of each plot 1 $\sigma$ error is shown. For clarity purpose successive Stokes \textit{V} profiles are shifted vertically and the scale is expanded by a factor of 2.}
\label{lsdtimeseries}
\end{figure*}

 \begin{figure*}
 \centering
\captionsetup[subfigure]{position=top,singlelinecheck=off} 
\captionsetup[subfigure]{labelformat=empty}
 \begin{tabular}{ccc}
  \subfloat[2007.59]{\includegraphics[scale = 0.33]{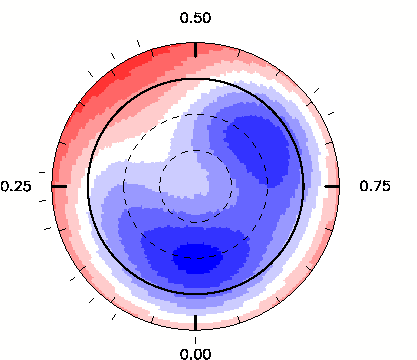}}&
\subfloat[] {\includegraphics[scale = 0.33]{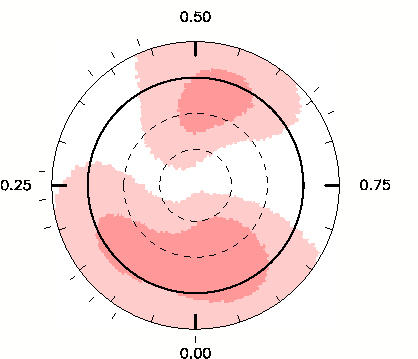}}&
\subfloat[] {\includegraphics[scale = 0.33]{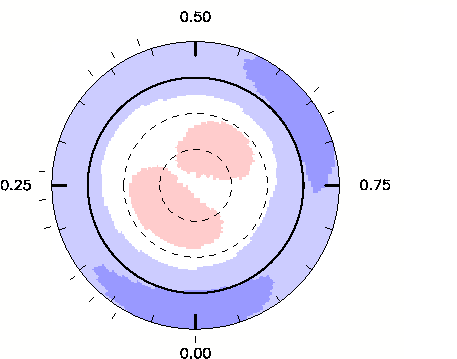}}\\
   \subfloat[2008.64]{\includegraphics[scale = 0.33]{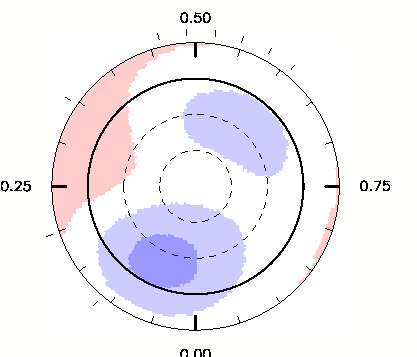}}&
\subfloat[]{\includegraphics[scale = 0.33]{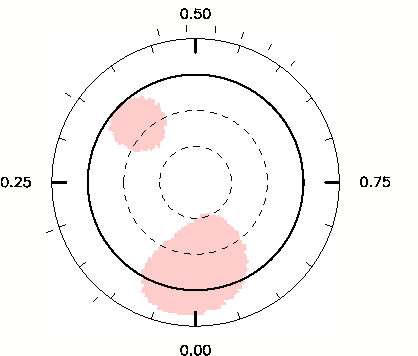}}&
\subfloat[] {\includegraphics[scale = 0.33]{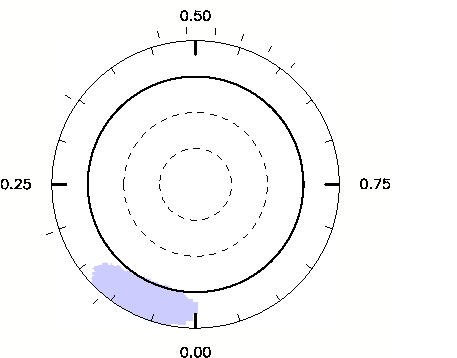}}\\
   \subfloat[2010.55]{\includegraphics[scale = 0.33]{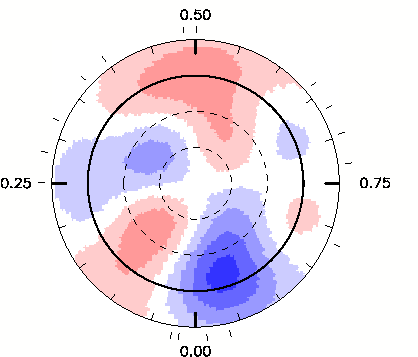}}&
\subfloat[]{\includegraphics[scale = 0.33]{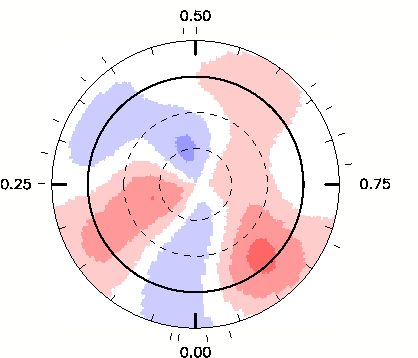}}&
\subfloat[] {\includegraphics[scale = 0.33]{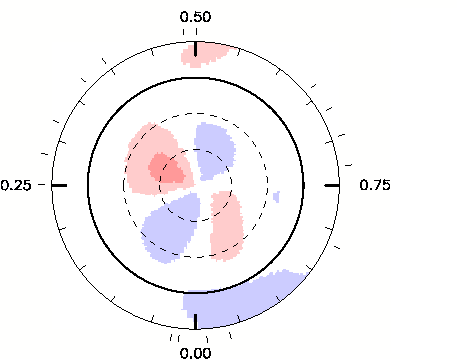}}\\
 \subfloat[2013.61]{\includegraphics[scale = 0.33]{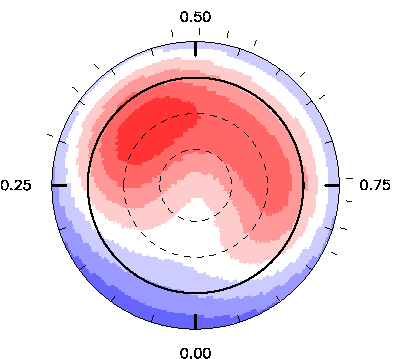}}&
\subfloat[]{\includegraphics[scale = 0.33]{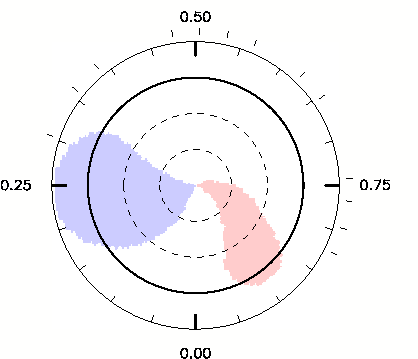}}&
\subfloat[] {\includegraphics[scale = 0.33]{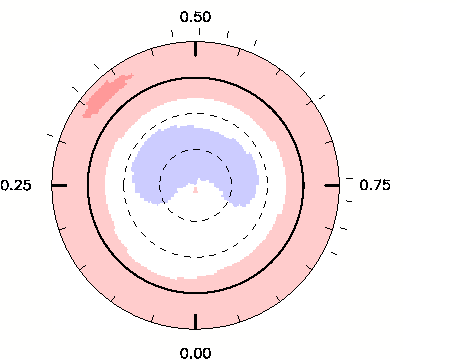}}\\
  \subfloat[2014.61]{\includegraphics[scale = 0.33]{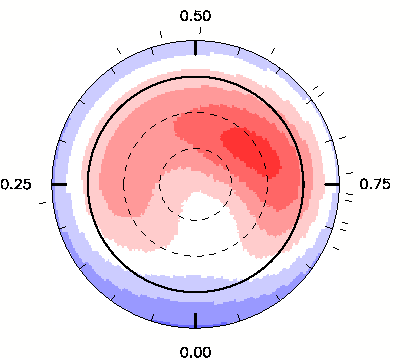}}&
\subfloat[]{\includegraphics[scale = 0.33]{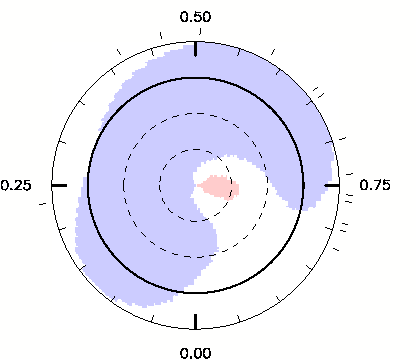}}&
\subfloat[] {\includegraphics[scale = 0.33]{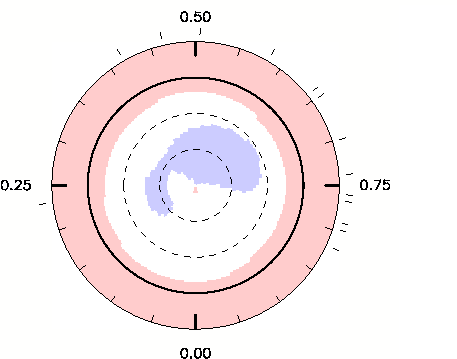}}\\
  \subfloat[2015.54]{\includegraphics[scale = 0.33]{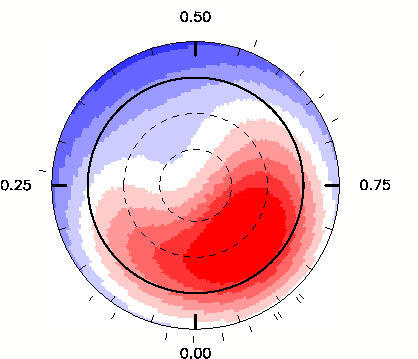}}&
\subfloat[] {\includegraphics[scale = 0.33]{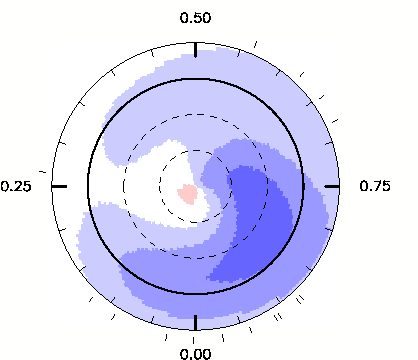}}&
\subfloat[]{\includegraphics[scale = 0.33]{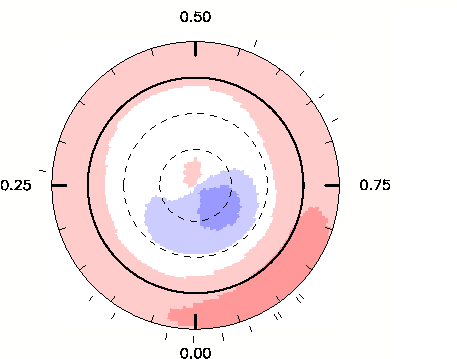}}\\
 \end{tabular}
\includegraphics[scale=0.32]{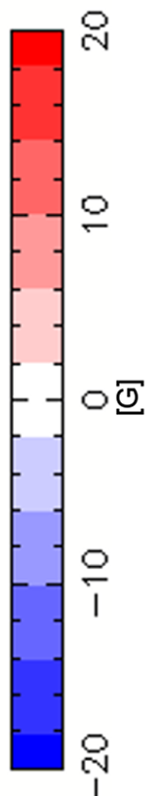}
 \caption{Surface magnetic geometry of 61 Cyg A over six observational epochs (2007.59, 2008.64, 2010.55, 2013.61, 2014.61 and 2015.54). The three columns represent the radial, azimuthal, and the meridional component of the large-scale 
field. The star is shown in flattened polar projection down to latitudes of -30${^\circ}$, where the dark line depicts the stellar equator.}
\label{radial_maps}
 \end{figure*}

\paragraph{}
For 61 Cyg A we use a stellar model of 5000 grid points, where each grid point is associated with both a local Stokes \textit{V} and Stokes \textit{I} profile. A solar-type limb darkening law is used in the stellar model and synthetic Stokes \textit{V} profiles 
are computed for all rotational phases. Each of these local Stokes \textit{V} profiles are calculated under the weak field assumption and uniform Stokes \textit{I} profile is assumed over the entire photosphere. Using maximum entropy as a regularisation 
scheme \citep{skillingandbryan84}, ZDI reconstructs  $\alpha_{l,m}$, $\beta_{l,m}$, and $\gamma_{l,m}$ by iteratively comparing the synthetic Stokes \textit{V} profiles to the observed Stokes \textit{V} profiles.

\subsection{Vector magnetic field}
61 Cyg A exhibits chromospheric variability on rotational timescales, indicating a high stellar inclination. We choose an inclination angle of 70$^\circ$ which together with a rotation period of 34.2 days, a stellar radius of 
0.665 $R_{\sun}$ results in a $vsini$ of 0.92 $kms^{-1}$. This $vsin i$ is consistent with the  $v_e$ of 0.94 kms$^{-1}$ mentioned in Section 2. Hence, the $v\sin i$ of 0.92 kms$^{-1}$ and an inclination angle of 70$^{\circ}$ was used in
our magnetic field reconstructions. We tested a range of spherical harmonics orders $l_\mathrm{max}$={5,7,9,11} to reconstruct the magnetic maps. For all epochs of our analysis except 2010.55 and 2015.54, the  target $\chi^2$ was 
achieved for $l_\mathrm{max}$ = 7. Using $l_\mathrm{max}\geq7$ does not improve the $\chi^2$ or the fit to the spectra at epochs 2007.59, 2008.64, 2013.61 and 2014.61. However for epoch 2010.55 and 2015.54 the best $\chi^2$ fit 
was obtained for $l_\mathrm{max}$ $\geq$ 11. To make comparisons between the epochs we thus kept the $l_\mathrm{max}$ to 11 throughout our analysis. We also include differential rotation in the large-scale field reconstructions (see Section 5.3 for more details).
\paragraph{}
Fig \ref{lsdtimeseries} shows the Stokes \textit{V} LSD time series of 61 Cyg A for our observational time-span. The fit between the modelled Stokes \textit{V} profiles and observed Stokes \textit{V} profiles was obtained with a reduced $\chi^2$ of 1.0 
for epoch 2007.59. A reduced $\chi^2$ of 1.1 was achieved for epochs 2008.64, 2013.61 and 2014.61. For epochs 2010.55 and 2015.54 the observed fit was obtained with a reduced $\chi^2$ of 1.2 and 1.5 respectively. It is evident from 
Fig \ref{lsdtimeseries} that during epoch 2007.59, which is around activity minimum, the time series of Stokes \textit{V} profiles is dominated by a single sign. The Stokes \textit{V} profiles in epoch 2010.55 are not dominated by a single orientation. 
The high variability shown by the Stokes \textit{V} profile in 2010.55 is caused by a complex magnetic field geometry. This complexity of the field is likely to make the impact of differential rotation more apparent in this epoch.  In the 
following three epochs (2013.61, 2014.61, and 2015.54) the Stokes \textit{V} profiles flips sign and are dominated again by a single orientation. This evolution of the Stokes \textit{V} time series indicates that the surface magnetic field geometry is 
strongly evolving with single polarity field during low activity and a more complex field during strong activity.
\paragraph{}
Since the maximum entropy reconstruction does not allow error calculations of our magnetic maps we employ the technique described by \citet{petit08}, where a range of magnetic maps were calculated for the wide range of input stellar 
parameters used in our ZDI code. These input parameters are varied within their error bars and the resulting dispersion is taken as error bars as shown in Table \ref{magnetic_energy}.

\subsubsection{Epoch 2007.59}
The large-scale surface magnetic field of 61 Cyg A at epoch 2007.59 exhibits a simple dipolar magnetic geometry. Fig \ref{radial_maps} shows this large-scale magnetic field geometry reconstructed into three components: radial,
azimuthal and meridional. The radial component of the magnetic field exhibits strong negative polarity magnetic regions at higher latitudes and positive polarity magnetic field at lower latitudes. On the other 
hand the azimuthal field component is dominated by two positive polarity magnetic field regions between the pole and lower latitudes. The $B_\mathrm{mean}$, which is the average of the surface magnetic field from the reconstructed maps, 
is 12$\pm$3 G. The poloidal magnetic field dominates, accounting for 93$\pm$5 $\%$ of the magnetic energy seen in Stokes \textit{V}. The fraction of the magnetic energy in other field components is shown in Table \ref{magnetic_energy}. As shown 
in Fig \ref{benergy}, almost 100$\%$ of the poloidal magnetic energy is stored in the lower order spherical harmonics modes, $l\leq$ 3. 
\subsubsection{Epoch 2008.64}
In epoch 2008.64 the large-scale magnetic field in the radial component exhibits a similar  magnetic geometry as reconstructed for epoch 2007.59, shown in Fig \ref{radial_maps}. However, the mean magnetic field ($B_\mathrm{mean}
$) is weaker than in the previous epoch as shown in Table \ref{magnetic_energy}, where $B_\mathrm{mean}$ is 3$\pm$1 G. In the azimuthal field component the positive polarity magnetic regions seen in the previous epoch are almost 
non-existent as shown in Fig \ref{radial_maps}. The magnetic field is predominantly poloidal, constituting 92$\pm$2 $\%$ of the magnetic energy. The dipolar component of the poloidal field accounts for 56$\pm$5 $\%$ of the 
poloidal magnetic energy. Similar to the previous epoch, the lower order modes ($l\leq$ 3) accounts for almost 100$\%$ of the poloidal magnetic energy budget as shown in Fig \ref{benergy}.
\subsubsection{Epoch 2010.55}
The large-scale magnetic field of 61 Cyg A changes dramatically in epoch 2010.55. The high latitude negative polarity field reconstructed in epoch 2007.59 is absent from this epoch and the radial field component is dominated by a more 
complex magnetic geometry between the equator and the poles as shown in Fig \ref{radial_maps}. The azimuthal field component also exhibits a more complex geometry compared to previous epochs. The averaged $B_\mathrm{mean}$ for epoch 
2010.55 is 5$\pm$2 G, which is still weaker than epoch 2007.59 as shown in Table \ref{magnetic_energy}. The poloidal field dominates and constitutes 87$\pm$3 $\%$ of the magnetic energy. Only 19$\pm$6 $\%$ of the poloidal field is 
dipolar, 21$\pm$5 $\%$ is quadrupolar and 44$\pm$13 $\%$ of the poloidal field is reconstructed into its octopolar component as shown in Table \ref{magnetic_energy}. The field is also the least axisymmetric in this epoch. 
Unlike previous epochs, a few $\%$ of the poloidal magnetic field is also distributed in the higher order spherical harmonics modes as shown in Fig \ref{benergy}. The percentage of the magnetic field reconstructed into its toroidal
component is higher at this epoch when compared to the previous epochs.
\begin{table*}
\caption{Magnetic properties of 61 Cyg A extracted from the ZDI maps. The columns represent fractional dates, number of observations, radial velocity ($v_r$), mean magnetic field strength (B$_\mathrm{mean}$), fraction of magnetic energy
reconstructed as the poloidal component, fraction of poloidal magnetic field stored as dipole, quadrupole, and octopole modes. Finally it also shows the fraction of the total magnetic energy in the axisymmetric component of the magnetic 
field, and the poloidal axisymmetric fraction (for $m\leq\frac{l}{2}$). The differential rotation parameters $\Omega_\mathrm{eq}$ and d$\Omega$ is also shown in the last two columns.}             
\label{magnetic_energy}      
\centering                                      
\begin{tabular}{c c c c c c c c c c c c c}          
\hline                        
Dates&no of&$v_{r}$ &$B_\mathrm{mean}$& poloidal&dipole&quad&oct&axi&axi&$\Omega_\mathrm{eq}$&d$\Omega$\\
&obs&(km s$^{-1}$)&(G)&($\%$tot)&($\%$pol)&($\%$pol)&($\%$pol)&($\%$tot)&($\%$pol)&(rad d$^{-1}$)&(rad d$^{-1}$)\\\\
\hline                                   
2007.59&12&-65.47$\pm$0.05&12$\pm$3& 93$\pm$5&77$\pm$9&17$\pm$7&5$\pm$3&80$\pm$13&80$\pm$17&&\\
2008.64&9&-65.48$\pm$0.05&3$\pm$1&92$\pm$2&56$\pm$5&36$\pm$5&8$\pm$2&56$\pm$10&53$\pm$11&&\\
2010.55&16&-65.48$\pm$0.05&5$\pm$2&87$\pm$3&19$\pm$6&21$\pm$5&44$\pm$13&7$\pm$2&2$\pm$4&0.18$\pm$0.03&0.04$\pm$0.02\\
2013.61&13&-65.49$\pm$0.05&9$\pm$6&99${^{+1}_{-3}}$&80$\pm$9&16$\pm$7&3$\pm$1&77$\pm$6&77$\pm$6&&\\
2014.61&14&-65.50$\pm$0.05&8$\pm$4&93$\pm$4&78$\pm$8&18$\pm$6&4$\pm$3&81$\pm$7&81$\pm$9&&\\
2015.54&15&-65.52$\pm$0.05&12$\pm$5&87$\pm$5&85$\pm$9&12$\pm$6&2$\pm$1&59$\pm$12&55$\pm$20&&\\
    \hline                                             
\end{tabular}
\end{table*} 

\subsubsection{Epoch 2013.61}
The large-scale magnetic field geometry at epoch 2013.61 is shown in Fig \ref{radial_maps}, where higher latitude magnetic regions are reconstructed in the radial field component with opposite polarity as seen in epochs 2007.59 and 
2008.64. The azimuthal field component exhibits mixed polarity but it is weak and almost non-existent as shown in Fig \ref{radial_maps}.  The mean magnetic field derived from the magnetic maps, ($B_\mathrm{mean}$ = 9$\pm$6 G) is 
stronger than the previous two epochs. The poloidal component comprises of 99${^{+1}_{-3}}$ $\%$ of the magnetic energy as shown in Table \ref{magnetic_energy}. The poloidal magnetic field is 80$\pm$9 $\%$ dipolar. The distribution of 
the magnetic field is primarily concentrated at the lower spherical harmonics modes (dipolar, quadrupolar, and octopolar) as shown in Fig \ref{benergy}.
\subsubsection{Epoch 2014.61}
\begin{figure}
 \centering
 \includegraphics[scale= 0.3]{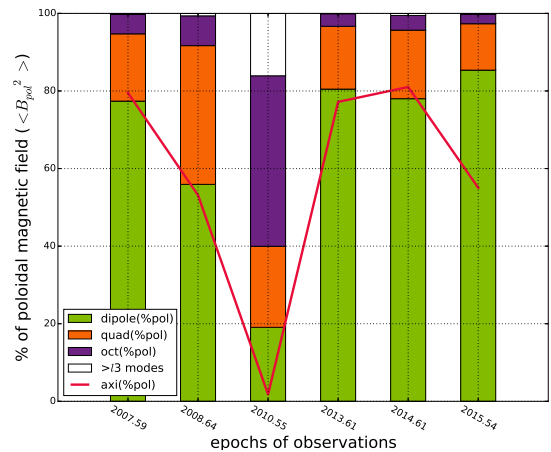}
 \caption{Poloidal magnetic field distribution of 61 Cyg A for six epoch of observations. }
 \label{benergy}
 \end{figure}

The strong positive field in epoch 2013.61 is also reconstructed in the radial component in epoch 2014.61 as shown in Fig \ref{radial_maps}. The azimuthal field as shown in Fig \ref{radial_maps} is reconstructed as a band 
of negative polarity magnetic field at the surface of the star, where the field is of opposite polarity as seen in epoch 2007.59. Table \ref{magnetic_energy} shows the mean magnetic field $B_\mathrm{mean}$ of epoch 2014.61, where 
the field strength is 8$\pm$4 G. The fraction of the magnetic field distributed into the different field components is shown in Table \ref{magnetic_energy}, where 93$\pm$4 $\%$ of the magnetic field is reconstructed into its 
poloidal component. 78$\pm$8 $\%$ of the poloidal field is in its dipolar component. The magnetic energy budget is spread across the lower spherical harmonics mode $l\leq$ 3, as shown in Fig \ref{benergy}.
\subsubsection{Epoch 2015.54}
During epoch 2015.54 the radial field component exhibits a positive polarity magnetic region between the poles and the equator together with a negative polarity magnetic field at the equator as shown in Fig \ref{radial_maps}. A band of 
negative polarity magnetic field is reconstructed in the azimuthal component of the magnetic field as shown in Fig \ref{radial_maps}. The mean magnetic field $B_\mathrm{mean}$ of 12$\pm$5 G is similar to what was measured in 
epoch 2007.59. 87$\pm$5 $\%$ of the magnetic field is in the poloidal component as shown in Table \ref{magnetic_energy}. The percentage of the poloidal field reconstructed into dipolar component is 85$\pm$9 $\%$ and  as shown in Fig 
\ref{benergy} the poloidal field is distributed across the lower spherical harmonics modes. 
\subsection{Evolution of the different multipolar modes over the magnetic cycle}
\begin{figure}
\centering
\includegraphics[scale=0.45]{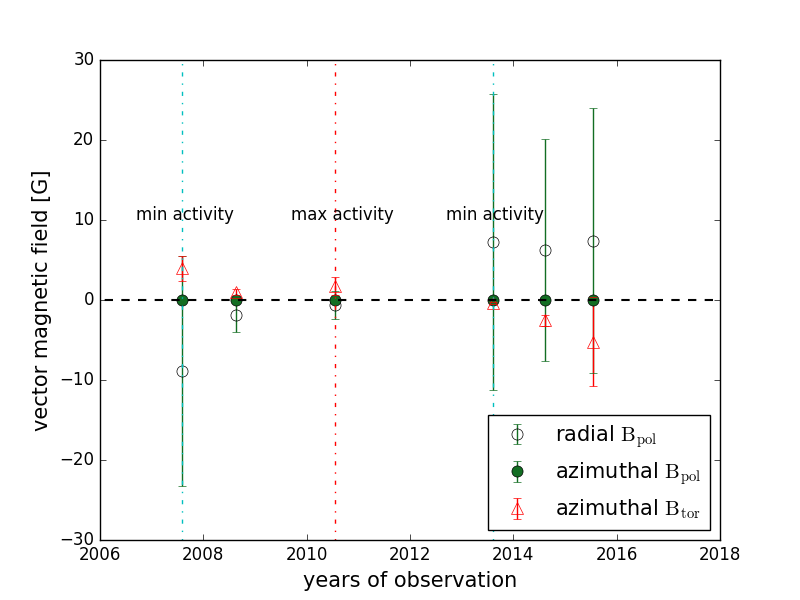}
\caption{Plot showing the vector component of the poloidal and toroidal magnetic field strength averaged over the stellar surface. The field strengths in the radial and azimuthal components are shown. The horizontal dashed line shows 
where the field strength is 0. The cyan vertical dashed lines show the epochs with minimum activity and the red vertical dashed line shows the epoch with maximum chromospheric activity.}
\label{Bwhat}
\end{figure}

\begin{figure}
\centering
\includegraphics[scale=0.45]{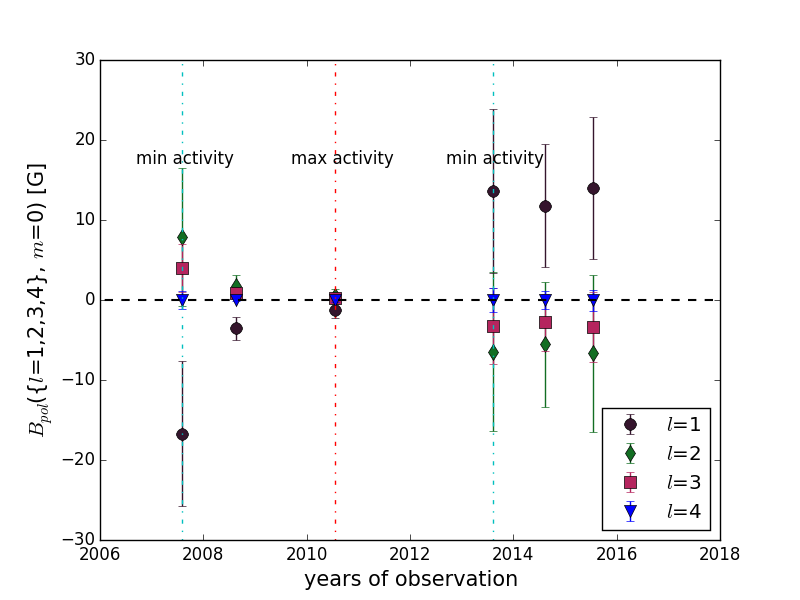}
\caption{Magnetic field strength B$_\mathrm{r}$ at the magnetic pole for the different axisymmetric modes ($l=\{$1,2,3$\}$). The horizontal dashed line represents B$_\mathrm{r}=0$, and the vertical dashed lines represents the epochs 
with minimum and maximum activity as shown in Fig \ref{Bwhat}.}
\label{Br}
\end{figure}

\begin{figure}
\centering
\includegraphics[scale=0.45]{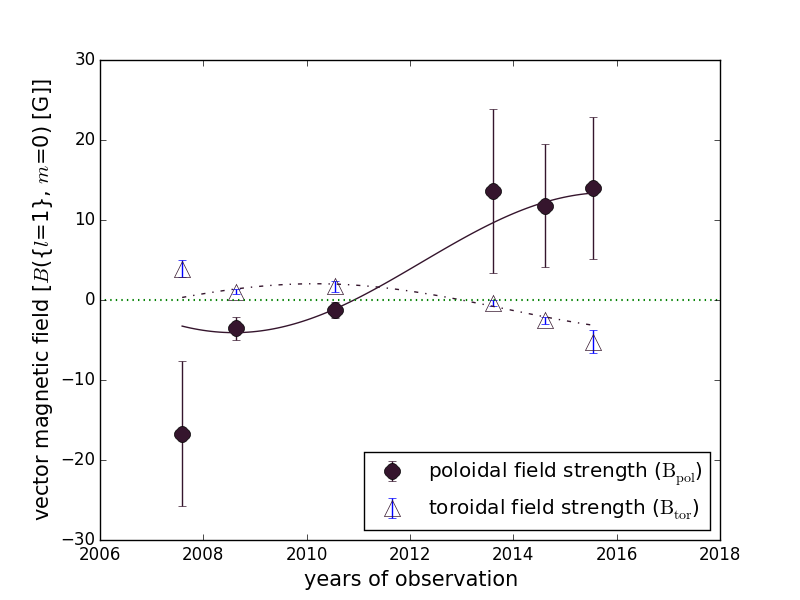}
\caption{Magnetic field strength B at the magnetic pole for the axisymmetric \textit{l}=1 mode for the poloidal component, which is similar to \textit{l}=1 in Fig \ref{Br}. For the toroidal component B represents the strength of the 
latitudinal or azimuthal field. The green horizontal dashed line represents B=0. The fits are obtained with a magnetic cycle period of 14.4 years that is twice the activity cycle period of 7.2 years.}
\label{Bl0}
\end{figure}

In order to assess the long-term variability of the magnetic field of 61 Cyg A, we consider several ways of parameterising the reconstructed magnetic field. First we measure the vector component of the magnetic field for each epoch, 
averaged over the stellar surface. The vector radial field component, as mentioned in earlier in this Section, is entirely poloidal in nature and the signed value of this component is shown in Fig \ref{Bwhat}. The field is averaged over 
the stellar surface for co-latitudes less than 30, which is close to the rotational pole. Here co-latitudes are surface co-latitudes relative to the stellar rotational pole in spherical co-ordinates. The vector azimuthal field 
on the other hand, comprises of both poloidal and toroidal field components in this case. Hence the signed azimuthal poloidal field and the signed azimuthal toroidal field is treated separately and calculated by averaging over the surface for co-latitudes less than 30. As seen from Fig \ref{Bwhat}, the poloidal field strength $B_\mathrm{pol}$ in the radial component is the weakest in epoch 2010.55 and is stronger during epochs of low chromospheric activity. The 
poloidal component of the azimuthal field is almost negligible for the total azimuthal field. From Fig \ref{Bwhat} it is clear that the poloidal field strength on the stellar surface, for co-latitudes $\leq 30^\circ$, is mostly radial. 
The toroidal field strength $B_\mathrm{tor}$ exhibits weak anti correlation to the radial field strength $B_\mathrm{pol}$.
\paragraph{}
Since the axisymmetric modes are of greater interest for existing dynamo models, we investigate the time evolution of both even and odd axisymmetric modes ($m=0$) for $l$= {1,2,3,4}. Fig \ref{Br} shows the field strength, at the pole, 
of the axisymmetric modes of the poloidal field, for the first 4 orders of spherical harmonics. The poloidal field strength in the axisymmetric modes is the signed (radial) field strength at the visible rotational pole of the star. The 
first odd mode \textit{l}=1, which is the dipolar mode, exhibits similar behaviour to the full averaged $B_\mathrm{pol}$ in Fig \ref{Bwhat}. The quadrupolar \textit{l}=2 mode is anti correlated with the dipolar mode. The 
octopolar mode (\textit{l}=3) exhibits correlation with the quadrupolar mode. 
\paragraph{}
Fig \ref{Bl0} shows the field strength of only the lowest order ($l=1$) axisymmetric ($m=0$) poloidal and toroidal modes. The poloidal field strength is the same as shown in Fig \ref{Br} but only the dipolar mode is plotted. The 
evolution of the dipolar field is consistent with the polarity reversals of the large-scale radial field shown in Fig \ref{radial_maps} and shows strong correlation with the averaged radial poloidal field strength in Fig 
\ref{Bwhat}. The signed toroidal field strength is calculated by measuring the strength of the azimuthal component of the toroidal ($l=0, m=0$) field, evaluated at the maximum of the toroidal band. The toroidal field strength is 
anti-correlated with the poloidal field strength in Fig \ref{Bl0}.

\paragraph{}
The uncertainties were estimated using the same technique as described in Section 5.1.
\subsection{Differential rotation}
\begin{figure}
 \centering
 \includegraphics[scale=0.5]{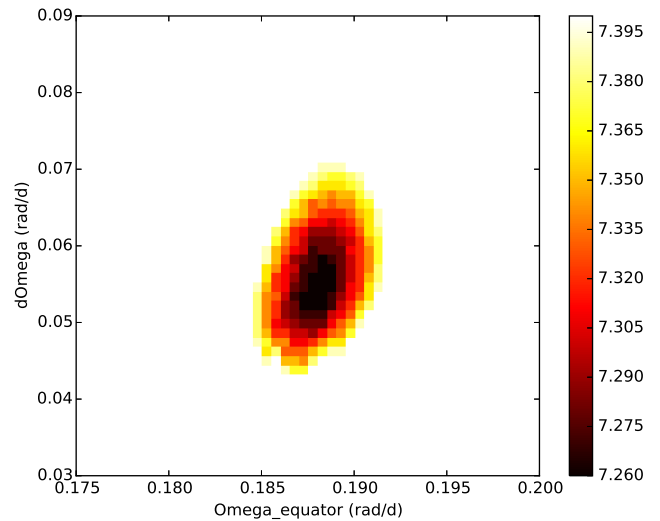}
 \caption{Best fit $B_\mathrm{mean}$ map obtained by varying the differential parameters for epoch 2010.55. The $\Omega_\mathrm{eq}$ and d$\Omega$ values obtained from this map are  0.18 rad d$^{-1}$ and 0.04 rad d$^{-1}$ respectively.}
 \label{diffrot14}
\end{figure}

Each epoch spanned over multiple weeks, hence we investigate the effect of differential rotation on our data. In order to investigate the differential rotation parameters of 61 Cyg A, we incorporated a simple solar-type differential 
rotation law into our ZDI code as described by \citet{petit02},

\begin{equation}
\mathrm{\Omega}(l) = \mathrm{\Omega_{eq}-d\Omega \sin^2}l
\label{diffrot}
\end{equation}
where $\mathrm{\Omega}(l)$ represents the stellar rotation at latitude \textit{l}, $\mathrm{\Omega_{eq}}$ is the equatorial rotation rate and $\mathrm{d\Omega}$ is the difference between the rotation rate at the equator and the 
poles. 
\paragraph{}
We set the magnetic field strength to a fixed value and carry out ZDI reconstruction for a wide range of differential rotation parameters $\Omega_\mathrm{eq}$ and $\mathrm{d\Omega}$ in a 2D parameter space. The set of differential 
rotation parameters for which the $\chi^2$ minimum is achieved is then selected as shown in Fig \ref{diffrot14}. The uncertainties were calculated by the bootstrap technique, where the input stellar parameters were varied within their 
error bars and the differential rotation was measured for this wide range of input parameters. The resulting dispersion in the obtained values are considered as the associated uncertainties and is shown in Table \ref{magnetic_energy}.
\paragraph{}
To obtain reliable differential rotation parameters it is necessary to have observations with good phase coverage spread over multiple rotation periods of the star \citep{petit02,morgenthaler12}. Consequently, the differential 
parameters for 61 Cyg A could only be obtained for epoch 2010.55, as shown in Table \ref{magnetic_energy}. For epoch 2010.55 we measured a $\mathrm{\Omega_\mathrm{eq}}$=0.18 rad d$^{-1}$ and  $\mathrm{d\Omega}$ = 0.04 rad d$^{-1}$. We 
obtained a rotation period $P_{rot}$ of  34.2$\pm$3.7 days. This rotation period is in agreement with the rotation period obtained from chromospheric measurements as shown in Table \ref{table:1}. Using this $d\Omega$ and $\Omega$
value we determine the relative horizontal shear ($\alpha$) of 0.2 which is equivalent to the $\alpha_\sun = 0.2$.
\paragraph{}
Since, the phase coverage of the other epochs was sparse we could not obtain differential rotation measurements as reliable as in epoch 2010.55. The magnetic field geometry is also simple with single polarity dominating the surface in 
all epochs except 2010.55 making it harder to determine the differential rotation for those epochs. Hence, we adopted the differential rotation parameters of 2010.55 for the rest of the observational epochs.

\section{Long-term evolution of the magnetic and activity cycle}
\begin{figure}
 \centering
 \includegraphics[scale= 0.45]{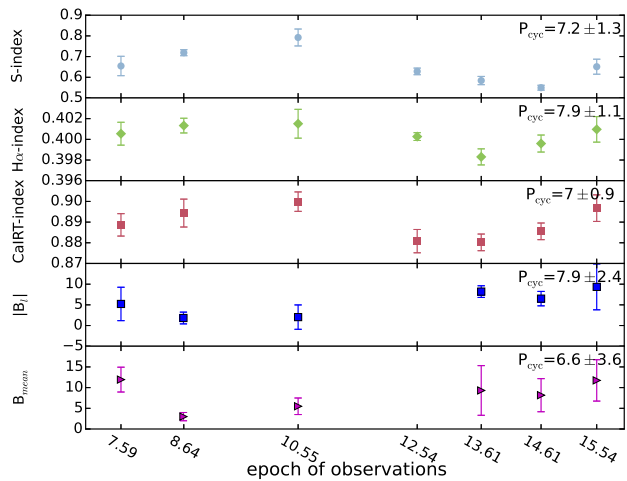}
\caption{Long-term evolution of the average values of the activity indices, the average longitudinal field, and the mean magnetic field measured from the ZDI maps.}
 \label{mean_activity}
 \end{figure}

The long-term evolution of the chromospheric indicators and the magnetic field strength is shown in Fig \ref{mean_activity}. The average value of the different chromospheric activity indicators exhibits good correlation over a long 
timescale. Fig \ref{mean_activity} also shows the mean longitudinal magnetic field of 61 Cyg A averaged over each epoch, where the error bars represent the standard deviation in $B_{l}$ per observational epoch. Strong temporal 
variations are detected for the average longitudinal field, with stronger field strength in epochs 2007.59, 2013.61, 2014.61 and 2015.54. The field strength is the weakest in epochs 2008.64 and 2010.55. However, it is interesting to 
note that epoch 2007.59 and 2010.55 exhibit stronger dispersion in $B_{l}$ compared to the other epochs. Long-term observations of 61 Cyg A exhibits a weak anti-correlation between the mean magnetic field strength and the different 
chromospheric activity indicators.
\paragraph{}
The mean magnetic field obtained from the reconstructed ZDI maps averaged over each epoch is also shown in Fig \ref{mean_activity}. $B_\mathrm{mean}$ follows a similar trend as $B_{l}$, however $B_\mathrm{mean}$ is weakest in epoch 2008.64 and does 
not show any pronounced anti-correlation with S-index.
\section{Discussion}
The large-scale magnetic field geometry of the solar-type K5 dwarf 61 Cyg A was reconstructed using ZDI for five observational epoch spread over nine years covering a complete chromospheric activity cycle. It exhibits polarity reversals 
of its large-scale magnetic field geometry. 
\subsection{Large-scale magnetic field}
\subsubsection{Polarity reversal of the large-scale field}
We report polarity reversals of the large-scale magnetic field of 61 Cyg A, where for the first time a strong correlation is detected between the magnetic cycle and the chromospheric activity cycle. Polarity reversals have been 
previously detected in the large-scale magnetic field reconstructions of several cool stars such as $\tau$ Boo, HD 190771 and HD 78366. However the polarity reversals detected in previous targets do not exhibit any 
correlation with the star's chromospheric activity cycle. \paragraph{}
The $vsin i$ of 61 Cyg A is comparatively lower than other ZDI targets. A previous numerical study by \citet{kochukhov02} has shown that although reconstructions of the surface geometry from spectropolarimetric data can be 
carried out for slow rotators, the resolution of the surface reconstruction goes down with lower $vsin i$. Since the $v sin i$ of this star is low we can only resolve down the surface elements into lower order spherical harmonics 
mode. In spite of this shortcoming, the polarity reversal of 61 Cyg A is clearly seen in the Stokes \textit{V} spectra itself. Fig \ref{lsdtimeseries} clearly demonstrates that the Stokes \textit{V} profiles change with rotational phase 
and also changes with the chromospheric activity cycle. As shown in Fig \ref{lsdtimeseries} the form of the Stokes \textit{V} profile in epoch 2007.59 flips in epoch 2013.61, indicating that the large-scale field also flips polarity. The lack of any dominant Stokes \textit{V} shape in epoch 2010.55 also indicates that the large-scale field is more complex during that epoch. The time series of Stokes \textit{V} profile reconfirms that the polarity reversals of the large-scale field are in fact 
independent of any artefact.
\subsubsection{Evolution of the multipolar modes of the large-scale field}
The magnetic field of 61 Cyg A's large-scale field is strongly poloidal at all epochs of our observations, which we quantify through energy fractions ($\textlangle{B^2}\textrangle$). Particularly, at epoch 2013.61, which is the epoch 
around which the field geometry flips polarity, the poloidal magnetic field constitutes to 99$\%$ of the total magnetic energy budget as seen in Stokes \textit{V}. During epoch 2010.55 which is coincidentally the chromospheric activity 
maximum, the percentage of the toroidal magnetic field increases compared to other epochs. The axisymmetry of the poloidal field component also reaches its minimum in epoch 2010.55, as shown in Fig \ref{benergy}. It indicates that 
the field is complex during activity maximum similar to the Sun. The poloidal field is dominated by dipolar component except during activity maximum, when the other higher order modes dominate. 
\paragraph{}
We investigate the lower order spherical harmonics modes of the star and interpret these results in the context of the Sun. Throughout our observational time span the vector magnetic field, averaged over the surface, is stronger in the 
poloidal component except during epoch 2010.55. During activity maximum the vector component of the both the poloidal and toroidal field averaged over the surface of the star is close to zero. During epochs of minimum activity a weak 
anti-correlation is detected between the poloidal and toroidal field. 
\paragraph{}
The low order axisymmetric modes of the poloidal field indicate that the dipolar component is strongest compared to the quadrupolar and octopolar modes throughout our observational timespan. The poloidal dipolar component (\textit{l}=1) 
is strongest during epochs of low activity and is almost negligible during activity maximum in epoch 2010.55. The other higher order modes are almost negligible throughout our observations. The large-scale solar magnetic field of the 
Sun is also more dipolar during epochs of low activity \citep{derosa12}. The quadrupolar component of the solar large-scale field dominates during activity maximum which is not seen in 61 Cyg A. One explanation for this disparity between
the modes of 61 Cyg A might be caused by cross-talk between the odd and even sets of modes. To investigate that we suppressed the even modes and measured the field coefficients again. Although the odd modes are favoured over the even 
modes, no improvements were detected. An alternate explanation might be that the ZDI reconstruction technique has different sensitivities to the different spherical harmonics.

\subsection{Chromospheric activity}
The chromospheric activity of 61 Cyg A was determined by using three activity proxies: the Ca II H$\&$K, H$\alpha$, and Ca II IRT lines. From the combined S-index data we measure a cycle period of 7.2 $\pm$1.3 years. An activity cycle 
period of 7.3$\pm$0.1 years was also previously determined by the long-term Mount Wilson survey \citep{baliunas95}.
\paragraph{}
A correlation is detected between the S-index and H$\alpha$-index of 61 Cyg A as shown in Fig \ref{corr_plot}. Such correlation is also observed in the Sun, where the correlation between the two indices follows the activity 
cycle \citep{livingston07}. Correlations were also observed for other BCool targets such as $\xi$ Boo A \citep{morgenthaler12} and HN Peg \citep{borosaikia15}. The correlation between the fluxes in Ca II H$\&$K lines and the H$\alpha$
lines was investigated by \citet{cincunegui07} for 109 cool stars. They observed strong dispersion in correlations with both positive and negative correlations. They also suggested that the correlation does not depend on activity 
and spectral type but on stellar colour.
\paragraph{}
The correlation between a star's S-index and H$\alpha$-index has been further studied by \citet{meunier09}, who suggest that one of the main factors affecting the correlation of these two indices is the filling factor of plages 
and filaments on the stellar surfaces. Since plages affect both Ca II $\&$K and H$\alpha$ lines, and filaments are mostly observed as absorption in H$\alpha$, they claim that the correlations depended strongly on the spatial and 
temporal distribution of filaments on the stellar surface. They also observe an activity dependence on the correlations where stronger correlation was observed near the cycle maxima. They also suggest that the correlations might also 
depend on time span of the observations and stellar inclination.
\paragraph{}
A correlation was also observed between 61 Cyg A's S-index and IRT-index as show in Fig \ref{corr_plot} and between H$\alpha$-index and IRT-index as shown in Fig \ref{corr_plot}. A stronger correlation is observed between S-index and 
IRT-index. This is not surprising as the Ca II IRT lines also arises from plage dominated areas on the stellar surface \citep{andretta05,busa07}. In the BCool snapshot study \citet{marsden14} also reports a good correlation between these
two indices. 
\paragraph{}
A weak anti-correlation was observed between the S-index and the longitudinal magnetic field. In previous ZDI targets no correlation was observed between the chromospheric activity and longitudinal magnetic field B$_\mathrm{l}$,
 e.g., in $\xi$ Boo A \citep{morgenthaler12} and HN Peg \citep{borosaikia15}. The lack of correlation is expected due to the cancellation of small scale magnetic features in the mean field measurements. However, for 61 Cyg A the 
longitudinal magnetic field B$_\mathrm{l}$ is weaker during activity maximum and stronger during activity minimum. No anti correlation was detected, though, between the chromospheric S-index and the mean magnetic field 
($B_\mathrm{mean}$) measured from the magnetic maps.  

\section{Summary}
In this paper we reconstruct the large-scale surface geometry of 61 Cyg A for six epochs over nine years of observations. We report the presence of a possible magnetic cycle which is twice the length of the activity cycle. This is the 
first detection of a cool star apart from the Sun where the magnetic cycle is in phase with its chromospheric cycle. The large-scale surface magnetic field geometry of 61 Cyg A flips its polarity and is highly variable with the radial 
field showing strong polar field during activity minimum and a complex field during activity maximum. Throughout our observations, the large-scale field of 61 Cyg A is strongly poloidal except during activity minimum when the poloidal 
field is relatively weaker. The dipolar component of the poloidal field is strongest close to activity minimum and is weakest during activity maximum. During activity maximum higher order modes such as quadrupolar and octopolar modes 
dominate over the dipolar mode. The evolution of the large-scale field of 61 Cyg A over the activity cycle shows close resemblance to the solar large-scale field, which has never been seen before in cool stars. 
\begin{acknowledgements}
 This work was carried out as part of Project A16 funded by  the Deutsche Forschungsgemeinschaft (DFG) under SFB 963. AAV acknowledges support from an Ambizione Fellowship of the Swiss National Science Foundation. CPF was supported by 
the grant ANR 2011 Blanc SIMI5-6 020 01 ``Toupies: Towards understanding the spin evolution of stars''. Part of the work was also supported by the COST action MP1104 \textit{Polarisation as a tool to study the Solar-system and beyond}. 
We also thank Dr Jan Robrade of Hamburger Sternwarte for his help in data analysis of the \textit{XMM-Newton} data. Finally we thank the anonymous referee for their detailed constructive comments.
\end{acknowledgements}
\bibliographystyle{aa}
 \bibliography{61cyga_arxiv}

\begin{thebibliography}{62}
\expandafter\ifx\csname natexlab\endcsname\relax\def\natexlab#1{#1}\fi

\bibitem[{{Andretta} {et~al.}(2005){Andretta}, {Bus{\`a}}, {Gomez}, \&
  {Terranegra}}]{andretta05}
{Andretta}, V., {Bus{\`a}}, I., {Gomez}, M.~T., \& {Terranegra}, L. 2005, \aap,
  430, 669

\bibitem[{{Auri{\`e}re}(2003)}]{auriere03}
{Auri{\`e}re}, M. 2003, in EAS Publications Series, Vol.~9, EAS Publications
  Series, ed. J.~{Arnaud} \& N.~{Meunier}, 105

\bibitem[{{Ayres}(2015)}]{ayres15}
{Ayres}, T.~R. 2015, \aj, 149, 58

\bibitem[{{Baliunas} {et~al.}(1995){Baliunas}, {Donahue}, {Soon}, {Horne},
  {Frazer}, {Woodard-Eklund}, {Bradford}, {Rao}, {Wilson}, {Zhang}, {Bennett},
  {Briggs}, {Carroll}, {Duncan}, {Figueroa}, {Lanning}, {Misch}, {Mueller},
  {Noyes}, {Poppe}, {Porter}, {Robinson}, {Russell}, {Shelton}, {Soyumer},
  {Vaughan}, \& {Whitney}}]{baliunas95}
{Baliunas}, S.~L., {Donahue}, R.~A., {Soon}, W.~H., {et~al.} 1995, \apj, 438,
  269

\bibitem[{{Barnes}(2007)}]{barnes07}
{Barnes}, S.~A. 2007, \apj, 669, 1167

\bibitem[{{Bessel}(1838)}]{bessel38}
{Bessel}, F.~W. 1838, \mnras, 4, 152

\bibitem[{{Boro Saikia} {et~al.}(2015){Boro Saikia}, {Jeffers}, {Petit},
  {Marsden}, {Morin}, \& {Folsom}}]{borosaikia15}
{Boro Saikia}, S., {Jeffers}, S.~V., {Petit}, P., {et~al.} 2015, \aap, 573, A17

\bibitem[{{Bus{\`a}} {et~al.}(2007){Bus{\`a}}, {Aznar Cuadrado}, {Terranegra},
  {Andretta}, \& {Gomez}}]{busa07}
{Bus{\`a}}, I., {Aznar Cuadrado}, R., {Terranegra}, L., {Andretta}, V., \&
  {Gomez}, M.~T. 2007, \aap, 466, 1089

\bibitem[{{Cincunegui} {et~al.}(2007){Cincunegui}, {D{\'{\i}}az}, \&
  {Mauas}}]{cincunegui07}
{Cincunegui}, C., {D{\'{\i}}az}, R.~F., \& {Mauas}, P.~J.~D. 2007, \aap, 469,
  309

\bibitem[{{DeRosa} {et~al.}(2012){DeRosa}, {Brun}, \& {Hoeksema}}]{derosa12}
{DeRosa}, M.~L., {Brun}, A.~S., \& {Hoeksema}, J.~T. 2012, \apj, 757, 96

\bibitem[{{Donati}(2003)}]{donati_esp03}
{Donati}, J.-F. 2003, in Astronomical Society of the Pacific Conference Series,
  Vol. 307, Solar Polarization, ed. J.~{Trujillo-Bueno} \& J.~{Sanchez
  Almeida}, 41

\bibitem[{{Donati} {et~al.}(2012){Donati}, {Gregory}, {Alencar}, {Hussain},
  {Bouvier}, {Dougados}, {Jardine}, {M{\'e}nard}, \& {Romanova}}]{donati12}
{Donati}, J.-F., {Gregory}, S.~G., {Alencar}, S.~H.~P., {et~al.} 2012, \mnras,
  425, 2948

\bibitem[{{Donati} {et~al.}(2006){Donati}, {Howarth}, {Jardine}, {Petit},
  {Catala}, {Landstreet}, {Bouret}, {Alecian}, {Barnes}, {Forveille},
  {Paletou}, \& {Manset}}]{donati06}
{Donati}, J.-F., {Howarth}, I.~D., {Jardine}, M.~M., {et~al.} 2006, \mnras,
  370, 629

\bibitem[{{Donati} {et~al.}(2008){Donati}, {Moutou}, {Far{\`e}s}, {Bohlender},
  {Catala}, {Deleuil}, {Shkolnik}, {Collier Cameron}, {Jardine}, \&
  {Walker}}]{donati08}
{Donati}, J.-F., {Moutou}, C., {Far{\`e}s}, R., {et~al.} 2008, \mnras, 385,
  1179

\bibitem[{{Donati} {et~al.}(1997){Donati}, {Semel}, {Carter}, {Rees}, \&
  {Collier Cameron}}]{donati97}
{Donati}, J.-F., {Semel}, M., {Carter}, B.~D., {Rees}, D.~E., \& {Collier
  Cameron}, A. 1997, \mnras, 291, 658

\bibitem[{{Duncan} {et~al.}(1991){Duncan}, {Vaughan}, {Wilson}, {Preston},
  {Frazer}, {Lanning}, {Misch}, {Mueller}, {Soyumer}, {Woodard}, {Baliunas},
  {Noyes}, {Hartmann}, {Porter}, {Zwaan}, {Middelkoop}, {Rutten}, \&
  {Mihalas}}]{duncan91}
{Duncan}, D.~K., {Vaughan}, A.~H., {Wilson}, O.~C., {et~al.} 1991, \apjs, 76,
  383

\bibitem[{{Fares} {et~al.}(2009){Fares}, {Donati}, {Moutou}, {Bohlender},
  {Catala}, {Deleuil}, {Shkolnik}, {Collier Cameron}, {Jardine}, \&
  {Walker}}]{fares09}
{Fares}, R., {Donati}, J.-F., {Moutou}, C., {et~al.} 2009, \mnras, 398, 1383

\bibitem[{{Fares} {et~al.}(2013){Fares}, {Moutou}, {Donati}, {Catala},
  {Shkolnik}, {Jardine}, {Cameron}, \& {Deleuil}}]{fares13}
{Fares}, R., {Moutou}, C., {Donati}, J.-F., {et~al.} 2013, \mnras, 435, 1451

\bibitem[{{Favata} {et~al.}(2008){Favata}, {Micela}, {Orlando}, {Schmitt},
  {Sciortino}, \& {Hall}}]{favata08}
{Favata}, F., {Micela}, G., {Orlando}, S., {et~al.} 2008, \aap, 490, 1121

\bibitem[{{Gizis} {et~al.}(2002){Gizis}, {Reid}, \& {Hawley}}]{gizis02}
{Gizis}, J.~E., {Reid}, I.~N., \& {Hawley}, S.~L. 2002, \aj, 123, 3356

\bibitem[{{G{\"u}del}(2004)}]{gudel04}
{G{\"u}del}, M. 2004, \aapr, 12, 71

\bibitem[{{Hall}(2008)}]{hall08}
{Hall}, J.~C. 2008, Living Reviews in Solar Physics, 5, 2

\bibitem[{{Hall} {et~al.}(2007){Hall}, {Lockwood}, \& {Skiff}}]{hall07}
{Hall}, J.~C., {Lockwood}, G.~W., \& {Skiff}, B.~A. 2007, \aj, 133, 862

\bibitem[{{Hathaway}(2010)}]{hathaway10}
{Hathaway}, D.~H. 2010, Living Reviews in Solar Physics, 7, 1

\bibitem[{{Hempelmann} {et~al.}(2006){Hempelmann}, {Robrade}, {Schmitt},
  {Favata}, {Baliunas}, \& {Hall}}]{hempelmann06}
{Hempelmann}, A., {Robrade}, J., {Schmitt}, J.~H.~M.~M., {et~al.} 2006, \aap,
  460, 261

\bibitem[{{Hempelmann} {et~al.}(1996){Hempelmann}, {Schmitt}, \& {St{\c
  e}pie{\'n}}}]{hempelmann96}
{Hempelmann}, A., {Schmitt}, J.~H.~M.~M., \& {St{\c e}pie{\'n}}, K. 1996, \aap,
  305, 284

\bibitem[{{Jefferies} \& {Thomas}(1959)}]{jefferies59}
{Jefferies}, J.~T. \& {Thomas}, R.~N. 1959, \apj, 129, 401

\bibitem[{{Jeffers} {et~al.}(2014){Jeffers}, {Petit}, {Marsden}, {Morin},
  {Donati}, \& {Folsom}}]{jeffers14}
{Jeffers}, S.~V., {Petit}, P., {Marsden}, S.~C., {et~al.} 2014, \aap, 569, A79

\bibitem[{{Kervella} {et~al.}(2008){Kervella}, {M{\'e}rand}, {Pichon},
  {Th{\'e}venin}, {Heiter}, {Bigot}, {ten Brummelaar}, {McAlister}, {Ridgway},
  {Turner}, {Sturmann}, {Sturmann}, {Goldfinger}, \& {Farrington}}]{kervella08}
{Kervella}, P., {M{\'e}rand}, A., {Pichon}, B., {et~al.} 2008, \aap, 488, 667

\bibitem[{{Kochukhov} {et~al.}(2010){Kochukhov}, {Makaganiuk}, \&
  {Piskunov}}]{kochukhov10}
{Kochukhov}, O., {Makaganiuk}, V., \& {Piskunov}, N. 2010, \aap, 524, A5

\bibitem[{{Kochukhov} \& {Piskunov}(2002)}]{kochukhov02}
{Kochukhov}, O. \& {Piskunov}, N. 2002, \aap, 388, 868

\bibitem[{{Livingston} {et~al.}(2007){Livingston}, {Wallace}, {White}, \&
  {Giampapa}}]{livingston07}
{Livingston}, W., {Wallace}, L., {White}, O.~R., \& {Giampapa}, M.~S. 2007,
  \apj, 657, 1137

\bibitem[{{Lockwood} {et~al.}(2007){Lockwood}, {Skiff}, {Henry}, {Henry},
  {Radick}, {Baliunas}, {Donahue}, \& {Soon}}]{lockwood07}
{Lockwood}, G.~W., {Skiff}, B.~A., {Henry}, G.~W., {et~al.} 2007, \apjs, 171,
  260

\bibitem[{{Lomb}(1976)}]{lomb76}
{Lomb}, N.~R. 1976, \apss, 39, 447

\bibitem[{{Malkov} {et~al.}(2012){Malkov}, {Tamazian}, {Docobo}, \&
  {Chulkov}}]{malkov12}
{Malkov}, O.~Y., {Tamazian}, V.~S., {Docobo}, J.~A., \& {Chulkov}, D.~A. 2012,
  \aap, 546, A69

\bibitem[{{Mamajek} \& {Hillenbrand}(2008)}]{mamajek08}
{Mamajek}, E.~E. \& {Hillenbrand}, L.~A. 2008, \apj, 687, 1264

\bibitem[{{Mann} {et~al.}(2013){Mann}, {Gaidos}, \& {Ansdell}}]{mann13}
{Mann}, A.~W., {Gaidos}, E., \& {Ansdell}, M. 2013, \apj, 779, 188

\bibitem[{{Marsden} {et~al.}(2014){Marsden}, {Petit}, {Jeffers}, {Morin},
  {Fares}, {Reiners}, {do Nascimento}, {Auri{\`e}re}, {Bouvier}, {Carter},
  {Catala}, {Dintrans}, {Donati}, {Gastine}, {Jardine}, {Konstantinova-Antova},
  {Lanoux}, {Ligni{\`e}res}, {Morgenthaler}, {Ram{\`i}rez-V{\`e}lez},
  {Th{\'e}ado}, {Van Grootel}, \& {BCool Collaboration}}]{marsden14}
{Marsden}, S.~C., {Petit}, P., {Jeffers}, S.~V., {et~al.} 2014, \mnras, 444,
  3517

\bibitem[{{Mengel} {et~al.}(2016){Mengel}, {Fares}, {Marsden}, {Carter},
  {Jeffers}, {Petit}, {Donati}, {Folsom}, \& {the BCool
  Collaboration}}]{mengel16}
{Mengel}, M.~W., {Fares}, R., {Marsden}, S.~C., {et~al.} 2016, ArXiv e-prints

\bibitem[{{Meunier} \& {Delfosse}(2009)}]{meunier09}
{Meunier}, N. \& {Delfosse}, X. 2009, \aap, 501, 1103

\bibitem[{{Morgenthaler} {et~al.}(2011){Morgenthaler}, {Petit}, {Morin},
  {Auri{\`e}re}, {Dintrans}, {Konstantinova-Antova}, \&
  {Marsden}}]{morgenthaler11}
{Morgenthaler}, A., {Petit}, P., {Morin}, J., {et~al.} 2011, Astronomische
  Nachrichten, 332, 866

\bibitem[{{Morgenthaler} {et~al.}(2012){Morgenthaler}, {Petit}, {Saar},
  {Solanki}, {Morin}, {Marsden}, {Auri{\`e}re}, {Dintrans}, {Fares}, {Gastine},
  {Lanoux}, {Ligni{\`e}res}, {Paletou}, {Ram{\'{\i}}rez V{\'e}lez},
  {Th{\'e}ado}, \& {Van Grootel}}]{morgenthaler12}
{Morgenthaler}, A., {Petit}, P., {Saar}, S., {et~al.} 2012, \aap, 540, A138

\bibitem[{{Perryman} {et~al.}(1997){Perryman}, {Lindegren}, {Kovalevsky},
  {Hoeg}, {Bastian}, {Bernacca}, {Cr{\'e}z{\'e}}, {Donati}, {Grenon},
  {Grewing}, {van Leeuwen}, {van der Marel}, {Mignard}, {Murray}, {Le Poole},
  {Schrijver}, {Turon}, {Arenou}, {Froeschl{\'e}}, \& {Petersen}}]{perryman97}
{Perryman}, M.~A.~C., {Lindegren}, L., {Kovalevsky}, J., {et~al.} 1997, \aap,
  323, L49

\bibitem[{{Petit} {et~al.}(2009){Petit}, {Dintrans}, {Morgenthaler}, {Van
  Grootel}, {Morin}, {Lanoux}, {Auri{\`e}re}, \&
  {Konstantinova-Antova}}]{petit09}
{Petit}, P., {Dintrans}, B., {Morgenthaler}, A., {et~al.} 2009, \aap, 508, L9

\bibitem[{{Petit} {et~al.}(2008){Petit}, {Dintrans}, {Solanki}, {Donati},
  {Auri{\`e}re}, {Ligni{\`e}res}, {Morin}, {Paletou}, {Ramirez Velez},
  {Catala}, \& {Fares}}]{petit08}
{Petit}, P., {Dintrans}, B., {Solanki}, S.~K., {et~al.} 2008, \mnras, 388, 80

\bibitem[{{Petit} {et~al.}(2002){Petit}, {Donati}, \& {Collier
  Cameron}}]{petit02}
{Petit}, P., {Donati}, J.-F., \& {Collier Cameron}, A. 2002, \mnras, 334, 374

\bibitem[{{Pevtsov} {et~al.}(2003){Pevtsov}, {Fisher}, {Acton}, {Longcope},
  {Johns-Krull}, {Kankelborg}, \& {Metcalf}}]{pevtsov03}
{Pevtsov}, A.~A., {Fisher}, G.~H., {Acton}, L.~W., {et~al.} 2003, \apj, 598,
  1387

\bibitem[{{Reiners} \& {Basri}(2006)}]{reiners06}
{Reiners}, A. \& {Basri}, G. 2006, \apj, 644, 497

\bibitem[{{Robinson}(1980)}]{robinson80}
{Robinson}, Jr., R.~D. 1980, \apj, 239, 961

\bibitem[{{Robrade} {et~al.}(2012){Robrade}, {Schmitt}, \&
  {Favata}}]{robrade12}
{Robrade}, J., {Schmitt}, J.~H.~M.~M., \& {Favata}, F. 2012, \aap, 543, A84

\bibitem[{{Saar}(1996)}]{saar96}
{Saar}, S.~H. 1996, in IAU Symposium, Vol. 176, Stellar Surface Structure, ed.
  K.~G. {Strassmeier} \& J.~L. {Linsky}, 237

\bibitem[{{Sanderson} {et~al.}(2003){Sanderson}, {Appourchaux}, {Hoeksema}, \&
  {Harvey}}]{sanderson03}
{Sanderson}, T.~R., {Appourchaux}, T., {Hoeksema}, J.~T., \& {Harvey}, K.~L.
  2003, Journal of Geophysical Research (Space Physics), 108, 1035

\bibitem[{{Sanz-Forcada} {et~al.}(2013){Sanz-Forcada}, {Stelzer}, \&
  {Metcalfe}}]{sanz-forcada13}
{Sanz-Forcada}, J., {Stelzer}, B., \& {Metcalfe}, T.~S. 2013, \aap, 553, L6

\bibitem[{{Scargle}(1982)}]{scargle82}
{Scargle}, J.~D. 1982, \apj, 263, 835

\bibitem[{{Schr{\"o}der} {et~al.}(2013){Schr{\"o}der}, {Mittag}, {Hempelmann},
  {Gonz{\'a}lez-P{\'e}rez}, \& {Schmitt}}]{schroeder13}
{Schr{\"o}der}, K.-P., {Mittag}, M., {Hempelmann}, A.,
  {Gonz{\'a}lez-P{\'e}rez}, J.~N., \& {Schmitt}, J.~H.~M.~M. 2013, \aap, 554,
  A50

\bibitem[{{Semel}(1989)}]{semel89}
{Semel}, M. 1989, \aap, 225, 456

\bibitem[{{Skilling} \& {Bryan}(1984)}]{skillingandbryan84}
{Skilling}, J. \& {Bryan}, R.~K. 1984, \mnras, 211, 111

\bibitem[{{Takeda} {et~al.}(2007){Takeda}, {Ford}, {Sills}, {Rasio}, {Fischer},
  \& {Valenti}}]{takeda07}
{Takeda}, G., {Ford}, E.~B., {Sills}, A., {et~al.} 2007, \apjs, 168, 297

\bibitem[{{Valenti} \& {Fischer}(2005)}]{valentifischer05}
{Valenti}, J.~A. \& {Fischer}, D.~A. 2005, \apjs, 159, 141

\bibitem[{{Wilson}(1978)}]{wilson78}
{Wilson}, O.~C. 1978, \apj, 226, 379

\bibitem[{{Wright} {et~al.}(2004){Wright}, {Marcy}, {Butler}, \&
  {Vogt}}]{wright04}
{Wright}, J.~T., {Marcy}, G.~W., {Butler}, R.~P., \& {Vogt}, S.~S. 2004, \apjs,
  152, 261

\bibitem[{{Zechmeister} \& {K{\"u}rster}(2009)}]{zechmeister09}
{Zechmeister}, M. \& {K{\"u}rster}, M. 2009, \aap, 496, 577

\end{thebibliography}
\end{document}